\documentclass[onecolumn,noshowpacs,superscriptaddress,nobibnotes,nofootinbib,12pt,showkeys,preprintnumbers]{revtex4-1}
\UseRawInputEncoding
\usepackage{float}
\usepackage{amsmath,amssymb}
\usepackage{bm}
\usepackage{bbold}
\usepackage{slashed}
\usepackage{epsfig}
\usepackage{hyperref}
\usepackage{xcolor}
\usepackage{caption}
\usepackage{subcaption}
\usepackage{ragged2e}

\hypersetup{
    colorlinks=true,
    linkcolor=blue,
    filecolor=magenta,      
    urlcolor=blue,
    citecolor=blue
}
\usepackage[compat=1.1.0]{tikz-feynman}
\usepackage{physics}

\usepackage{cleveref}
\crefname{section}{Sec.}{Secs.}
\crefname{paragraph}{Sec.}{Secs.}
\crefname{figure}{Fig.}{Figs.}
\crefname{appendix}{Appendix}{Appendices}
\crefname{equation}{Eq.}{Eqs.}
\crefname{table}{Table}{Tables}
\Crefname{section}{Section}{Sections}
\Crefname{paragraph}{Section}{Sections}
\Crefname{figure}{Figure}{Figures}
\Crefname{appendix}{Appendix}{Appendices}
\Crefname{equation}{Equation}{Equations}
\Crefname{table}{Table}{Tables}

\newcommand{\GeV}{{{\ }\textrm{GeV}}}
\newcommand{\fm}{{{\ }\textrm{fm}}}

\begin{document}

\title{Factorization and virtuality evolution of jet functions 
\\in heavy-ion collisions}

\author{Carlos Lamas}
\email{carloslamas.rodriguez@usc.es}
\affiliation{Instituto Galego de Fisica de Altas Enerxias (IGFAE), Universidade de Santiago de Compostela, E-15782 Galicia, Spain}

\author{Carlos A. Salgado}
\email{carlos.salgado@usc.es}
\affiliation{Instituto Galego de Fisica de Altas Enerxias (IGFAE), Universidade de Santiago de Compostela, E-15782 Galicia, Spain}
\affiliation{Axencia Galega de Innovaci\'on (GAIN), Xunta de Galicia, Galicia, Spain}

\author{Bin Wu}
\email{bin.wu@usc.es}
\affiliation{Instituto Galego de Fisica de Altas Enerxias (IGFAE), Universidade de Santiago de Compostela, E-15782 Galicia, Spain}

\begin{abstract}
We develop a new framework to perform a virtuality-resolved description of jet propagation in heavy-ion collisions using perturbative techniques. The role of virtuality is rigorously identified by first deriving a factorized jet cross section that defines an in-medium jet function. This new definition allows to extend the usual BDMPS-Z formalism to include the virtuality-differential transverse momentum broadening distribution and medium-induced soft gluon spectrum. We study both the cases of a medium which is created with some delay time $\tau_0$ after the hard collision as well as the case in which $\tau_0\to 0$. Integration over the complete virtuality range recovers the standard BDMPS-Z results, while finite virtuality reveals the interplay between vacuum-like evolution and medium-induced dynamics. We identify the ratio between the medium formation time and the jet formation time as the variable controlling which initial virtualities contribute appreciably to the jet function. 

\end{abstract}
\maketitle

\section{Introduction}
QCD factorization plays a central role in hadron collider physics~\cite{Collins:1989gx, Sterman:1995fz}. It is commonly assumed to hold when applying perturbative QCD to study hard processes in collider physics, with non-perturbative physics factorized into universal quantities such as parton distribution functions (PDFs)~\cite{Ellis:1996mzs, Becher:2014oda}. Proofs of factorization theorems to all orders in $\alpha_s$ have to be established on a process-by-process basis. They have been achieved for relatively simple processes, such as deep inelastic scattering and the Drell-Yan process in hadron-hadron collisions, while proving factorization for more complicated processes remains very challenging~\cite{Collins:2011zzd}. Moreover,  factorization  allows to systematically improve the perturbative cross section calculations with higher orders in the coupling constants, resummations of logarithmically-enhanced contributions, etc.

QCD factorization is also one of the fundamental assumptions underlying theoretical studies of hard processes, such as the production of QCD jets, using perturbative QCD in high-energy nuclear collisions, as recently reviewed in~\cite{Cao:2020wlm, Apolinario:2022vzg}. The phenomenological successes summarized in these reviews have been achieved using different theoretical formalisms for jet quenching, such as the BDMPS-Z~\cite{Baier:1996kr, Zakharov:1996fv, Zakharov:1997uu, Baier:1998kq, Baier:2000mf}, GLV~\cite{Gyulassy:2000fs, Gyulassy:2000er}, ASW~\cite{Wiedemann:2000za,Salgado:2003gb,Armesto:2003jh}, 
higher-twist~\cite{Guo:2000nz, Wang:2001ifa}, and AMY~\cite{Arnold:2002ja} formalisms. However, extractions of medium properties, such as the transport coefficient (jet quenching parameter) $\hat{q}$~\cite{Baier:1996sk}, from experimental data on jet quenching using these different approaches have yielded different values of $\hat{q}$~\cite{Apolinario:2022vzg}.

Such a discrepancy can be partially attributed to differences in how these formalisms describe the coupling of jets to the QCD bulk matter. At a more fundamental level, it also stems from the lack of a proper treatment of the interplay between vacuum and medium-induced radiation when theoretical descriptions of medium-induced radiation are incorporated into phenomenological applications, as explored in, e.g.,~\cite{Baier:2001yt, Vitev:2002pf, Salgado:2003gb, Schenke:2009gb, Ovanesyan:2011xy, Mehtar-Tani:2011hma, Zapp:2012ak, Casalderrey-Solana:2014bpa, He:2015pra, Cao:2017qpx, Caucal:2018dla, Putschke:2019yrg, Duan:2026nvr, Andres:2024egc}. This issue is closely related to the lack of a consensus on a systematic factorization framework for heavy-ion collisions. Moreover, a rigorous justification of factorization in heavy-ion collisions has largely been lacking, with existing treatments relying primarily on physically motivated assumptions or on extensions of established frameworks for simpler processes -- in particular, several Monte Carlo implementations are available: HYDJET \cite{Lokhtin:2005px}, QPYTHIA \cite{Armesto:2007dt,Armesto:2009fj}, JEWEL \cite{Zapp:2008gi,Zapp:2012ak,Zapp:2026cqf}, MARTINI \cite{Schenke:2009gb}, MATTER \cite{Majumder:2013re,JETSCAPE:2017eso}. This contrasts with other types of collisions, such as DIS or $pp$ collisions~\cite{Collins:2011zzd}, where different approaches essentially share a common theoretical foundation.

In this work, we aim to investigate, at least partially, the conditions under which a factorized jet cross section can be obtained for heavy-ion collisions at fixed orders in $\alpha_s$. As a useful working example, we focus on the factorization of the jet function for a specific process, the $\gamma$-jet production cross section. This rigorous definition allows us to then identify the relevant terms where virtuality plays a role and how they can be modified by the medium. For this purpose, we treat the colliding nuclei as collections of uncorrelated nucleons~\cite{Kovchegov:2013cva, Wu:2017rry, Armesto:2024rtl}, as in the Glauber models used in heavy-ion collisions~\cite{Miller:2007ri} and in parton saturation physics~\cite{Kovchegov:2012mbw}. In this case, the cold nuclear effects have been shown to factorize at leading order in $\alpha_s$, the hard momentum scale $Q$, and the nuclear length scale~\cite{Armesto:2024rtl}, identical to that in DIS and $pA$ collisions at the same order~\cite{Kovchegov:1998bi, Kovchegov:2012mbw}. Specifically, the final-state effects can be organized into a jet function.

The purpose of this work is then twofold. First, we derive a factorized formula for jet cross sections, focusing on initial-virtuality-dependent jet functions. This is a needed step to identify how to include virtuality in heavy-ion collisions, with new scales that are absent or irrelevant in the simplest $pp$ collisions -- we include here an initial formation time of the medium, $\tau_0$, the jet quenching parameter, $\hat q$, and the medium length $L$.
The derivation is generic, within the Glauber modeling of the nuclei as the starting point, while the calculation is performed within perturbative QCD. Second, we study how the virtuality of an initial parton, as encoded in the jet functions, contributes to the formation of a final-state quenched jet. Specifically, we carry out detailed calculations using the BDMPS-Z formalism, although our jet functions are defined independently of the specific formalism. We hope to provide a common working ground for comparing different models of energy loss in heavy-ion collisions.

Even within the simplifying BDMPS-Z approach, in which the full virtuality evolution of the parton shower through the medium cannot be dynamically described, our findings clearly identify the dominance of vacuum radiation over medium-induced radiation when the jet formation time is much smaller than the time $\tau_0$ at which the medium is created. In contrast, in the limit $\tau_0 \to 0$, the mean free path emerges as the relevant time scale for this comparison. This observation is particularly important for calculations of jet quenching during the initial stages, before thermalization takes place, and especially in small systems, where this early phase is expected to play a comparatively larger role than in heavy-ion collisions. Considerable effort has been devoted in recent years to coupling this out-of-equilibrium stage to jet-quenching dynamics \cite{Ipp:2020mjc,Carrington:2021dvw,Carrington:2022bnv,Avramescu:2023qvv,Barata:2024xwy,Barata:2025agq,Avramescu:2026fgv,Avramescu:2026qro,Boguslavski:2024ezg,Altenburger:2025iqa,Barata:2025zku} and to assessing the phenomenological relevance of such a coupling for the description of experimental data \cite{Andres:2019eus,Zigic:2019sth,Adhya:2021kws,Pablos:2025cli} .

The paper is organized as follows. In Sec.~\ref{sec:derivation}, we present a detailed derivation of the factorized formula for jet cross sections in heavy-ion collisions and introduce the corresponding virtuality-dependent jet functions. We then adapt the Feynman rules of the BDMPS-Z formalism to the calculation of jet functions in Sec.~\ref{sec:BDMPS}. In Sec.~\ref{sec:jetFuncLO}, we calculate the jet functions at LO and study their dependence on the virtuality of the initial parton. In Sec.~\ref{sec:jetFuncNLO}, we evaluate the next-to-leading-order (NLO) jet functions in the soft radiation limit, specifically for the LPM spectrum in the BDMPS-Z formalism differential in the initial jet virtuality. Numerical results are presented in Sec.~\ref{sec:Results}, including the effects of virtuality dependence on medium-induced radiation. Finally, we summarize our results and discuss possible extensions and future directions in Sec.~\ref{sec:Conclusions}.

\section{Factorized jet cross sections in heavy-ion collisions}

In this section, we present the factorization formula for jet cross sections in the collision of nuclei $A_1$ and $A_2$, exemplified by the process of $\gamma$+jet production:
\begin{align}
\label{eq:X}
A_1 + A_2 \rightarrow \gamma + \text{Jet} + \mathcal{X},
\end{align}
where $\mathcal{X}$ denotes any particles other than the photon and the jet.

\subsection{Factorized jet cross sections and jet functions}
\label{sec:derivation}

Unlike in $pp$ collisions, in heavy-ion collisions it is of interest to discuss the factorization of cross sections for different collision geometries. For this purpose, we focus on the factorization of the impact-parameter-dependent cross section.

\subsubsection{The impact-parameter dependent cross section}

In this work, we focus on hard processes and consider the impact-parameter-dependent cross section, defined as~\cite{Wu:2021ril}
\begin{align}\label{eq:dsigmadb}
        \frac{d\sigma}{d^2{\mathbf b} dO} = & \int \prod_f \left[d\Gamma_{p_f}\right] \delta(O - O(\{p_f\})) \langle \phi_1 \phi_2 | \hat{S}^\dagger | \{ p_f \} \rangle \langle \{ p_f \} | \hat{S} | \phi_1 \phi_2 \rangle,
\end{align}
where  $\hat{S}$ denotes the $S$-matrix with the identity contribution omitted, $O$ is an observable defined by the final-state momenta $\{p_f\}$, $\phi_i$ represents the wave packet of nucleus $i$ as a whole. We consider the transverse spatial spread of the center of mass distribution to be much smaller than the impact parameter $b \equiv |\mathbf{b}|$. Here, the phase-space measure for a particle with momentum $p$ and mass $m$ is defined as
\begin{align}\label{eq:Gammapf}
    \int d\Gamma_{p} \equiv \int \frac{d^4p}{(2\pi)^4} (2\pi) \delta(p^2 - m^2) \theta(p^0).
\end{align}
The impact parameter is related to centrality, which is determined through the distribution of an observable constructed from a subset of the particles that make up $\mathcal{X}$~\cite{Miller:2007ri}.

We further adopt the Glauber model, describing the colliding nuclei as ensembles of uncorrelated nucleons and treating the nucleon Wigner distributions classically. These functions describe the phase-space distribution of nucleons within the nuclei~\cite{Kovchegov:2013cva, Wu:2017rry}, leading to~\cite{Armesto:2024rtl}
\begin{align}\label{eq:dsigmadbJet}
        \frac{d\sigma}{d^2{\mathbf b} dO} = & \int \prod_f \left[d\Gamma_{p_f}\right] \delta(O - O(\{p_f\})) \notag \\
        & \times \prod_{i=1}^{A_1} \frac{1}{2P^+_{1}} \int d^2 \mathbf{b}_i \, db_i^- \hat{\rho}_{A_1}(b_i^-, \mathbf{b}_i) \int \frac{dq_i^+ d^2 \mathbf{q}_i}{(2\pi)^3} e^{i q_i^+ b_i^- - i \mathbf{q}_i \cdot \mathbf{b}_i} \notag \\
        & \times \prod_{j=1}^{A_2} \frac{1}{2P^-_{2}} \int d^2 \mathbf{b}'_j \, db_j'^+ \hat{\rho}_{A_2}(b_j'^+, \mathbf{b}'_j - \mathbf{b}) \int \frac{dq_j'^- d^2 \mathbf{q}'_j}{(2\pi)^3} e^{i q_j'^- b_j'^+ - i \mathbf{q}'_j \cdot \mathbf{b}'_j} \notag \\
        & \times \langle \{P_{1} - \frac{q_i}{2}\}, \{P_{2} - \frac{q'_j}{2}\} | \hat{S}^\dagger | \{ p_f \} \rangle \langle \{ p_f \} | \hat{S} | \{P_{1} + \frac{q_i}{2}\}, \{P_{2} + \frac{q'_j}{2}\} \rangle,
\end{align}
where the momenta of the nucleons within the two nuclei are given by $P_{1}^{\mu} = \frac{P_{1}^+}{\sqrt{2}} n_1^{\mu}$ and $P_{2}^{\mu} = \frac{P_{2}^-}{\sqrt{2}} n_2^{\mu}$, with the beam directions defined as $n_1^\mu = (1, 0, 0, 1)$ and $n_2^\mu = (1, 0, 0, -1)$. Here, $\hat{\rho}_{A_i} \equiv \frac{\rho_{A_i}}{A_i}$, where $\rho_{A_i}$ is the nucleon density and $A_i$ is the mass number, represents the normalized nucleon density in nucleus $i$. The longitudinal components of a four-vector $V^\mu$ are defined as $V^\pm \equiv (V^0 \pm V^3)/\sqrt{2}$, and the corresponding transverse components are denoted in bold. Note that the $\pm$ and transverse momentum components here are defined with respect to the beam directions, which are different from those associated with the jet direction, as discussed in Sec.~\ref{sec:BDMPS} and the subsequent sections.

\begin{figure}[h]
    \centering
    \includegraphics[width=0.45\textwidth]{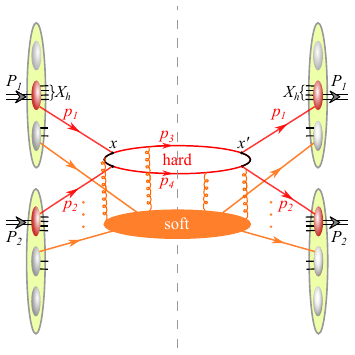}
    \caption{\justifying A generic diagram for jet cross sections in perturbative descriptions of heavy-ion collisions. The hard process is initiated by a binary collision between two nucleons with momenta $P_1$ and $P_2$ from the two nuclei. The remaining $A_1-1$ and $A_2-1$ nucleons (shown in gray) in nuclei $A_1$ and $A_2$ generate partons that are attached to the hard subprocess in all possible ways.}
\label{fig:jetsProduct}
\end{figure}

As illustrated in \cref{fig:jetsProduct}, for processes involving a hard scale, the hard process is most probably initiated by a binary nucleon-nucleon collision. As each pair of nucleons from different nuclei is equally likely to collide, one has
\begin{align}\label{eq:dsigmadbJetInter}
        \frac{d\sigma}{d^2{\mathbf b} dO}     =&\int\prod\limits_f\left[d\Gamma_{p_f}\right]\delta(O-O(\{p_f\}))\notag\\
        &\times\,\frac{1}{2 s_{NN}}\int\, d^2\mathbf{b}_1\,db_1^-{\rho}_{A_1}(b_1^-, \mathbf{b}_1)\int\frac{dq^+d^2\mathbf{q}}{(2\pi)^3} e^{i q^+ b_1^-\,- i \mathbf{q}\cdot\mathbf{b}_1}
        \notag\\
        &\times\int\, d^2\mathbf{b}'_1\,db'^{+}_1{\rho}_{A_2}(b'^{+}_1,\mathbf{b}'_1-\mathbf{b})\int\frac{dq^{\prime -}d^2\mathbf{q}'}{(2\pi)^3} e^{i  q'^-\,b_1'^+\,- i \mathbf{q}'\cdot\mathbf{b}'_1}\notag\\
        &\times \langle\langle P_{1}-{q}/{2}, P_{2}-{q'}/{2}|\hat{S}^\dagger|\{ p_f\}\rangle\langle \{ p_f\}|\hat{S}|P_{1}+{q}/{2}, P_{2}+{q'}/{2}\rangle\rangle,
\end{align}
where $s_{NN}={2P_1^+P_2^-}$ denotes the nucleon--nucleon center-of-mass energy squared, and the double brackets denote the average over the remaining $(A_1+A_2-2)$ nucleons. Their scatterings contribute to the QCD medium with which the produced jet interacts. The explicit form of this average can be read out from \cref{eq:dsigmadbJet}.

\subsubsection{Factorized jet cross sections}
\label{sec:jetFunctions}

Even under the Glauber modeling of nuclei, the jet cross section is not a priori factorizable. For example, as illustrated in \cref{fig:jetsProduct}, the produced medium (soft) degrees of freedom may couple to an incoming parton before it participates in the hard process in a complete calculation. In this work, we expand these contributions at large $Q$ and to leading order in the medium length, building upon the observation in~\cite{Armesto:2024rtl} that cold nuclear effects factorize at leading order in $\alpha_s$ in heavy-ion collisions under this expansion.

\paragraph{\centering The kinematics\\}
\label{sec:Kinematics}

We study the production of QCD jets measured with transverse momentum $\mathbf{p}_{T}$ and rapidity $y_J$ with respect to the beam direction. Here, the jet transverse momentum sets the hard scale $Q = |\mathbf{p_T}| = p_T$ in our analysis. In particular, in the lab frame, the four-momentum of the jet can be expressed as
\begin{align}
\label{eq:pJ}
p_J^\mu &= \sum_{j\in\text{jet}} p_j^\mu = (\sqrt{p_T^2 \csc^2\theta_J + m_J^2},  p_T\cos\phi_J, p_T\sin\phi_J, p_T \cot\theta_J) \notag\\
&= (\sqrt{p_T^2+m_J^2}\cosh y_J, p_T\cos\phi_J, p_T\sin\phi_J, \sqrt{p_T^2+m_J^2}\sinh y_J),
\end{align}
where $p_j$ is the four-momentum of the $j^{\text{th}}$ jet constituent, while $y_J$, $\theta_J$, and $\phi_J$ denote the jet's rapidity, polar angle, and azimuthal angle, respectively. Accordingly, the jet direction is labeled as
\begin{align} \label{eq:LightVectorJet}
n_J^\mu = (1, \vec{n}_J)=(1, \vec{p}_J/|\vec{p}_J|).
\end{align}
The corresponding jet state is defined as
\begin{align}\label{eq:pJstate}
|p_J\rangle \langle p_J|\equiv
\int\bigg(\prod_{i\in\text{jet}} d\Gamma_{p_i}\bigg) |\{p_i\}\rangle \langle \{p_i\}|\int dm_J^2(2\pi)^3\delta^{(4)}(p_J - \sum\limits_{j\in\text{jet}} p_j),
\end{align}
where $|\{p_i\}\rangle$ denotes a multiparticle state with color and spin summed over. The phase-space measure for the jet state is given by
\begin{align}
\int d\Gamma_{p_J}\equiv \int \frac{d^4p_J}{(2\pi)^3} \theta(p_J^0)\delta(p_J^2 - m_J^2) = \int \frac{dy_J d^2\mathbf{p}_T}{2(2\pi)^3}.
\end{align}

In the presence of a QCD medium, the momentum of the jet emerging from the hard vertex is, in general, different from the measured jet momentum. From an experimental point of view, one measures a jet with a certain final-state direction, as discussed above, while the initial jet momentum, denoted as $p_I^\mu$, does not need to have vanishing transverse momentum with respect to $n_J$ and $\bar{n}_J\equiv(1, -\vec{n}_J)$. This initial jet momentum is integrated over in order to obtain the jet cross section.

On the other hand, from a theoretical point of view, it is more convenient to describe the broadening or deviation of the measured jet from the initial jet direction, defined by $p_I^\mu$. Accordingly, in the calculation of the jet functions in QCD matter, one can work in the light-cone coordinate system defined by the following two light-like vectors:
\begin{align} \label{eq:JetInitialDirectionVector}
n^\mu \equiv (1, \vec{n}), \qquad \bar{n}^\mu \equiv (1, -\vec{n}), \qquad \text{with } \vec{n} \equiv \vec{p}_I / |\vec{p}_I|.
\end{align}
Note that in $pp$ collisions, one can identify the initial and final jet directions, so that $n$ and $n_J$ do not need to be distinguished when using the so-called standard jet axis~\cite{Chien:2019gyf}.

\paragraph{\centering Derivation of factorized jet cross sections\\}

\begin{figure}[H]
    \centering
    \includegraphics[width=0.45\textwidth]{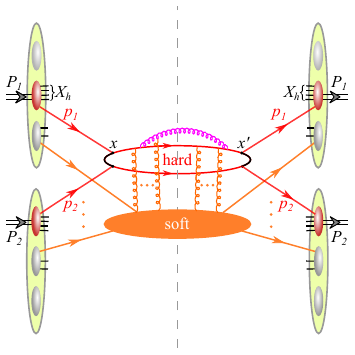}
    \caption{\justifying Diagrams defining the jet function. As illustrated in this diagram, the produced jet couples to the QCD medium (soft degrees of freedom) generated by the scattering of nucleons that do not participate in the hard collision producing the jet. The internal propagators of partons with virtuality $\sim Q^2$ are point-like (black) in the large-$Q$ expansion, with their coordinates denoted by $x$ and $x'$ in the amplitude and conjugate amplitude, respectively. Consequently, background nucleons only couple to the jet lines.}
\label{fig:jets}
\end{figure}

As illustrated in Fig.~\ref{fig:jets}, we shall neglect the diagrams that couple the QCD medium to initial-state partons, whose effects can be absorbed into cold nuclear effects on the nuclear parton distributions at lowest order in $\alpha_s $\cite{Armesto:2024rtl}, although a higher-order justification is still lacking. The diagrams that couple the medium to the internal propagators in the hard process are also subleading, since the hard process is point-like in the large-$Q$ expansion and these contributions are therefore not enhanced by the medium length. The remaining diagrams contribute to the jet functions. 

Specifically, we focus on jet functions defined beyond leading order in $\alpha_s$ by considering only collinear (with respect to the jet direction) and soft radiation, which gives rise to infrared singularities in vacuum. Medium effects are treated as corrections of order $\mathcal{O}(\alpha_s^0)$. We further restrict the hard scattering to the leading-order parton-level $2\leftrightarrow2$ process in the following derivation, which can be achieved by using a physical gauge, as discussed below. The derivation can be extended to include higher-order corrections in $\alpha_s$ to the hard scattering, as well as radiation collinear to the beam directions. A detailed discussion of how to derive a factorized formula with higher-order corrections to all the different components of the factorization formula is left to future work.

\begin{figure}[tp]
    \centering
    \resizebox{0.9\textwidth}{!}{%
\begin{tikzpicture}

\node (ME) at (0,0)
    {$\langle p_q p_{\bar{q}} p_{\gamma}|i\hat{M}(x)|p_1 p_2\rangle$};
\node (EQ) [right=0.5cm of ME]
    {$=$};

\begin{feynman}

\node (aux) [right=1.0cm of EQ];
\vertex (i1) [above=1.0cm of aux, label=$p_1$];
\vertex [right=of i1] (f1) [label=$x$];
\vertex [below=1cm of aux] (i2) [label=$p_2$];
\vertex [right=of i2] (f2);

\vertex [right=of f1] (b1); 
\vertex [right=of f2] (b2) [label=$p_\gamma$];
\vertex [above right=of b1] (c1) [label=$p_q$];
\vertex [below right=of b1] (c2) [label=below:$p_{\bar{q}}$];

\draw[fermion] (i1) -- (f1);
\draw[fermion] (f1) -- (f2);
\draw[fermion] (f2) -- (i2);

\draw[gluon] (f1) -- (b1);
\draw[photon] (f2) -- (b2);
\draw[fermion] (c2) -- (b1);
\draw[fermion] (b1) -- (c1);


\node (pl) [right=5cm of aux]
    {$+$};

\node (aux2) [right=1.0cm of pl];
\vertex [above=1cm of aux2] (i1b) [label=$p_1$];
\vertex [right=of i1b] (f1b);
\vertex [below=1cm of aux2] (i2b) [label=$p_2$];
\vertex [right=of i2b] (f2b) [label=below:$x$];

\vertex [right=of f1b] (b1b) [label=$p_\gamma$];
\vertex [right=of f2b] (b2b); 
\vertex [above right=of b2b] (c1b) [label=$p_q$];
\vertex [below right=of b2b] (c2b) [label=below:$p_{\bar{q}}$];

\draw[fermion] (i1b) -- (f1b);
\draw[fermion] (f1b) -- (f2b);
\draw[fermion] (f2b) -- (i2b);

\draw[photon] (f1b) -- (b1b);
\draw[gluon] (f2b) -- (b2b);
\draw[fermion] (c2b) -- (b2b);
\draw[fermion] (b2b) -- (c1b);

\end{feynman}
\end{tikzpicture}%
}
    \caption{\justifying The parton-level $q\bar{q}\to \gamma q\bar{q}$ process. The operator $\hat{M}(x)$, when evaluated at $x=0$, is defined to give the amplitude for this process when sandwiched between the initial and final partonic states.}
\label{fig:hard diagrams}
\end{figure}
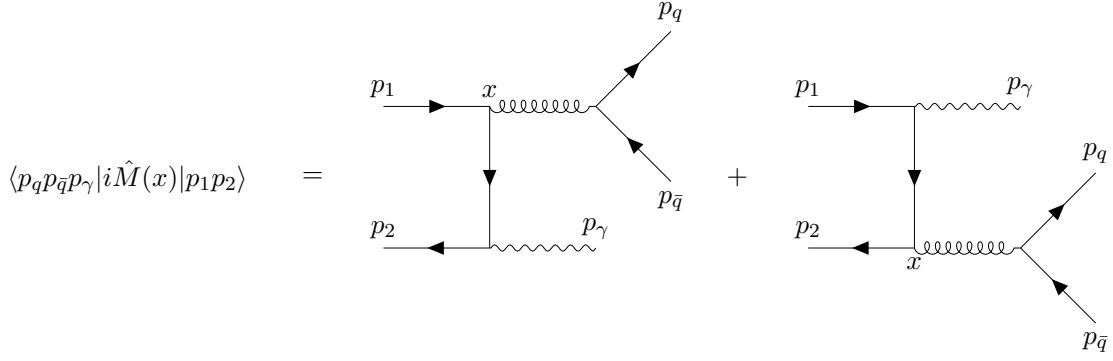

We work with Feynman rules in coordinate space and express the interacting part of the $S$-matrix as
\begin{align}
\label{eq:iM_def}
\hat{S}
\equiv
i \int d^4 x\,\hat{M}(x),
\end{align}
where $x$ denotes the spacetime coordinate of the vertex at which a parton is created to initiate the jet, as illustrated in \cref{fig:hard diagrams}. Using the displacement relation
\begin{align}
\label{eq:displacement}
\hat{M}(x) = e^{i\hat{p}\cdot x}\hat{M}(0)e^{-i\hat{p}\cdot x},
\end{align}
the integration of $x$ in \cref{eq:iM_def} imposes the conservation of the overall four momentum. Accordingly, the operator $\hat{M}(0)$, when sandwiched between the initial and final states, gives the corresponding amplitude:
\begin{align}
\label{eq:amplitude}
i M(\{p_f\}, \{p_i\}) = \langle\{p_f\}| i \hat{M}(0) |\{p_i\}\rangle.
\end{align}

By expressing \cref{eq:dsigmadbJetInter} in terms of $\hat{M}$ and integrating out $q$ and $q'$, the impact parameter dependent cross section reads
\begin{align}\label{eq:dsigmadbJetAmplitude}
        \frac{d\sigma}{d^2{\mathbf b} dO}     =&\int\prod\limits_f\left[d\Gamma_{p_f}\right]\delta(O-O(\{p_f\})) \frac{1}{2 s_{NN}} \int d^4 X \rho_{A_1}(X^-, \mathbf{X}) \rho_{A_2}(X^+, \mathbf{X-\mathbf{b}})\notag\\
        &\times\, \int d^4 x \langle \left\langle P_1, P_2\right| \hat{M}^\dagger (x/2) \left| \{p_f\} \right\rangle \left\langle \{p_f\} \right| \hat{M} (-x/2) \left|P_1, P_2 \right\rangle\rangle \, ,
\end{align}
where $X = (x_1 + x_2)/2$ is the average creation point and $x = x_1 - x_2$ denotes the difference in the creation point between the amplitude and the conjugate amplitude. Here, we have dropped the dependence on $q$ and $q'$ in the second line after the dependence on $X$ is eliminated using the displacement relation in \cref{eq:displacement}, because they are smaller than any perturbative scales in the squared amplitude. As both nuclei are highly Lorentz contracted along the beam direction, one can further neglect the dependence of the amplitude on $X^\pm$ to write
\footnote{If cold nuclear effects are important for the observables of interest, their contributions can be retained in terms of saturation momentum, as nucleon densities are enhanced as well~\cite{Armesto:2024rtl}.
}
\begin{align}\label{eq:dsigmadbJetThickness}
        \frac{d\sigma}{d^2{\mathbf b} dO}     =&\int\prod\limits_f\left[d\Gamma_{p_f}\right]\delta(O-O(\{p_f\})) \frac{1}{2 s_{NN}} \int d^2 \mathbf{X}\, T_{A_1}(\mathbf{X}) 
        \,
        T_{A_2}(\mathbf{X-\mathbf{b}})\notag\\
        &\times\, \int d^4 x \langle \left\langle P_1, P_2\right| \hat{M}^\dagger (x/2) \left| \{p_f\} \right\rangle \left\langle \{p_f\} \right| \hat{M} (-x/2) \left|P_1, P_2 \right\rangle\rangle \, ,
\end{align}
where we have introduced the nuclear thickness function $T_{A_i}(\mathbf{s})\equiv\int\,dz\rho_{A_i}(z,\mathbf{s})$~\cite{Miller:2007ri}.

Now, we specify the observable as the transverse momenta and rapidities of the photon and the jet. For these observables, in terms of the partonic degrees of freedom, the impact parameter-dependent cross section can be expressed as
 \begin{align}\label{eq:dsig_LO}
    & \frac{d\sigma}{d^2{\mathbf b} dy_J d^2{\mathbf p}_T d \eta_\gamma d^2{\mathbf p}_\gamma} =\frac{1}{4(2\pi)^6} \sum\limits_{ij}\int\frac{d\xi}{\xi}\frac{d\xi'}{\xi'} \int~d^2 \mathbf{X} \mathcal{T}_{i/A_1}(\mathbf{X}, \xi)
    \mathcal{T}_{j/A_2}(\mathbf{X} - \mathbf{b}, \xi') \frac{1}{2 s_{NN}} \frac{1}{2d_i}\frac{1}{2d_j}
    \notag\\
    & \hspace{20 pt}\times \int d^4x\, \int \prod_{l \in \mathcal{X}} [d \Gamma_{k_l}]
    \langle\langle  p_{1},  p_{2}|\hat{M}^\dagger(x/2)|p_{\gamma}, p_J, \{k_l\} 
    \rangle\langle p_{\gamma}, p_J, \{k_l\}
    |\hat{M}(-x/2)| p_{1},  p_{2}\rangle\rangle
\end{align}
with $p_1= \xi P_1$, $p_2 = \xi' P_2$, and $d_i$ denoting the dimension of the color representation of parton $i$. Here, the colors and spins of all incoming and outgoing particles are implicitly summed over. $|\{k_l\}\rangle$ represents the multiparticle state of the final-state particles besides the photon and the particles clustered into the jet by a jet algorithm, that is, the particles comprising $\mathcal{X}$ in the process definition in \cref{eq:X}. The measure $d \Gamma_{k_l}$ denotes the phase-space measure associated with the momentum $k_l$ of particle $l\in\mathcal{X}$. The initial parton distributions $\mathcal{T}_{i/A}(\mathbf{X}, \xi)$ are defined as
\begin{align}
    \mathcal{T}_{i/A}(\mathbf{X}, \xi) \equiv T_{A}(\mathbf{X}) f_i(\xi)
\end{align}
with $f_i(\xi)$ being the parton distribution function of species $i$. They are the same as the thickness beam functions $\mathcal{T}_{i/A}(\mathbf{X}, \xi, \mathbf{x})$, as defined in Ref.~\cite{Wu:2021ril}, evaluated at $\mathbf{x}=0$ in the Glauber modeling of the nuclei, and are subject to modifications due to cold nuclear effects.

Because the nucleons that do not initiate the hard process may only connect to the jet and not to the other parts of the diagrams, as mentioned above and illustrated in \cref{fig:jets}, all the diagrams can be separated at $x$ into two parts: the jet part, which can be generated by the corresponding field coupled to soft diagrams associated with other nucleons, and the hard part. As justified below, this separation can be performed after choosing a suitable physical gauge. We then have
\begin{align}\label{eq:S_matrix1_LO}
    &\langle{p_\gamma, p_J, \{k_l\}
    }| \hat{M}(x) \ket{p_{1}, p_{2}}\otimes|A_1-1, A_2-1\rangle \notag\\
    &= \langle p_J, \{k_l\}
    | T \Phi^{\dagger a}_\alpha(x)\ket{0}\otimes|A_1-1, A_2-1\rangle \langle p_\gamma | \hat{M}_h^{a\alpha}(x) |p_1, p_2 \rangle \notag \\
    & \equiv \langle p_J, \{k_l\}
    | T \Phi^{\dagger a}_\alpha(x)\ket{0} \rangle e^{i(p_\gamma - p_1 - p_2) \cdot x} M_h^{a \alpha}(p_\gamma; p_1, p_2)\, ,
\end{align}
where $|A_1-1, A_2-1\rangle$ represents the state of the remaining nucleons, and $\otimes$ denotes the tensor product. The field $\Phi^a_\alpha$, with $a$ and $\alpha$ denoting the color and Lorentz/spinor indices, corresponds to $\psi^a_\alpha$, $\bar{\psi}^a_\alpha$, and $A^a_\alpha$ for quark, antiquark, and gluon jets, respectively. Here, $\hat{M}_h^{a\alpha}(x)$ denotes the hard-scattering operator, with the field $\Phi^a_\alpha$ connecting the hard process to the jet sector, and $M_h^{a\alpha}$, referred to as the amputated hard amplitude, denotes the corresponding hard amplitude with the time-ordered propagator of the $\Phi^a$ field, as well as the rest of the diagram attached to it, removed at $x$.  

Now, let us denote
\begin{align}
\label{eq:Jab}
    J^{ab}_{\alpha\beta} & (x_0', x_0 ; p_J, \{k_l\};
    \mathbf{X}) \equiv\langle\langle0|\bar{T}\Phi^{a}_\alpha(x'_0)|p_J, \{k_l\}
    \rangle \langle p_J, \{k_l\} 
    |T\Phi^{\dagger b}_\beta(x_0)|0\rangle\rangle
    \notag\\
    &=\frac{\delta^{ab}}{d_C}\langle\langle 0|\bar{T}\Phi^{c}_\alpha(x'_0)|p_J, \{k_l\} 
    \rangle\langle  p_J, \{k_l\}
    |T\Phi^{\dagger c}_\beta(x_0)|0\rangle\rangle\equiv \delta^{ab} J_{\alpha\beta}(x_0', x_0; p_J, \{k_l\};
    \mathbf{X}),
\end{align}
where $d_C$ stands for the dimension of the color representation of the field, and we have employed color neutrality to arrive at the second line, as the colors of the final-state partons are summed over.

In terms of $J_{\alpha\beta}$, after using \cref{eq:S_matrix1_LO}, one has
\begin{align}
        \frac{d\sigma}{d^2{\mathbf b} dy_J d^2{\mathbf p}_T d \eta_\gamma d^2{\mathbf p}_\gamma}    & =\frac{1}{4(2\pi)^6}\sum\limits_{ij}\int\frac{d\xi}{\xi}\frac{d\xi'}{\xi'}  \int~d^2 \mathbf{X} \mathcal{T}_{i/A_1}(\mathbf{X}, \xi)
    \mathcal{T}_{j/A_2}(\mathbf{X} - \mathbf{b}, \xi')
        \notag\\
        &\times\int d^4 x e^{i ( p_1 + p_2 - p_\gamma )\cdot x} \int \prod_{l 
        \in\mathcal{X}
        } [d \Gamma_{k_l}] J_{\alpha\beta}(x/2, -x/2; p_J, \{k_l\};  \mathbf{X})\notag\\
        &\times\frac{1}{2s_{NN}} \frac{1}{2d_i}\frac{1}{2d_j} M_h^{*c\alpha} (p_{1}, p_{2}; p_\gamma) M_h^{c\beta} (p_{1}, p_{2}; p_\gamma) \, .
\end{align}
Let us now introduce the jet initial momentum $p_I = p_1 + p_2 - p_\gamma$, as discussed in Sec.~\ref{sec:Kinematics}. We can then write
\begin{align} \label{eq:dsigmaSepparated}
        \frac{d\sigma}{d^2{\mathbf b} dy_J d^2{\mathbf p}_T d \eta_\gamma d^2{\mathbf p}_\gamma}     =&\frac{1}{4(2\pi)^6}\sum\limits_{ij}\int\frac{d\xi}{\xi}\frac{d\xi'}{\xi'}\int~d^2 \mathbf{X} \mathcal{T}_{i/A_1}(\mathbf{X}, \xi)
    \mathcal{T}_{j/A_2}(\mathbf{X} - \mathbf{b}, \xi')
        \notag\\
        & \hspace{-40 pt} \times \frac{1}{2s_{NN}}\frac{1}{2d_i}\frac{1}{2d_j} \int \frac{d^4 p_I}{(2\pi)^4} J_{\alpha\beta}(p_I; p_J; \mathbf{X})\notag\\
        & \hspace{-40 pt} \times (2\pi)^4 \delta^{(4)}(p_1 + p_2 - p_\gamma - p_I) M_h^{*c\alpha} (p_{1}, p_{2}; p_\gamma) M_h^{c\beta} (p_{1}, p_{2}; p_\gamma),
\end{align}
where we define
\begin{align} \label{eq:MomSPaceJetFunct}
J_{\alpha\beta}(p_I; p_J; \mathbf{X}) \equiv \int d^4 x\, e^{i p_I \cdot x}  \int \prod_{l 
\in\mathcal{X}} [d \Gamma_{k_l}] J_{\alpha\beta}(x/2, -x/2; p_J, \{k_l\};  \mathbf{X}).
\end{align}
So far, we have only rewritten the impact-parameter-dependent cross section in terms of the jet part and the hard part. We need further to choose a proper gauge and perform the expansion to leading order in $1/Q$ in order to factorize the cross section into a convolution of the jet function, which describes the evolution of the jet in the medium and should be gauge invariant and (ideally) independent of the underlying hard process, and the cross section for the hard process involving only on-shell incoming and outgoing partons.

In our calculations, we always use physically polarized final states. In this case, according to the Ward identity~\cite{Peskin:1995ev, Wu:2026iao}, $J_{\alpha\beta}(p_I; p_J; \mathbf{X})$ for quark jets is gauge invariant, as we consider the limit in which the four-momentum squared $p_I^2\ll |\mathbf{p}_T|^2$. Keeping only leading terms in $\bar{n}\cdot p_I \sim Q$, one has the projection
\begin{align}
\label{eq:Jq}
J_{\alpha\beta} = {P_q}_{\alpha\beta}(p_I) \mathcal{J}_q
\end{align}
with
\begin{align}
\label{eq:Pq}
    {P_q}(p_I)\equiv\frac{\bar{n}\cdot p_I}{2}\slashed{n}.
\end{align}
This can be justified using old-fashioned perturbation theory or, explicitly, in the BDMPS-Z formalism as presented in the next subsection.

For gluon jets,  we adopt the light-cone gauge condition $\bar{n} \cdot A = 0$. In this case, one has $J_{\alpha\beta} n^{\alpha}=0=J_{\alpha\beta} \bar{n}^{\alpha}$ and, accordingly, the projection
\begin{align}
    \label{eq:Jg}
    J_{\alpha\beta} = {P_g}_{\alpha\beta}(p_I)\mathcal{J}_g
\end{align}
with
\begin{align}
\label{eq:Pg}
    {P_g}_{\alpha\beta}(p_I) = -g_{\perp\alpha\beta} \equiv \frac{n_{\alpha} \bar{n}_{\beta} + n_{\beta} \bar{n}_{\alpha}}{2} - g_{\alpha\beta}.
\end{align}
Here, $J_{\alpha\beta} n^{\alpha}=0$ is a consequence of the Ward identity $p_I^{\alpha} J_{\alpha\beta} = 0$ up to power-suppressed terms in $1/\bar{n}\cdot p_I$.
Consequently, one has
\begin{align}
    \label{eq:J_def}
    \mathcal{J}_q = \frac{1}{2 \bar{n}\cdot p_I}\text{Tr}\bigg(\frac{\slashed{\bar{n}}}{2}J\bigg) ~~\text{for $q/\bar{q}$}\qquad\text{and}\qquad
    \mathcal{J}_g = -\frac{1}{2} g_\perp^{{\alpha\beta}} J_{\alpha\beta}~~\text{for $g$}.
\end{align}

The projectors $P_q$ and $P_g$ need to be contracted with the amputated hard amplitude squared to obtain the amplitudes squared for the hard process $pp\to \gamma$ + parton of momentum $(\bar{n}\cdot p_I) n^\mu/2$. Here, we replace the initial momentum $p_I^{\mu}$ of the parton that initiates the jet with an on-shell momentum, in the spirit of the large-$Q$ expansion. As a result, we finally obtain
 \begin{align}\label{eq:dsig_Fact}
        \frac{d\sigma}{d^2{\mathbf b} dy_J d^2{\mathbf p}_T d \eta_\gamma d^2{\mathbf p}_\gamma}     =&\sum\limits_{ijk}\int{d\xi}{d\xi'}\int~d^2 \mathbf{X} \mathcal{T}_{i/A_1}(\mathbf{X}, \xi)
    \mathcal{T}_{j/A_2}(\mathbf{X} - \mathbf{b}, \xi')
        \notag\\
        & \hspace{-40 pt} \times \int \frac{d^4 p_I}{(2\pi)^4} \mathcal{J}_k(p_I; p_J, \mathbf{X})\frac{d\hat{\sigma}_{ij\to \gamma k}}{ d\eta_I d^2{\mathbf p}_I d \eta_\gamma d^2{\mathbf p}_\gamma} (\xi P_1, \xi' P_2\to p_\gamma, p_k),
\end{align}
where $p_k^\mu = (\bar{n}\cdot p_I) n^\mu/2$, $\eta_I$ represents the pseudorapidity corresponding to $p_k$, and $\hat{\sigma}_{ij\to \gamma k}$ denotes the cross section for the partonic process: $ij\to \gamma k$:
\begin{align}
\frac{d\hat{\sigma}_{ij\to \gamma k}}{ d\eta_I d^2{\mathbf p}_I d \eta_\gamma d^2{\mathbf p}_\gamma} (p_1, p_2\to p_\gamma, p_k)=&\frac{1}{4(2\pi)^6}\frac{1}{2\xi\xi's_{NN}}(2\pi)^4 \delta^{(4)}(p_1 + p_2 - p_\gamma - p_k)\notag\\
&\times \frac{1}{2d_i}\frac{1}{2d_j} M_h^{*c\alpha} (p_{1}, p_{2}; p_\gamma) M_h^{c\beta} (p_{1}, p_{2}; p_\gamma)P_{k\alpha\beta}(p_k).
\end{align}
This factorized cross section generally holds, independent of the modeling of the interactions between the medium and the jet. With a proper medium treatment it can be used to perform phenomenological estimations of experimental observables in heavy-ion collisions. Remarkably, the jet cross section depends on the jet's initial virtuality $m_I^2 \equiv p_I^2$, a feature missing in most of the jet quenching calculations, so it can be used to study the virtuality evolution of jets inside QCD matter.

As a check of the normalization, at LO without final-state interactions, as in $pp$ collisions, we have, for both quark and gluon jets,
\begin{align}
    \mathcal{J}_k(p_I; p_J) &=\int dm_J^2\int d\Gamma_{p_1}  \int{d^4 x}\,e^{i (p_I-p_1)\cdot x} (2\pi)^3 \delta^{(4)}(p_J-p_1)= (2\pi)^4\delta^{(4)}(p_I-p_J).
\end{align}
Here, we have used the fact that the single jet state in \cref{eq:pJstate} consists of one parton:
\begin{align}
\label{eq:jetState_1}
    |p_J\rangle \langle p_J| = \int d\Gamma_{p_1} |\{p_1\}\rangle \langle \{p_1\}|(2\pi)^3\delta^{(4)}(p_J - p_1) = |\{p_J\}\rangle \langle \{p_J\}|
\end{align}
with $|\{p_J\}\rangle \langle \{p_J\}|$ now denoting a single-parton state.

\subsection{Feynman rules for jet functions in the BDMPS-Z formalism}
\label{sec:BDMPS}

The jet propagation inside the medium is determined by the jet function $\mathcal{J}_k (p_I; p_J, \mathbf{X})$. We here present the analytical techniques to evaluate the jet functions and use them to compute physical observables.

\subsubsection{Light-cone coordinates for jets}

In the rest of this paper, we will work in the light-cone coordinate system defined by the jet directions $n$ and $\bar{n}$ as defined in \cref{eq:JetInitialDirectionVector}, which is natural for jet propagation. The light-cone $\pm$ components of a four-vector $V^\mu$ are redefined with respect to the jet directions as
\begin{align}
V^\pm \equiv (V^0 \pm V^3)/\sqrt{2},
\end{align}
while its transverse components, denoted by the bold symbol $\mathbf{V}$, are obtained by projecting with the transverse metric $g_\perp^{\mu\nu}$, as defined in \cref{eq:Pg}. Specifically, for the initial jet momentum $p_I^\mu$, one has
\begin{align} \label{eq:JetLCCoordinates}
p_I^+ \equiv \bar{n} \cdot p_I/\sqrt{2}, \qquad
p_I^- \equiv n \cdot p_I/\sqrt{2} = \frac{m_I^2}{2 p_I^+}
\end{align}
with the initial virtuality $m_I^2 = p_I^2$. Accordingly, it can be expressed as
\begin{align}
p_I^\mu = \left( p_I^+, \frac{m_I^2}{2p_I^+}, \mathbf{0} \right).
\end{align}
For phenomenological calculations, one may compute the jet functions in this coordinate system and then transform back to the original system referred to the beam before plugging them into the factorized cross section in \cref{eq:dsig_Fact}.

\subsubsection{Feynman rules for jet functions in the BDMPS-Z formalism}
The jet function as defined in \cref{eq:J_def} requires a detailed treatment of the bulk medium. It incorporates a unified description of hard and soft degrees of freedom of the problem. On the other hand, it complicates a full calculation.
To simplify the analysis, we adopt the BDMPS-Z formalism~\cite{ Baier:1996kr, Zakharov:1996fv, Baier:1996sk, Baier:1998kq, Baier:2000mf} to model the bulk medium with background gluon fields $A^{a\mu}(x^+, \mathbf{x})$ and retain only leading-order contributions in $p^+$. The latter is consistent with the large-$Q$ expansion.

In this formalism, there is no exchange of the + momentum between the jet and the medium. As a result, one can use the displacement operator to eliminate the $x^-$ dependence in the defining fields in $J_{\alpha\beta}$, as expressed in \cref{eq:Jab}, and Fourier transform it with respect to the $x^-$ coordinate to convert it into an overall delta function imposing the conservation of the + momenta of all the final-state partons produced via the splitting of the initial parton associated with the jet-defining field. Among these partons, those that are not clustered inside the jet by the jet algorithm are therefore considered as part of $\mathcal{X}$. The $+$ momenta of the remaining particles in $\mathcal{X}$ do not enter this momentum-conservation constraint. Consequently, we can factor out the delta function from the jet function and define
\begin{align}
\label{eq:Jk_3D}
\mathcal{J}_{k}(p_I; p_J, \mathbf{X}) \equiv \int d^3 x\, e^{ip_I\cdot x} \int \prod_{l \notin jet} [d \Gamma_{k_l}]& \, 2\pi \delta(p_I^+ - p_J^+ - \sum_{l \notin jet} k_l^+) \notag\\
&\times
\tilde{\mathcal{J}}_{k}(x/2, -x/2; p_J, \{k_l\}; \mathbf{X}),
\end{align}
where $k_l$ now represents the momentum of the out-of-jet radiation from the initial parton. Here, $\tilde{\mathcal{J}}_k$ depend on the vectors living in three-dimensional Minkowski space defined as
\begin{align}
{x}^\mu=(x^+, \mathbf{x}),\qquad {p}^\mu = (p^-, \mathbf{p}),
\end{align}
which are denoted by the same symbols as the corresponding four-dimensional vectors. Their dimensions can be figured out from the context.

The jet function $\tilde{\mathcal{J}}_k$ can be calculated in coordinate space according to the following Feynman rules in 1+2 dimensions, which are equivalent to those in the mixed coordinates with $(p^+, x^+, \mathbf{x})$ in 1+3 dimensions. This is achieved by integrating out the - momentum in the propagators of the internal particles to write~\cite{Wiedemann:2000za, Casalderrey-Solana:2007knd, Iqbal:2026ytv}
\begin{align}
\label{eq:DF}
    D_F(x) = \frac{1}{2p^+}\mathcal{P}(p^+, -i \nabla_\perp) G_F(x^+, \mathbf{x}; p^+), 
\end{align}
where the projector $\mathcal{P}(p^+, \mathbf{p}) = u^s(p)\bar{u}^s(p), v^s(p)\bar{v}^s(p), $ or $\epsilon_{\lambda}^{\mu}(p^)\epsilon_{\lambda}^{*{\nu}}(p)$ for $q$, $\bar{q}$ or $g$, respectively, and the free propagator is defined as
\begin{align}\label{eq:GF}
    G_F(x
    ; p^+) &\equiv  \int\frac{d^2\mathbf{p}}{(2\pi)^2} e^{-i\frac{\mathbf{p}^2 + m^2-i\epsilon}{2p^+}x^+ + i\mathbf{p}\cdot\mathbf{x}}\theta(x^+)
    =e^{-i\frac{m^2}{2p^+}}\int \mathcal{D} \mathbf{r} e^{i\frac{p^+}{2}\int dt \dot{\mathbf{r}}^2}\theta(x^+)
\end{align}
with the time derivative denoted by an overdot and $m$ denoting the parton mass. In the following discussion, the three-dimensional coordinates, e.g., $x$ on the left-hand side of the above equation, are sometimes also written explicitly in terms of the $+$ time and the transverse coordinates if we need to specify the time explicitly. If the propagator is not connected to a splitting vertex, one can simply replace $p$ with $p^+ n/\sqrt{2}$ in $\mathcal{P}(p^+, \mathbf{p})$.

Each field and its conjugation in the definition of $\mathcal{J}_{\alpha\beta}$ in \cref{eq:Jab} contracts either with an external state or with another field to form a propagator as expressed above. In both cases, each field is associated with a spinor or polarization vector. One can hence combine them with the projectors in \cref{eq:J_def} to obtain
\begin{align}
\label{eq:J3D_spin_average}
\frac{1}{2\sqrt{2}p_I^+} \frac{1}{2} \bar{u}^s \slashed{\bar{n}} u^{s'} = \frac{\delta^{ss'}}{2},\qquad -\frac{g_{\perp\mu\nu}}{2} \epsilon_{\lambda'}^{\mu} \epsilon_{\lambda}^{*{\nu}} = \frac{\delta^{\lambda \lambda'}}{2}
\end{align}
for $q/\bar{q}$ and $g$, respectively. Accordingly, one only needs to average over the initial spins or polarizations in the calculation of $\tilde{\mathcal{J}}_k$ and $\mathcal{J}_k$, while omitting the associated spinors or polarization vectors.

When a parton generated by the defining fields of the jet function enters the QCD medium at an initial time $\tau_0$ and at a transverse location $\mathbf{x}$, one can break its free propagator $G_F$ into two at this point and integrate over $\mathbf{x}$. Afterwards, if the parton enters a vertex coupled to the background field, the $1/2p^+$ times the spinor or polarization vector in the propagator \cref{eq:DF}, combined with the vertex and another spinor or polarization vector after the vertex, effectively yields the same structure as in \cref{eq:J3D_spin_average}, leaving us with the following rules for constructing a parton propagator in a QCD medium:
\begin{equation}
  \raisebox{-0.18\height}{%
    \begin{tikzpicture}
      \begin{feynman}
        \vertex (xi) {$y$};
        \vertex[right=2cm of xi] (xf) {$x$};
        \draw (xi) -- (xf);
        \vertex[right=1cm of xi, label=$p^+$];
      \end{feynman}
    \end{tikzpicture}%
  } = G_F(x-y; p^+) \, , \qquad
  \raisebox{-0.65\height}{%
    \begin{tikzpicture}
      \begin{feynman}
        \vertex (xi);
        \vertex[right=2cm of xi] (xf);
        \vertex[right=1cm of xi, label=$x$] (aux);
        \node[dot, below=1cm of aux, label={[xshift=0.3 cm] $a$}](s);
        \draw (xi) -- (aux);
        \draw (aux) -- (xf);
        \draw[gluon] (s) -- (aux);
      \end{feynman}
    \end{tikzpicture}%
  } = -ig \hat{T}^a A^{a -}(x
  )\, ,
\end{equation}
where all partons are represented by solid lines, and $\hat{T}^a$ stands for the color generator: $\hat{T}^a_{bc}=t^a_{bc}$ for quarks, $\hat{T}^a_{bc}=-t^a_{cb}$ for antiquarks, and $\hat{T}^a_{bc}=-if^{abc}$ for gluons. One needs to integrate over $x^\mu=(x^+, \mathbf{x})$ at each vertex.

Accordingly, the full propagator in the medium from $y$ to $x$ formally takes the form
\begin{align}\label{eq:DressedPropagator}
        G(x;y; p^+)
        = G(x^+, \mathbf{x};y^+, \mathbf{y}; p^+)
        = \int\limits_{\mathbf{r}(y^+)=\mathbf{y}}
        ^{\mathbf{r}(x^+)=\mathbf{x}}\mathcal{D} \mathbf{r} \exp\bigg\{i \frac{p^+}{2}\int_{y^+}^{x^+} dt \dot{\mathbf{r}}^2\bigg\} W(\mathbf{r}; x^+, y^+)
\end{align}
with the Wilson line defined as
\begin{align}\label{eq:WilsonLine}
    W(\mathbf{r}; t_2, t_1)\equiv \mathcal{P} \exp{-ig \int_{t_1}^{t_2} d x^+ A^-(x^+, \mathbf{r})},
\end{align}
where $A^- = \hat{T}^a A^{a -}$, which represents a rotation in color space. These rules also apply to dressing propagators with the background fields connecting to an external outgoing state while exiting the QCD medium~\cite{Armesto:2024rtl}.  The fully dressed propagators satisfy
\begin{align}
\label{eq:GG2G}
&\int d^2\mathbf{y}\,G(x;y; p^+)G(y;z; p^+)
=G(x;z; p^+)
\qquad\text{for $x^+\geq y^+\geq z^+$},\notag\\
&\int d^2\mathbf{y}\,G^\dagger(y;x; p^+)G(y;z; p^+)
=G(x;z; p^+)
\qquad\text{for $y^+\geq x^+\geq z^+$},
\end{align}
which can be most easily justified using the equivalent two-dimensional quantum mechanics.

When the full propagators enter or exit a splitting kernel, one needs to retain their transverse-momentum dependence in the residual spinors or polarization vectors associated with each dressed propagator. As the interactions with the background fields do not change the parton spins, one effectively only needs to average or sum over the spins of the incoming or outgoing partons at the splitting kernel.

In a generic diagram with $n$ final-state partons with momentum $p_i$, when all the full propagators in the diagram exit the medium at time $\tau_f$, they may undergo further splitting to produce all these final-state partons before a time $\tau_{\infty}$. One can extend the contracted final states at their production vertices up to this time by free propagators. Let us denote their transverse locations $\mathbf{x}_i$ and $\mathbf{x}'_i$ in the amplitude and conjugate amplitude at $\tau_{\infty}$, respectively, leading to an integration over all the final-state momenta and the transverse locations:
\begin{align}\label{eq:integralsFinalStates}
\bigg(\prod_{i=1}^n \int d^2\mathbf{x}_i d^2\mathbf{x}'_i\int\frac{dp_i^+ d^2\mathbf{p}_i}{2 p_i^+(2\pi)^3} e^{i\mathbf{p}_i\cdot(\mathbf{x}'_i-\mathbf{x}_i)}\bigg)\delta(p_I^+-\sum\limits_i p_i^+)\int dm_J^2(2\pi)^3\delta^{(4)}(p_J - \sum\limits_{j\in\text{jet}} p_j).
\end{align}
Here, the delta function of the $+$ momentum is the same as that on the right-hand side of \cref{eq:Jk_3D}. Among these $n$ final-state partons,
for those which are not clustered inside the jet,
the integration over $\mathbf{p}_i$ simply gives $(2\pi)^2\delta^{(2)}(\mathbf{x}_i-\mathbf{x}'_i)$. After integrating over $\mathbf{x}_i$ and $\mathbf{x}'_i$, the propagators in the amplitude and conjugate amplitude corresponding to the parton are simply concatenated, according to \cref{eq:GG2G}. The same happens if we do not measure the jet transverse momentum $\mathbf{p_J}$ and integrate over it, fixing the center of mass of the jet in the amplitude and the conjugated amplitude to be in the same position.

After the diagrams are expressed in terms of the full propagators dressed with the background fields, one needs to carry out the ensemble average over the background field configurations, given by
\begin{equation}
\expval{\expval{A^{a-}(x^+_1, \mathbf{x}) A^{b-}(x^+_2, \mathbf{y})}} = \delta(x^+_1-x^+_2) \delta^{ab} n(x^+) \sigma(\mathbf{x}-\mathbf{y}),
\label{eq:field averages}
\end{equation}
where $n(x^+)$ is the number density of medium constituents and $\sigma$ is the dipole cross section. For cold nuclear effects, such an ensemble average is equivalent to the average in \cref{eq:dsigmadbJetInter}~\cite{Armesto:2024rtl}. In this work, we also assume this equivalence for a hot QCD medium, as in the conventional BDMPS-Z formalism.

One new quantity that uniquely shows up in our calculations below is given by
\begin{align} \label{eq:1pointWilson}
\frac{1}{d_R} \Tr \langle G(x;y; p^+) \rangle = \exp{- \frac{1}{2} \int_{y^+}^{x^+} \frac{dt}{\lambda_R(t)}}G_F(x-y;p^+),
\end{align}
where $d_R$ denotes the dimension of the color representation $R$, and we introduce the light-cone mean free path of the propagating particle
\begin{align} \label{eq:MeanFreePath}
\lambda_R(x^+) = \frac{1}{g^2 C_R n(x^+) \sigma(\mathbf{0})}
\end{align}
with $C_R$ the Casimir of the color representation $R$. The single-line correlator in \cref{eq:1pointWilson} is then an exponential suppression by the number of scatterings suffered by the propagating particle. This suppression can be understood as the probability that a particle in a given color state remains in the same state after propagating through the medium for a given amount of time, necessary to preserve color neutrality.
This completes the Feynman rules for our calculations of the jet functions.

\subsubsection{Soft gluon radiation and harmonic oscillator approximation}

In our calculation, we will assume the energy of the leading parton to be large enough that $\mathbf{p}^2 L^+/p^+ \ll 1$, where $\mathbf{p}^2$ is the squared transverse momentum transferred from the medium to the jet and $L^+$ is the medium length. We will also consider that, for the high energies considered, the quarks are approximately massless. We can therefore neglect the mass-dependent term accompanying $x^+$ in the exponential of \cref{eq:GF}, and in eikonal approximation the dressed propagator is reduced to the Wilson line
\begin{align}
\label{eq:Geik}
    G(x^+, \mathbf{x}; y^+, \mathbf{y}; p^+) = \delta^{(2)}(\mathbf{x}-\mathbf{y}) W(\mathbf{x}; x^+, y^+)\, .
\end{align}
The radiated gluons carry a small fraction of the total longitudinal momentum fraction of the leading parton. Therefore, we have to keep the $\mathbf{p}^2 x^+/k^+$ terms in the gluon propagators and work with the full dressed propagator in \cref{eq:DressedPropagator}. We will also assume that $m_I^2$ can be much larger that $\mathbf{p}^2$ and therefore keep the $m_I^2\, x^+/p_J^+$ phase in the propagator of the initial parton. In this limit, besides the transverse momentum transfer, we can hence keep track of the - momentum transfer from the medium to the jets via the radiated soft gluon lines.

The Wilson line in \cref{eq:WilsonLine} is valid for a given medium configuration and must be averaged over all the possible configurations of the background fields. 
For the dipole amplitude we will use the harmonic oscillator approximation
and the dipole cross section is related to the jet quenching parameter $\hat{q}_R$~\cite{Baier:1996kr}:
\begin{align}
    g^2 C_R n(x^+) [\sigma(\mathbf{0})-\sigma(\mathbf{x})]\approx\frac{1}{4 \sqrt{2}}\hat{q}_R(x^+) \mathbf{x}^2 \, ,
\end{align}
for a parton in the color representation $R$. This relation contains all the dynamical information about the interaction between the jet and the background fields. The factor $1/\sqrt{2}$ is due to our light-cone coordinate convention, so that the jet quenching parameter corresponds to the average transverse momentum transferred from the medium to the jet per unit of time, $\hat{q} \sim \delta \mathbf{p}^2/ \delta t$.
The configuration average of two Wilson lines follows as
\begin{align} \label{eq:2pointWilson}
    \frac{1}{d_R} \Tr \langle W^\dagger(\mathbf{y}) W(\mathbf{x})\rangle = \exp{-  \frac{1}{4 \sqrt{2}} \int dx^+ \hat{q}_R(x^+) (\mathbf{y}-\mathbf{x})^2}.
\end{align}

Beyond LO we have to incorporate the splitting vertex to the formalism. In this work, we will, for simplicity, illustrate the calculation for the case of a high energy quark jet which radiates soft gluons in the following sections. We therefore focus only on the soft gluon emission vertex
\begin{equation} \label{eq:SoftEmissionVertex}
  \raisebox{-0.18\height}{%
    \begin{tikzpicture}
      \begin{feynman}
        \vertex (xi) {$p_i$};
        \vertex[right=1.5cm of xi, label={[yshift=-0.5 cm] $x$}] (s);
        \vertex[right=1.5cm of s] (xf) {$p_f$};
        \vertex[above=1 cm of xf] (yf) {$k$};
        \draw[fermion] (xi) -- (s);
        \draw[fermion] (s) -- (xf);
        \draw[gluon] (s) -- (yf);
      \end{feynman}
    \end{tikzpicture}%
  } = \frac{1}{2p^+} \bar{u}_\lambda(p_f) (-igt^a \gamma^\mu \epsilon_\mu(k)) u_{\lambda^\prime}(p_i) \simeq -igt^a \frac{\mathbf{k} \cdot \boldsymbol{\epsilon}}{k^+} \delta_{\lambda\lambda^\prime}\, ,
\end{equation}
where we only retained the leading contribution in $p^+$ and used that $\epsilon^- = \mathbf{k} \cdot \boldsymbol{\epsilon}/k^+$. The above rule for the splitting vertex applies also to the emission of a soft gluon off a high-energy gluon (with $t^{a}_{bc}$ replaced by $-if^{abc}$). When the partonic lines connecting to the vertex correspond to internal propagators it is convenient to rewrite the expression in coordinate space using that $\mathbf{k} = i \nabla_{\mathbf{r}}$, with $\mathbf{r}$ the coordinate of the gluon coupling to the vertex, so that we can integrate out all the momenta in the propagators.

For a quark jet, let us mention here some useful relations that will be used in the latter sections. We will denote by $G_A$ and $W_A$ the propagators and Wilson lines in the adjoint representation and leave the quantities without a subindex for the fundamental representation. First, the Wilson lines in the fundamental and adjoint representation are related through
\begin{align} \label{eq:RelAdjointFundamental}
    2 \Tr[t^{c}W^\dagger(\mathbf{x}; t_2, t_1) t^{c'} W(\mathbf{x}; t_2, t_1)] = W_A^{cc'}(\mathbf{x}; t_2, t_1) \, ,
\end{align}
which allows us to simplify the Wilson line products and rewrite the calculation in terms of adjoint representation field correlators. An important quantity to define is the splitting kernel, which in the harmonic oscillator approximation reads
\begin{align} \label{eq:SplittingKernel}
    \begin{split}
    \mathcal{K}(t_2, \mathbf{x}; t_1, \mathbf{y}; p^+) & \equiv \frac{1}{N_C^2-1} \Tr\langle G_A(t_2, \mathbf{x}; t_1, \mathbf{y}; p^+) W_A^\dagger(\mathbf{0}; t_2, t_1) \rangle \\
    & = \int_{\mathbf{r}(t_1)=\mathbf{y}}^{\mathbf{r}(t_2)=\mathbf{x}} \mathcal{D} \mathbf{r} \exp{i \frac{p^+}{2} \int_{t_1}^{t_2} dt \left[\dot{\mathbf{r}}^2 +i \frac{\hat{q}_A(t)}{2\sqrt{2}\, p^+} \mathbf{r}^2 \right] }\, .
    \end{split}
 \end{align}
It corresponds to the path integral of an harmonic oscillator of complex frequency
\begin{align} \label{eq:OscilatorFrequency}
    \Omega^2 \equiv - i \frac{\hat{q}_A(t)}{2\sqrt{2}\, p^+}\, .
\end{align}
It is also useful to introduce the correlator of a $qg$ pair coming from a quark splitting
\begin{align}\label{eq:qgCorrelator}
    t^{a_1}_{c_1b} & \left\langle G_A^{aa_1}(t_2, \mathbf{x}; t_1, \mathbf{y}; p^+) W_{cc_1}(\mathbf{0}; t_2, t_1) \right\rangle = t^{a}_{bc} \exp{-\frac{1}{2}\int_{t_1}^{t_2} \frac{dx^+}{\lambda_F}} \tilde{\mathcal{K}}(t_2, \mathbf{x}; t_1, \mathbf{y}; p^+)\, ,
\end{align}
where $\tilde{\mathcal{K}}(t_2, \mathbf{x}; t_1, \mathbf{y}; p^+)$ is the path integral of an harmonic oscillator of frequency $\tilde{\Omega} = \Omega/\sqrt{2}$. The correlator in \cref{eq:qgCorrelator} then corresponds to  the propagator of the $qg$ dipole multiplied by the probability that the dipole, which is in triplet state, does not change its global color configuration while traveling through the medium. Another useful relations as the convolutions of propagators in different kinematic regimes are shown in \cref{app:convolutions}. Note that \cref{eq:SplittingKernel} and \cref{eq:qgCorrelator} also hold for soft gluon radiation off a high-energy gluon, with $\lambda_F$ and $t^a$ replaced accordingly in \cref{eq:qgCorrelator}.

\section{Virtuality dependent jet functions}

The jet functions defined in \cref{sec:jetFunctions} contain all the information about the evolution of the jet through the QCD medium. They depend on the virtuality of the parton which initiates the jet so they can be used to track the virtuality evolution of the partons traversing the bulk, and can help construct a medium induced parton shower. 

In this section, we study how medium induced phenomena depend on the initial virtuality of the jet in the BDMPS-Z formalism. Specifically, instead of carrying out phenomenological studies using the jet functions, we would like to explore the initial virtuality dependence in the $p_\perp$ distribution~\cite{Baier:1996sk} and soft radiation spectrum~\cite{Baier:1996kr, Zakharov:1997uu} in this formalism.
In the jet coordinate system where $\mathbf{p}_I = \mathbf{0}$, after integration over $p_I^+$, the jet functions are related to the transverse momentum distribution via:
\begin{align} \label{eq:JetCrossSect}
    \frac{dI}{d^2 \mathbf{p_J}}  = \int d^3 x \int \prod_{l \notin jet} [d\Gamma_{k_l}] & \int \frac{dm_I^2}{(2\pi)^3\, 2p_I^+} e^{i \frac{m_I^2}{2 p_I^+} x^+} \notag \\
    & \times \tilde{\mathcal{J}}_k (x/2, -x/2; p_J, \{k_l\}; \mathbf{X})|_{p_I^+=p_J^+ + \sum_{l \notin jet} k_l^+}\, .
\end{align}
Note that $\mathbf{p}_J$ is the final-state transverse momentum of the jet with respect to the initial jet direction, not the jet $p_T$ in the factorized cross section \cref{eq:dsig_Fact}.
Starting from \cref{eq:JetCrossSect}, we shall compute the distribution function $dI$ differential in $m_I^2$ and study how the aforementioned jet quenching observables depend on this quantity. In particular we want to show which regions of the virtuality phase space are more relevant to jet quenching and which are dominated by vacuum radiation. We also study which are the dominant diagrams in each region.

We will consider a generic scenario where the parton that initiates the jet starts to interact with the medium at an initial time $\tau_0$ after the hard process occurs at $\tau\approx 0$ and that the medium lives until a time $\tau_f$, so that $L^+ = \tau_f - \tau_0$. We will illustrate our calculation for an homogeneous medium, where the number density $n$ is independent of $x^+$. Remarkably, as our jet function is differential in virtuality, the uncertainty principle tells us that the creation point of the jet is not perfectly localized and can be different in the amplitude and the conjugated amplitude. Therefore, although the average creation point of the jet is $\tau \sim 0$, we can have diagrams where the jet is created inside the medium, or even after the medium, in one of the amplitudes.

\subsection{Jet functions at LO}
\label{sec:jetFuncLO}

In this section we compute the spectrum of the transverse momentum transferred from the medium to the jet, the so-called transverse momentum broadening, as a function of the initial jet virtuality. This observable follows from the distribution in \cref{eq:JetCrossSect} evaluating it at LO in $\alpha_s$. In this case, the high-energy parton defines the jet and there is no radiation so $p_I^+ = p_J^+$ and the virtuality differential cross section can be expressed as
\begin{align}
    \frac{dI^{LO}_k}{dm_I^2 d^2 \mathbf{p}_J} & =\int\frac{d^3x}{(2\pi)^3 2p_J^+}e^{i \frac{m_I^2}{2 p_J^+} x^+} \tilde{\mathcal{J}}^{LO}_k(x/2, -x/2; p_J, \mathbf{X}),
\end{align}
The average vertex location $\mathbf{X}$ only enters the calculation through the density of scattering centers. We will here for simplicity assume an homogeneous medium and omit $\mathbf{X}$.
Note that, as virtuality can be exchanged between the jet and the medium, here $m_I^2$ can take negative values. If one integrates over $m_I^2$ obtains a $\delta$ function of $x^+$, and it reduces to the transverse momentum broadening spectrum in the conventional BDMPS-Z formalism~\cite{Baier:1996sk}. As detailed in Sec.~\ref{sec:BDMPS}, $\tilde{\mathcal{J}}^{LO}_k(x/2, -x/2; p_J)$ can be easily evaluated in 1+2 dimensions using either the dipole picture~\cite{Liou:2013qya, Wu:2014nca} or the path integral approach~\cite{Casalderrey-Solana:2007knd}, as it only additionally depends on $x^+$. 

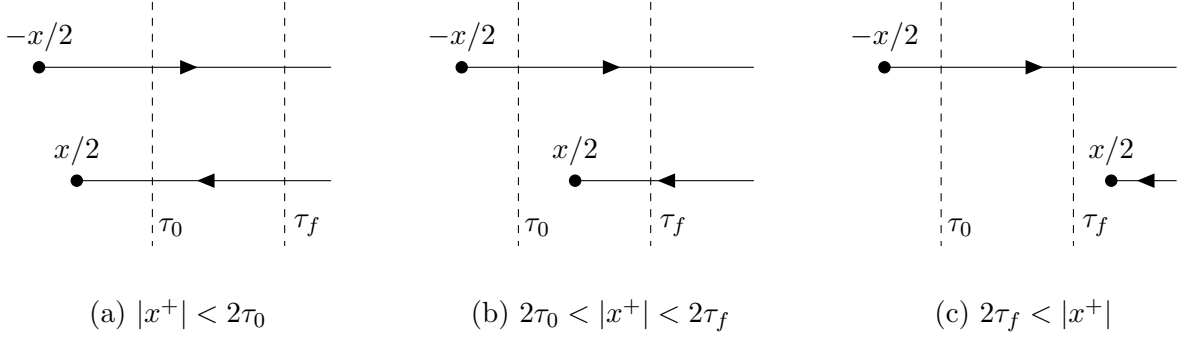
\begin{figure} [tp]
    \centering
    \begin{subfigure}[b]{0.32\textwidth}
        \centering
           \begin{tikzpicture}
  \begin{feynman}
  
    \node[dot, label = $-x/2$] (x);
    \node[right= 4 cm of x] (xf);
    
    \draw [fermion] (x) --  (xf);

    \node[below = 1.5 cm of x] (aux);
    \node[right = 0.5 cm of aux, dot, label=$x/2$] (xp);
    \node[right=3.5 cm of xp] (xfp);

    \draw[fermion] (xfp) -- (xp);

    \node[right= 1.5 cm of x] (aux2);
    \node[above=1 cm of aux2] (aux3);
    \node[below = 2.5 cm of aux2] (aux4) [label={[shift={(0.25, 0)}]$\tau_0$}];
    \node[right = 1.75 cm of aux3] (aux5);
    \node[right = 1.75 cm of aux4] (aux6) [label={[shift={(0.3, 0)}]$\tau_f$}];

    \draw[dashed] (aux3) -- (aux4);
    \draw[dashed] (aux5) -- (aux6);

  \end{feynman}
\end{tikzpicture}
        \caption{$\abs{x^+} < 2\tau_0$}
        \label{fig:LODiagram_a}
    \end{subfigure}
    \hfill
    \begin{subfigure}[b]{0.32\textwidth}
        \centering
        \begin{tikzpicture}
  \begin{feynman}
  
    \node[dot, label = $-x/2$] (x);
    \node[right= 4 cm of x] (xf);
    
    \draw [fermion] (x) --  (xf);

    \node[below = 1.5 cm of x] (aux);
    \node[right = 1.5 cm of aux, dot, label=$x/2$] (xp);
    \node[right=2.5 cm of xp] (xfp);

    \draw[fermion] (xfp) -- (xp);

    \node[right= 0.75 cm of x] (aux2);
    \node[above=1 cm of aux2] (aux3);
    \node[below = 2.5 cm of aux2] (aux4) [label={[shift={(0.25, 0)}]$\tau_0$}];
    \node[right = 1.75 cm of aux3] (aux5);
    \node[right = 1.75 cm of aux4] (aux6) [label={[shift={(0.3, 0)}]$\tau_f$}];

    \draw[dashed] (aux3) -- (aux4);
    \draw[dashed] (aux5) -- (aux6);

  \end{feynman}
\end{tikzpicture}
        \caption{$2\tau_0 < \abs{x^+} < 2\tau_f$}
        \label{fig:LODiagram_b}
    \end{subfigure}
    \hfill
    \begin{subfigure}[b]{0.32\textwidth}
        \centering
        \begin{tikzpicture}
  \begin{feynman}
  
    \node[dot, label = $-x/2$] (x);
    \node[right= 4 cm of x] (xf);
    
    \draw [fermion] (x) --  (xf);

    \node[below = 1.5 cm of x] (aux);
    \node[right = 3 cm of aux, dot, label=$x/2$] (xp);
    \node[right=1 cm of xp] (xfp);

    \draw[fermion] (xfp) -- (xp);

    \node[right= 0.75 cm of x] (aux2);
    \node[above=1 cm of aux2] (aux3);
    \node[below = 2.5 cm of aux2] (aux4) [label={[shift={(0.25, 0)}]$\tau_0$}];
    \node[right = 1.75 cm of aux3] (aux5);
    \node[right = 1.75 cm of aux4] (aux6) [label={[shift={(0.3, 0)}]$\tau_f$}];

    \draw[dashed] (aux3) -- (aux4);
    \draw[dashed] (aux5) -- (aux6);

  \end{feynman}
\end{tikzpicture}
        \caption{$2\tau_f < \abs{x^+}$}
        \label{fig:LODiagram_c}
    \end{subfigure}
    
\caption{\justifying Diagrams corresponding to the jet function at leading order. Three different situations are pictured: when the jet is created before the medium (left panel) in both the amplitude and the conjugate amplitude, when the jet is created inside the medium in one of the amplitudes (central panel), and when the jet is created after the medium in one of the amplitudes (right panel).}
\label{fig:LODiagrams}
\end{figure}

Let us specify the calculation for the case of a very high energy quark or gluon jet with $\mathbf{p}_J^2 \sim \hat{q}L^+ \ll p_J^+/L^+$ so that the dressed propagator is reduced to the Wilson line, as given in \cref{eq:Geik}. We will consider that the virtuality can be much larger than $\hat{q}L^+$, so we keep the $m_I^2$ term in the exponential of the initial parton propagator. There are three diagrams contributing to the jet function at LO, according to the creation point of the jet. This diagrams are depicted in \cref{fig:LODiagrams} for the case of a quark jet and similar diagrams will appear if we consider gluon jets.

The jet function associated with the diagram in \cref{fig:LODiagram_a} can be expressed, for an homogeneous medium, as 
\begin{align}
    \frac{dI_{|x^+|<2\tau_0}}{dm_I^2 d\mathbf{p}_J^2}
    & = 
    \int_{-2\tau_0}^{2\tau_0} \frac{d x^+}{2p_J^+(2\pi)}e^{i   \frac{m_I^2}{2 p_J^+} x^+} \frac{\sqrt{2}}{L^+ \hat{q}_R} \exp{-\frac{L^+ \hat{q}_R}{\sqrt{2}}\mathbf{p}_J^2}
    \, ,
\end{align}
where we used the two point function of the Wilson lines in \cref{eq:2pointWilson} and integrated over $\mathbf{x}$ and the azimuthal angle of the final particle momentum. Integration over $x^+$ is then trivial, yielding
\begin{align}\label{eq:LOBeforeResult}
    \begin{split}
    \frac{dI_{|x^+|<2\tau_0}}{dm_I^2 d\mathbf{p}_J^2}
    & = 
    \frac{1}{\pi m_I^2} \sin\left( \frac{m_I^2}{p_J^+} \tau_0 \right) \frac{\sqrt{2}}{\hat{q}_R L^+} \exp{-\frac{\sqrt{2}}{\hat{q}_R L^+} \mathbf{p}_J^2}
    \, .
    \end{split}
\end{align}
The result is equivalent to the BDMPS-Z transverse momentum broadening \cite{Baier:1996sk} with an additional oscillatory function that depends on the medium formation time $\tau_0$ and the jet formation time $t_J \equiv p_J^+/|m_I^2|$. This oscillation indicates that, if $\tau_0 \ll t_J$, the particle will enter the medium without radiating and we recover the BDMPS-Z results. However, if $t_J \ll \tau_0$, the highly virtual particle must radiate before entering the medium, so the contribution of this region of the phase space to the integrated LO jet function will be negligible. It is also easy to check that, if one integrates out the virtuality from $-\infty$ to $+\infty$, the BDMPS-Z result \cite{Baier:1996sk} is recovered exactly.

We can now consider the region $2\tau_0 < |x^+| < 2\tau_f$. There are two diagrams contributing, the one depicted in \cref{fig:LODiagram_b} and the equivalent diagram with $x^+ < 0$, which are related through complex conjugation. The corresponding jet function is then
\begin{align} \label{eq:LOInUnintegrated}
    \frac{dI_{2\tau_0 < |x^+|<2\tau_f}}{dm_I^2 d\mathbf{p}_J^2}
    = 2 \Re & \int_{2\tau_0}^{2\tau_f} \frac{d x^+}{2p_J^+(2\pi)}e^{i \frac{m_I^2}{2p_J^+} x^+} \notag \\
    & \times e^{-\frac{\frac{x^+}{2}-\tau_0}{2\lambda_R}} \frac{\sqrt{2}}{\hat{q}_R (\tau_f - \frac{x^+}{2})} \exp{- \frac{\sqrt{2}}{\hat{q}_R (\tau_f - \frac{x^+}{2})} \mathbf{p}_J^2}
    \, ,
\end{align}
where we used the 2-point function of the Wilson lines in \cref{eq:2pointWilson} for the region $\tau \in (x^+/2, \tau_f)$ and the 1-point function in \cref{eq:1pointWilson} for the region $\tau \in (\tau_0, x^+/2)$. The jet function contains the usual transverse momentum broadening gaussian factor from the region where the jet exits in both amplitudes, multiplied by an exponential suppression in the region where the jet only exists in one of the amplitudes. This suppression factor is interpreted as the probability that the color state of the jet in the amplitude is not modified by the medium interactions before the jet in the conjugated amplitude is created. Integration in $x^+$ is no longer trivial, but is admits a closed form in terms of generalized incomplete gamma functions
\begin{align} \label{eq:LOInResult}
    \frac{dI_{2\tau_0 < |x^+|<2\tau_f}}{dm_I^2 d\mathbf{p}_J^2}
     = \frac{\sqrt{2}}{\pi p_J^+ \hat{q}_R} & \Re\left\lbrace \exp{i \frac{m_I^2}{p_J^+}\tau_f - \frac{L^+}{2\lambda_R}} \right. \notag \\
    & \times \left. \Gamma\left(0, \frac{\sqrt{2}}{\hat{q}_R L^+} \mathbf{p}_J^2; - \frac{\sqrt{2}}{\hat{q}_R} \mathbf{p}_J^2 \left[ \frac{1}{2\lambda_R}-i \frac{m_I^2}{p_J^+} \right] \right)\right\rbrace
    \, ,
\end{align}
where the generalized upper incomplete gamma function is defined by
\begin{align}
    \Gamma(a, x; b) = \int_x^{\infty} dt t^{a-1} \exp{-t - \frac{b}{t}}\, .
\end{align}
\cref{eq:LOInResult} vanishes exactly if we integrate $m_I^2$ from $-\infty$ to $+\infty$. This is expected from the uncertainty principle $\Delta p_I^- \Delta x^+ \sim 1$, the integration over all the virtuality phase space completely delocalizes $p_I^-$ so the jet creation point is fixed $x^+ = 0$ and only the case where the jet is created before the medium in both amplitudes can contribute, which gives the BDMPS-Z result \cite{Baier:1996sk}. Numerical results for the jet function will be shown in \cref{sec:LOResults}.

In region pictured in \cref{fig:LODiagram_c} we also have two diagrams related by complex conjugation. The jet function reads
\begin{align} \label{eq:LOAfterUnintegrated}
    \frac{dI_{2\tau_f < |x^+|}}{dm_I^2 d\mathbf{p}_J^2}
    & = \delta(\mathbf{p}_J^2) e^{-\frac{L^+}{2\lambda_R}} 2\Re \int_{2\tau_f}^{+\infty} \frac{dx^+}{2p_J^+ (2\pi)} e^{i\frac{m_I^2 + i\varepsilon}{2p_J^+} x^+}
    \, .
\end{align}
Integrating over $x^+$ and using the Sokhotski-Plemelj theorem
\begin{align}\label{eq:SPTheorem}
    \frac{1}{x\pm i \varepsilon} = \mathrm{P.V} \left( \frac{1}{x} \right) \mp i\pi \delta(x)\, ,
\end{align}
where $\mathrm{P.V}$ indicates the principal value, it follows that
\begin{align}\label{eq:LOAfterResult}
    \begin{split}
    \frac{dI_{2\tau_f < |x^+|}}{dm_I^2 d\mathbf{p}_J^2}
    & = \delta(\mathbf{p}_J^2) e^{-\frac{L^+}{2\lambda_R}} \left[ \delta(m_I^2) - \frac{1}{\pi m_I^2 } \sin\left( \frac{m_I^2}{p_J^+} \tau_f \right) \right]
    \, .
    \end{split}
\end{align}
Again, due to the uncertainty principle, the contribution vanishes if we integrate over all the virtuality phase space. This diagram does not contribute to jet momentum broadening, as indicated by $\delta(\mathbf{p}_J^2)$, there is no net transverse momentum transfer from the medium to the jet, although virtuality can be exchanged. The jet function corresponds to vacuum propagation suppressed by the probability that the quark traverses the medium without getting its color state modified.

\subsection{Jet functions at NLO}
\label{sec:jetFuncNLO}

The above calculation can be straightforwardly generalized to the case with medium-induced radiation. We will again restrict ourselves to the case of soft gluon emission by an eikonal quark or gluon, such that $p_I^+ \approx p_J^+$. We will consider, for simplicity, that the jet is composed solely by the high-energy parton while the radiated gluon is always considered as energy lost outside of the jet. 
The formalism here developed can also be generalized to more complex jet algorithms useful for phenomenological comparisons to data, which we leave for future work. Also, as we are only interested in how the virtuality affects the probability of having a medium induced gluon emission, we integrate over the final transverse momenta of both the jet and the soft gluon and compute the energy spectrum of the radiated gluons as a function of the initial jet virtuality. The medium-induced soft gluon spectrum then reads
\begin{align}\label{eq:NLOSpectrum}
    k^+ \frac{dI^{NLO}}{ dk^+ dm_I^2}
    & = \frac{1}{4\pi} \int \frac{d^3 x}{(2\pi) 2p_J^+} e^{i \frac{m_I^2}{2p_J^+} x^+}\int \frac{d^2 \mathbf{p}_J}{(2\pi)^2} \frac{d^2 \mathbf{k}}{(2\pi)^2} [\tilde{\mathcal{J}}^{NLO}_q(x/2, -x/2; p_J, k) - (vac)]\, ,
\end{align}
where $(vac)$ represents the jet function in the absence of a QCD medium. We compute this spectrum using the BDMPS formalism and the harmonic oscillator approximation to model the medium.

As we mentioned in \cref{sec:jetFuncLO}, due to the uncertainty principle, when we compute virtuality differential jet functions, the jet creation point is not well localized, such that the jet can be created inside or after the medium in one of the amplitudes.
We will separate the three cases: when the jet is created before the medium in both amplitudes, when it is created inside the medium in one of them, and when it is created after the medium in one of them. For the NLO calculation there are several diagrams contributing to each of this cases. All the contributions will be finally put together in \cref{sec:NLOResults} to obtain the total spectrum. The calculation involves convolutions of different propagators, which are collected in \cref{app:convolutions}.

\subsubsection{Jet created before the medium in both amplitudes}

\begin{figure} [tp]
    \centering
    \begin{subfigure}[b]{0.45\textwidth}
        \centering
        \begin{tikzpicture}
  \begin{feynman}

    \node[dot, label = $-x/2$] (x0);
    \vertex[right=1.5 cm of x0] (x) [label={[yshift=-0.5cm, xshift=0.2cm] $x_s$}] ;
    \vertex[right=2.15 cm of x] (xf) [label={$p_J$}] ; 
    \vertex[above = 1.25 cm of xf] (yf) [label={$k$}];
    
    \draw [fermion] (x0) --  (x);
    \draw [fermion] (x) -- (xf);
    \draw [gluon] (x) -- (yf);

    \node[below=1 cm of x0] (aux);
    \node[right=0.75 cm of aux, dot, label={[yshift=-0.75 cm] $x/2$}] (x0bar);
    \vertex[right= 1.75 cm of x0bar] (xbar) [label=$y_s$];
    \vertex[right=1.15 cm of xbar] (xfbar) [label={$p_J$}] ; 
    \vertex[below = 0.75 cm of xfbar] (yfbar) [label={[yshift=-0.5cm] $k$}];

    \draw[fermion] (xbar) -- (x0bar);
    \draw[fermion] (xfbar) -- (xbar);
    \draw[gluon] (xbar) -- (yfbar);

    \node[right= -0.25 cm of x] (aux2);
    \node[above=2 cm of aux2] (aux3);
    \node[below = 3 cm of aux2] (aux4) [label={[shift={(0.25, 0)}]$\tau_0$}];
    \node[right = 1.75 cm of aux3] (aux5);
    \node[right = 1.75 cm of aux4] (aux6) [label={[shift={(0.25, -0.07)}]$\tau_f$}];

    \draw[dashed] (aux3) -- (aux4);
    \draw[dashed] (aux5) -- (aux6);

  \end{feynman}
\end{tikzpicture}
        \caption{In-in radiation}
        \label{fig:NLODiagram_before_inin}
    \end{subfigure}
    \hfill
    \begin{subfigure}[b]{0.45\textwidth}
        \centering
        \begin{tikzpicture}
  \begin{feynman}

    \node[dot, label = $-x/2$] (x0);
    \vertex[right=1.5 cm of x0] (x) [label={[yshift=-0.5cm, xshift=0.2cm] $x_s$}] ;
    \vertex[right=2.15 cm of x] (xf) [label={$p_J$}] ; 
    \vertex[above = 1.25 cm of xf] (yf) [label={$k$}];
    
    \draw [fermion] (x0) --  (x);
    \draw [fermion] (x) -- (xf);
    \draw [gluon] (x) -- (yf);

    \node[below=1 cm of x0] (aux);
    \node[right=0.75 cm of aux, dot, label={[yshift=-0.75 cm] $x/2$}] (x0bar);
    \vertex[right= 2 cm of x0bar] (xbar) [label=$y_s$];
    \vertex[right=0.9 cm of xbar] (xfbar) [label={$p_J$}] ; 
    \vertex[below = 0.75 cm of xfbar] (yfbar) [label={[yshift=-0.5cm] $k$}];

    \draw[fermion] (xbar) -- (x0bar);
    \draw[fermion] (xfbar) -- (xbar);
    \draw[gluon] (xbar) -- (yfbar);

    \node[right= -0.5 cm of x] (aux2);
    \node[above=2 cm of aux2] (aux3);
    \node[below = 3 cm of aux2] (aux4) [label={[shift={(0.25, 0)}]$\tau_0$}];
    \node[right = 1.25 cm of aux3] (aux5);
    \node[right = 1.25 cm of aux4] (aux6) [label={[shift={(0.25, -0.07)}]$\tau_f$}];

    \draw[dashed] (aux3) -- (aux4);
    \draw[dashed] (aux5) -- (aux6);

  \end{feynman}
\end{tikzpicture}
        \caption{In-af radiation}
        \label{fig:NLODiagram_before_inaf}
    \end{subfigure}
    \hfill
    \begin{subfigure}[b]{0.45\textwidth}
        \centering
        \begin{tikzpicture}
  \begin{feynman}

    \node[dot, label = $-x/2$] (x0);
    \vertex[right=1 cm of x0] (x) [label={[yshift=-0.5cm] $x_s$}] ;
    \vertex[right=2.75 cm of x] (xf) [label={$p_J$}] ; 
    \vertex[above = 1.25 cm of xf] (yf) [label={$k$}];
    
    \draw [fermion] (x0) --  (x);
    \draw [fermion] (x) -- (xf);
    \draw [gluon] (x) -- (yf);

    \node[below=1 cm of x0] (aux);
    \node[right=0.75 cm of aux, dot, label={[yshift=-0.75 cm] $x/2$}] (x0bar);
    \vertex[right= 1.5 cm of x0bar] (xbar) [label=$y_s$];
    \vertex[right=1.5 cm of xbar] (xfbar) [label={$p_J$}] ; 
    \vertex[below = 1 cm of xfbar] (yfbar) [label={[yshift=-0.5cm]$k$}];

    \draw[fermion] (xbar) -- (x0bar);
    \draw[fermion] (xfbar) -- (xbar);
    \draw[gluon] (xbar) -- (yfbar);

    \node[right= 0.55 cm of x] (aux2);
    \node[above=2 cm of aux2] (aux3);
    \node[below = 3 cm of aux2] (aux4) [label={[shift={(0.25, 0)}]$\tau_0$}];
    \node[right = 1.5 cm of aux3] (aux5);
    \node[right = 1.5 cm of aux4] (aux6) [label={[shift={(0.25, -0.07)}]$\tau_f$}];

    \draw[dashed] (aux3) -- (aux4);
    \draw[dashed] (aux5) -- (aux6);

  \end{feynman}
\end{tikzpicture}
        \caption{Bef-in radiation}
        \label{fig:NLODiagram_before_befin}
    \end{subfigure}
    \hfill
    \begin{subfigure}[b]{0.45\textwidth}
        \centering
        \begin{tikzpicture}
  \begin{feynman}

    \node[dot, label = $-x/2$] (x0);
    \vertex[right=1 cm of x0] (x) [label={[yshift=-0.5cm] $x_s$}] ;
    \vertex[right=2.75 cm of x] (xf) [label={$p_J$}] ; 
    \vertex[above = 1.25 cm of xf] (yf) [label={$k$}];
    
    \draw [fermion] (x0) --  (x);
    \draw [fermion] (x) -- (xf);
    \draw [gluon] (x) -- (yf);

    \node[below=1 cm of x0] (aux);
    \node[right=0.75 cm of aux, dot, label={[yshift=-0.75 cm] $x/2$}] (x0bar);
    \vertex[right= 2.25 cm of x0bar] (xbar) [label=$y_s$];
    \vertex[right=0.75 cm of xbar] (xfbar) [label={$p_J$}] ; 
    \vertex[below = 0.75 cm of xfbar] (yfbar) [label={[yshift=-0.5cm]$k$}];

    \draw[fermion] (xbar) -- (x0bar);
    \draw[fermion] (xfbar) -- (xbar);
    \draw[gluon] (xbar) -- (yfbar);

    \node[right= 0.5 cm of x] (aux2);
    \node[above=2 cm of aux2] (aux3);
    \node[below = 3 cm of aux2] (aux4) [label={[shift={(0.25, 0)}]$\tau_0$}];
    \node[right = 1 cm of aux3] (aux5);
    \node[right = 1 cm of aux4] (aux6) [label={[shift={(0.25, -0.07)}]$\tau_f$}];

    \draw[dashed] (aux3) -- (aux4);
    \draw[dashed] (aux5) -- (aux6);

  \end{feynman}
\end{tikzpicture}
        \caption{Bef-af radiation}
        \label{fig:LODiagram_befaf}
    \end{subfigure}
    
\caption{\justifying Diagrams corresponding to the case when the jet is created before the medium both in the amplitude and the conjugated amplitude, $|x^+| < 2\tau_0$. The different diagrams correspond to the different situations depending on whether the gluon is radiated before (bef), inside (in) or after (af) the medium.}
\label{fig:NLODiagrams_before}
\end{figure}
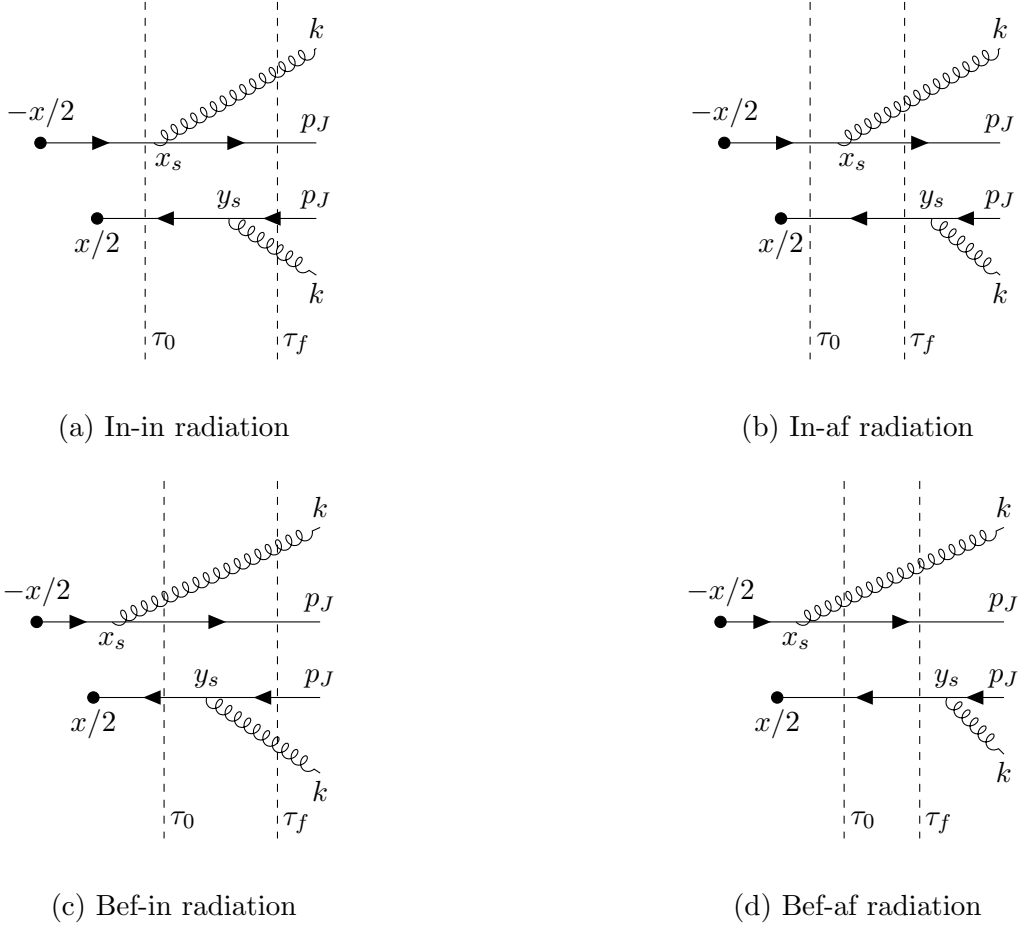

We start considering the case where the jet is created before the medium both in the amplitude and the conjugate amplitude, $|x^+| < 2\tau_0$. The diagrams contributing at NLO are shown in \cref{fig:NLODiagrams_before} (illustrated for the quark jet case) and the total contribution to the spectrum can be written as
\begin{align}
    dI_{|x^+|<2\tau_0} = dI_{|x^+|<2\tau_0}^{in-in} + dI_{|x^+|<2\tau_0}^{in-af} + dI_{|x^+|<2\tau_0}^{bef-in} + dI_{|x^+|<2\tau_0}^{bef-af}\, .
\end{align}
We have here used the fact that, after vacuum subtraction, the after-after and before-before jet functions do not contribute as they just correspond to vacuum emissions.

Let us start considering the diagram in \cref{fig:NLODiagram_before_inin}. The medium induced spectrum can be written as
\begin{align} \label{eq:SpectrumBeforeInin}
    k^+ \frac{dI_{|x^+|<2\tau_0}^{in-in}}{dk^+ dm_I^2} & =
    \frac{\alpha_s C_R}{(k^+)^2} 2\Re \int_{-2\tau_0}^{2\tau_0} \frac{dx^+}{2p_J^+ 2\pi} e^{i \frac{m_I^2}{2p_J^+}x^+} \int_{\tau_0}^{\tau_f} dy_s^+ \int_{\tau_0}^{y_s^+} dx_s^+ \notag \\
    & \times \nabla_\mathbf{u} \cdot \nabla_\mathbf{v} \left[\mathcal{K}(y_s^+, \mathbf{v}; x_s^+, \mathbf{u}; k^+) -  \mathcal{K}_0(y_s^+, \mathbf{v}; x_s^+, \mathbf{u}; k^+)\right]_{\mathbf{u=\mathbf{v=\mathbf{0}}}}\, .
\end{align}
If we consider the case of an homogeneous medium, the path integral can be solved analytically, yielding
\begin{align} \label{eq:RadiationKernelHarmonic}
    \mathcal{K}(t_2, \mathbf{x} & ; t_1, \mathbf{y}; p^+) \notag \\& = \frac{p^+ \Omega}{2\pi i \sin(\Omega \Delta t)} \exp{i \frac{p^+ \Omega}{2\sin(\Omega \Delta t)}\left[(\mathbf{x}^2 + \mathbf{y}^2) \cos(\Omega \Delta t) - 2 \mathbf{x} \cdot \mathbf{y} \right]}\, ,
\end{align}
where $\Delta t = t_2 - t_1$. Setting $\Omega \to 0$, we recover the coordinate space propagator of the free particle
\begin{align} \label{eq:FreePropagatorSolution}
    \mathcal{K}_0(t_2, \mathbf{x} & ; t_1, \mathbf{y}; p^+) = \frac{p^+}{2\pi i \Delta t} \exp{i \frac{p^+ }{2\Delta t}(\mathbf{x} - \mathbf{y})^2}\, .
\end{align}
Replacing this result in the jet function in \cref{eq:SpectrumBeforeInin}, it follows that
\begin{align} \label{eq:SpectrumBeforeIninHarmonic}
    k^+ \frac{dI_{|x^+|<2\tau_0}^{in-in}}{dk^+ dm_I^2} & = 
    \frac{2\alpha_s C_R}{\pi} \Re \int_{-2\tau_0}^{2\tau_0} \frac{dx^+}{2p_J^+ 2\pi} e^{i \frac{m_I^2}{2p_J^+}x^+} \log\left[ \frac{\sin(\Omega L^+)}{\Omega L^+} \right]\, .
\end{align}
It is easy to see that if one integrates $m_I^2$ over all the possible phase space, fixing $x^+ = 0$, then the BDMPS-Z result for the soft gluon spectrum for a quark approaching the medium from outside \cite{Baier:1998kq} is recovered.

In a similar way, the spectrum for the diagram in \cref{fig:NLODiagram_before_inaf} is
\begin{align} \label{eq:SpectrumBeforeInaf}
    k^+ & \frac{dI_{|x^+|<2\tau_0}^{in-af}}{dk^+ dm_I^2}  =
    \frac{\alpha_s C_R}{(k^+)^2} 2\Re \int_{-2\tau_0}^{2\tau_0} \frac{dx^+}{2p_J^+ 2\pi} e^{i \frac{m_I^2}{2p_J^+}x^+} \int_{\tau_0}^{\tau_f} dx_s^+ \int_{\tau_f}^{+\infty} dy_s^+ \nabla_\mathbf{u} \cdot \nabla_\mathbf{v} \notag \\
    & \times \left[\int d^2\mathbf{x_f} \mathcal{K}_0(y_s^+, \mathbf{v}; \tau_f, \mathbf{x_f}; k^+) \mathcal{K}(\tau_f, \mathbf{x_f}; x_s^+, \mathbf{u}; k^+) -  \mathcal{K}_0(y_s^+, \mathbf{v}; x_s^+, \mathbf{u}; k^+)\right]_{\mathbf{u=\mathbf{v=\mathbf{0}}}}\, ,
\end{align}
which, after solving the path integrals and performing the convolution of the propagators as in \cref{app:convolutions}, reads
\begin{align} \label{eq:SpectrumBeforeInafHarmonic}
    k^+ \frac{dI_{|x^+|<2\tau_0}^{in-af}}{dk^+ dm_I^2} & = 
    \frac{2\alpha_s C_R}{\pi} \Re \int_{-2\tau_0}^{2\tau_0} \frac{dx^+}{2p_J^+ 2\pi} e^{i \frac{m_I^2}{2p_J^+}x^+} \log\left[ \Omega L^+ \cot(\Omega L^+) \right]\, .
\end{align}
Again, the BDMPS-Z result for this diagram \cite{Baier:1996kr, Zakharov:1997uu, Baier:1998kq} is recover when we integrate over the initial jet virtuality.

For the diagram in \cref{fig:NLODiagram_before_befin}, we have
\begin{align} \label{eq:SpectrumBeforeBefin}
    k^+ & \frac{dI_{|x^+|<2\tau_0}^{bef-in}}{dk^+ dm_I^2} = 
    \frac{\alpha_s C_R}{(k^+)^2} 2\Re \int_{-2\tau_0}^{2\tau_0} \frac{dx^+}{2p_J^+ 2\pi} e^{i \frac{m_I^2}{2p_J^+}x^+} \int_{-x^+/2}^{\tau_0} dx_s^+ \int_{\tau_0}^{\tau_f} dy_s^+ \nabla_\mathbf{u} \cdot \nabla_\mathbf{v} \notag \\
    & \times \left[\int d^2\mathbf{x_i}\, \mathcal{K}(y_s^+, \mathbf{v}; \tau_0, \mathbf{x_i}; k^+) \mathcal{K}_0(\tau_0, \mathbf{x_i}; x_s^+, \mathbf{u}; k^+) -  \mathcal{K}_0(y_s^+, \mathbf{v}; x_s^+, \mathbf{u}; k^+)\right]_{\mathbf{u=\mathbf{v=\mathbf{0}}}}\, ,
\end{align}
which in the harmonic oscillator approximation, using the convolutions in \cref{app:convolutions}, reduces to
\begin{align} \label{eq:SpectrumBeforeBefinfHarmonic}
    k^+ \frac{dI_{|x^+|<2\tau_0}^{bef-in}}{dk^+ dm_I^2} =
    \frac{2\alpha_s C_R}{\pi} & \Re \int_{-2\tau_0}^{2\tau_0} \frac{dx^+}{2p_J^+ 2\pi} e^{i \frac{m_I^2}{2p_J^+}x^+} \notag \\
    & \times \log\left[ \frac{L^+}{\tau_f + \frac{x^+}{2}} \left(1 + \left(\tau_0 + \frac{x^+}{2} \right) \Omega \cot(\Omega L^+) \right) \right]\, .
\end{align}
If one integrates over $m_I^2$, fixing $x^+=0$, and takes the limit $\tau_0 \to \infty$, then one recovers
\begin{align}
    \left.  k^+ \frac{dI_{|x^+|<2\tau_0}^{bef-in}}{dk^+} \right|_{BDMPS} = \left.  k^+ \frac{dI_{|x^+|<2\tau_0}^{in-af}}{dk^+} \right|_{BDMPS} = \frac{2\alpha_s C_R}{\pi}  \log\left| \Omega L^+ \cot(\Omega L^+) \right| \, ,
\end{align}
as the radiation before and after the medium should be equivalent in this limit. 

Finally, we have to consider the diagram in \cref{fig:LODiagram_befaf}:
\begin{align} \label{eq:SpectrumBeforeBefaf}
    k^+ \frac{dI_{|x^+|<2\tau_0}^{bef-af}}{dk^+ dm_I^2} & = 
    \frac{\alpha_s C_R}{(k^+)^2} 2\Re \int_{-2\tau_0}^{2\tau_0} \frac{dx^+}{2p_J^+ 2\pi} e^{i \frac{m_I^2}{2p_J^+}x^+} \int_{-x^+/2}^{\tau_0} dx_s^+ \int_{\tau_f}^{+\infty} dy_s^+ \notag \\
    & \times \nabla_\mathbf{u} \cdot \nabla_\mathbf{v} \left[\int d^2\mathbf{x_i} d^2 \mathbf{x_f} \mathcal{K}_0(y_s^+, \mathbf{v}; \tau_f, \mathbf{x_f}; k^+) \mathcal{K}(\tau_f, \mathbf{x_f}; \tau_0, \mathbf{x_i}; k^+) \right. \notag \\
    & \times \left. \mathcal{K}_0(\tau_0, \mathbf{x_i}; x_s^+, \mathbf{u}; k^+) -  \mathcal{K}_0(y_s^+, \mathbf{v}; x_s^+, \mathbf{u}; k^+)\right]_{\mathbf{u=\mathbf{v=\mathbf{0}}}}\, .
\end{align}
Replacing the harmonic propagators and performing the double convolution as in \cref{app:convolutions}, one can obtain
\begin{align} \label{eq:SpectrumBeforeBefinfHarmonic}
    k^+ \frac{dI_{|x^+|<2\tau_0}^{bef-af}}{dk^+ dm_I^2} =
    \frac{2\alpha_s C_R}{\pi} & \Re \int_{-2\tau_0}^{2\tau_0} \frac{dx^+}{2p_J^+ 2\pi} e^{i \frac{m_I^2}{2p_J^+}x^+} \notag \\
    & \times \log\left[ \frac{\tau_f + \frac{x^+}{2}}{L^+}\, \tan(\Omega L^+) \frac{1 - \left(\tau_0 + \frac{x^+}{2}\right) \Omega \tan(\Omega L^+)}{\left(\tau_0 + \frac{x^+}{2}\right) \Omega +  \tan(\Omega L^+)} \right]\, .
\end{align}

We can now combine the four diagrams together to obtain the total spectrum of the medium induced emissions for the case when the jet is created before the medium in the amplitude and the conjugate amplitude, as a function of the initial jet virtuality. The spectrum reads
\begin{align} \label{eq:SpectrumBefore}
    k^+ \frac{dI_{|x^+|<2\tau_0}}{dk^+ dm_I^2}  = \frac{2\alpha_s C_R}{\pi} & \Re \int_{-2\tau_0}^{2\tau_0} \frac{dx^+}{2p_J^+ 2\pi} e^{i \frac{m_I^2}{2p_J^+}x^+} \notag \\
    & \times \log\left[ \cos(\Omega L^+) - \left(\tau_0 + \frac{x^+}{2} \right) \Omega \sin(\Omega L^+) \right]\, .
\end{align}
There exist large cancellations between the contributions of the different diagrams, in such a way that they cannot be understood independently, but the contribution of this piece of the phase space to the soft gluon radiation must be understood as a whole. If we now integrate $m_I^2$ from $-\infty$ to $+\infty$, fixing $x^+ = 0$, we recover the BDMPS-Z result with a delayed medium formation time \cite{Andres:2022bql}. When one takes $\tau_0= 0$, the soft gluon spectrum after integrating over $m_I^2$ is identical to that for the quark being produced in the medium~\cite{Baier:1998kq}. This indicates that the BDMPS-Z medium induced radiation is equivalent to our factorization formalism when the information about the initial virtuality of the jet is neglected and $m_I^2$ is integrated over. Moreover, our observation here and that under \cref{eq:SpectrumBeforeIninHarmonic} call into question many claims that the BDMPS-Z results only describe radiative energy loss of high-energy partons nearly on mass shell, whereas in our formalism the BDMPS-Z spectrum is obtained integrating over all the virtuality phase-space, with contributions predominantly given by initial partons with virtualities up to $m_I^2\lesssim p^+_I/\tau_0$. 

If we only had the cosine term in the logarithm, integration over $x^+$ would be trivial, yielding the same oscillatory factor as for the $\mathbf{p}_J$ distribution \cref{eq:LOBeforeResult} in Sec. \ref{sec:jetFuncLO}, depending on the medium and jet formation times. This indicates that, for high virtuality, the jet cannot enter the medium without radiating. Including second term integration is more complicated, but we can also express the result in a closed analytic form:
\begin{align} \label{eq:SpectrumBeforeIntegrated}
    k^+ \frac{dI_{|x^+|<2\tau_0}}{dk^+ dm_I^2} & =
    \frac{2\alpha_s C_R}{\pi} \Re \frac{1}{2\pi i} \frac{1}{m_I^2} \left\lbrace e^{i \frac{m_I^2}{p_J^+} \tau_0} \log\left(\cos(\Omega L^+) - 2\Omega \tau_0 \sin(\Omega L^+) \right) \right. \notag \\
    & \left. - e^{-i \frac{m_I^2}{p_J^+} \tau_0} \log(\cos(\Omega L^+)) - e^{-i \frac{m_I^2}{p_J^+} \left( \tau_0 - \frac{\cot(\Omega L^+)}{\Omega} \right)} \right. \notag \\
    & \left. \times \left[E_1\left(-i\frac{m_I^2}{p_J^+} \left( \frac{\cot(\Omega L^+)}{\Omega} - 2\tau_0 \right) \right) - E_1\left(-i\frac{m_I^2}{p_J^+} \frac{\cot(\Omega L^+)}{\Omega} \right) \right] \right\rbrace\, ,
\end{align}
where we introduced the principal-branch exponential integral
\begin{align} \label{eq:ExponentialIntegral}
    E_1(z) \equiv \Gamma(0,z) = \int_z^\infty \frac{e^{-t}}{t} dt\, .
\end{align}
The physical interpretation is not immediately transparent from this analytic formula, so it is useful to input in the result typical parameter values to study the phenomenological consequences.
Numerical results for this contribution to the spectrum will be provided in \ref{sec:NLOResults}.

\subsubsection{Jet created inside the medium in one of the amplitudes}

\begin{figure} [tp]
    \centering
    \begin{subfigure}[b]{0.32\textwidth}
        \centering
        \begin{tikzpicture}
  \begin{feynman}

    \node[dot, label = $-x/2$] (x0);
    \vertex[right=1.75 cm of x0] (x) [label={[yshift=-0.5cm, xshift=0.2cm] $x_s$}] ;
    \vertex[right=1.9 cm of x] (xf) [label={$p_J$}] ; 
    \vertex[above = 1.25 cm of xf] (yf) [label={$k$}];
    
    \draw [fermion] (x0) --  (x); 
    \draw [fermion] (x) -- (xf);
    \draw [gluon] (x) -- (yf);

    \node[below=1 cm of x0] (aux);
    \node[right=1.35 cm of aux, dot, label={[yshift=-0.75 cm] $x/2$}] (x0bar);
    \vertex[right= 0.9 cm of x0bar] (xbar) [label=$y_s$];
    \vertex[right=1.4 cm of xbar] (xfbar) [label={$p_J$}] ; 
    \vertex[below = 0.75 cm of xfbar] (yfbar) [label={[yshift=-0.5cm] $k$}];

    \draw[fermion] (xbar) -- (x0bar);
    \draw[fermion] (xfbar) -- (xbar);
    \draw[gluon] (xbar) -- (yfbar);

    \node[right= -0.95 cm of x] (aux2);
    \node[above=2 cm of aux2] (aux3);
    \node[below = 3 cm of aux2] (aux4) [label={[shift={(0.25, 0)}]$\tau_0$}];
    \node[right = 1.75 cm of aux3] (aux5);
    \node[right = 1.75 cm of aux4] (aux6) [label={[shift={(0.25, -0.07)}]$\tau_f$}];

    \draw[dashed] (aux3) -- (aux4);
    \draw[dashed] (aux5) -- (aux6);

  \end{feynman}
\end{tikzpicture}
        \caption{In-in radiation, $|x^+| < 2x_s^+$}
        \label{fig:NLODiagram_in_inin_a}
    \end{subfigure}
    \hfill
    \begin{subfigure}[b]{0.32\textwidth}
        \centering
        \begin{tikzpicture}
  \begin{feynman}

    \node[dot, label = $-x/2$] (x0);
    \vertex[right=1.5 cm of x0] (x) [label={[yshift=-0.5cm, xshift=0.2cm] $x_s$}] ;
    \vertex[right=2.15 cm of x] (xf) [label={$p_J$}] ; 
    \vertex[above = 1.25 cm of xf] (yf) [label={$k$}];
    
    \draw [fermion] (x0) --  (x);
    \draw [fermion] (x) -- (xf);
    \draw [gluon] (x) -- (yf);

    \node[below=1 cm of x0] (aux);
    \node[right=2.25 cm of aux, dot, label={[yshift=-0.75 cm] $x/2$}] (x0bar);
    \vertex[right= 0.6 cm of x0bar] (xbar) [label=$y_s$];
    \vertex[right=0.8 cm of xbar] (xfbar) [label={$p_J$}] ; 
    \vertex[below = 0.75 cm of xfbar] (yfbar) [label={[yshift=-0.5cm] $k$}];

    \draw[fermion] (xbar) -- (x0bar);
    \draw[fermion] (xfbar) -- (xbar);
    \draw[gluon] (xbar) -- (yfbar);

    \node[right= -0.25 cm of x] (aux2);
    \node[above=2 cm of aux2] (aux3);
    \node[below = 3 cm of aux2] (aux4) [label={[shift={(0.25, 0)}]$\tau_0$}];
    \node[right = 1.75 cm of aux3] (aux5);
    \node[right = 1.75 cm of aux4] (aux6) [label={[shift={(0.25, -0.07)}]$\tau_f$}];

    \draw[dashed] (aux3) -- (aux4);
    \draw[dashed] (aux5) -- (aux6);

  \end{feynman}
\end{tikzpicture}
        \caption{In-in radiation, $|x^+| > 2x_s^+$}
        \label{fig:NLODiagram_in_inin_b}
    \end{subfigure}
    \hfill
    \begin{subfigure}[b]{0.32\textwidth}
        \centering
        \begin{tikzpicture}
  \begin{feynman}

    \node[dot, label = $-x/2$] (x0);
    \vertex[right=1.5 cm of x0] (x) [label={[yshift=-0.5cm, xshift=0.2cm] $x_s$}] ;
    \vertex[right=2.15 cm of x] (xf) [label={$p_J$}] ; 
    \vertex[above = 1.25 cm of xf] (yf) [label={$k$}];
    
    \draw [fermion] (x0) --  (x);
    \draw [fermion] (x) -- (xf);
    \draw [gluon] (x) -- (yf);

    \node[below=1 cm of x0] (aux);
    \node[right=1.25 cm of aux, dot, label={[yshift=-0.75 cm] $x/2$}] (x0bar);
    \vertex[right= 1.5 cm of x0bar] (xbar) [label=$y_s$];
    \vertex[right=0.9 cm of xbar] (xfbar) [label={$p_J$}] ; 
    \vertex[below = 0.75 cm of xfbar] (yfbar) [label={[yshift=-0.5cm] $k$}];

    \draw[fermion] (xbar) -- (x0bar);
    \draw[fermion] (xfbar) -- (xbar);
    \draw[gluon] (xbar) -- (yfbar);

    \node[right= -0.75 cm of x] (aux2);
    \node[above=2 cm of aux2] (aux3);
    \node[below = 3 cm of aux2] (aux4) [label={[shift={(0.25, 0)}]$\tau_0$}];
    \node[right = 1.5 cm of aux3] (aux5);
    \node[right = 1.5 cm of aux4] (aux6) [label={[shift={(0.25, -0.07)}]$\tau_f$}];

    \draw[dashed] (aux3) -- (aux4);
    \draw[dashed] (aux5) -- (aux6);

  \end{feynman}
\end{tikzpicture}
        \caption{In-af radiation, $|x^+| < 2x_s^+$}
        \label{fig:NLODiagram_in_inaf_a}
    \end{subfigure}
    \hfill
    \begin{subfigure}[b]{0.32\textwidth}
        \centering
        \begin{tikzpicture}
  \begin{feynman}

    \node[dot, label = $-x/2$] (x0);
    \vertex[right=1 cm of x0] (x) [label={[yshift=-0.5cm, xshift=0.2cm] $x_s$}] ;
    \vertex[right=2.65 cm of x] (xf) [label={$p_J$}] ; 
    \vertex[above = 1.25 cm of xf] (yf) [label={$k$}];
    
    \draw [fermion] (x0) --  (x);
    \draw [fermion] (x) -- (xf);
    \draw [gluon] (x) -- (yf);

    \node[below=1 cm of x0] (aux);
    \node[right=1.85 cm of aux, dot, label={[yshift=-0.75 cm] $x/2$}] (x0bar);
    \vertex[right= 0.9 cm of x0bar] (xbar) [label=$y_s$];
    \vertex[right=0.9 cm of xbar] (xfbar) [label={$p_J$}] ; 
    \vertex[below = 0.75 cm of xfbar] (yfbar) [label={[yshift=-0.5cm] $k$}];

    \draw[fermion] (xbar) -- (x0bar);
    \draw[fermion] (xfbar) -- (xbar);
    \draw[gluon] (xbar) -- (yfbar);

    \node[right= -0.35 cm of x] (aux2);
    \node[above=2 cm of aux2] (aux3);
    \node[below = 3 cm of aux2] (aux4) [label={[shift={(0.25, 0)}]$\tau_0$}];
    \node[right = 1.5 cm of aux3] (aux5);
    \node[right = 1.5 cm of aux4] (aux6) [label={[shift={(0.25, -0.07)}]$\tau_f$}];

    \draw[dashed] (aux3) -- (aux4);
    \draw[dashed] (aux5) -- (aux6);

  \end{feynman}
\end{tikzpicture}
        \caption{In-af radiation, $|x^+| > 2x_s^+$}
        \label{fig:NLODiagram_in_inaf_b}
    \end{subfigure}
    \hfill
    \begin{subfigure}[b]{0.32\textwidth}
        \centering
        \begin{tikzpicture}
  \begin{feynman}

    \node[dot, label = $-x/2$] (x0);
    \vertex[right=0.75 cm of x0] (x) [label={[yshift=-0.5cm] $x_s$}] ;
    \vertex[right=3 cm of x] (xf) [label={$p_J$}] ; 
    \vertex[above = 1.25 cm of xf] (yf) [label={$k$}];
    
    \draw [fermion] (x0) --  (x);
    \draw [fermion] (x) -- (xf);
    \draw [gluon] (x) -- (yf);

    \node[below=1 cm of x0] (aux);
    \node[right=1.85 cm of aux, dot, label={[yshift=-0.75 cm] $x/2$}] (x0bar);
    \vertex[right= 0.65 cm of x0bar] (xbar) [label=$y_s$];
    \vertex[right=1.25 cm of xbar] (xfbar) [label={$p_J$}] ; 
    \vertex[below = 0.75 cm of xfbar] (yfbar) [label={[yshift=-0.5cm]$k$}];

    \draw[fermion] (xbar) -- (x0bar);
    \draw[fermion] (xfbar) -- (xbar);
    \draw[gluon] (xbar) -- (yfbar);

    \node[right= 0.35 cm of x] (aux2);
    \node[above=2 cm of aux2] (aux3);
    \node[below = 3 cm of aux2] (aux4) [label={[shift={(0.25, 0)}]$\tau_0$}];
    \node[right = 1.5 cm of aux3] (aux5);
    \node[right = 1.5 cm of aux4] (aux6) [label={[shift={(0.25, -0.07)}]$\tau_f$}];

    \draw[dashed] (aux3) -- (aux4);
    \draw[dashed] (aux5) -- (aux6);

  \end{feynman}
\end{tikzpicture}
        \caption{Bef-in radiation}
        \label{fig:NLODiagram_in_befin}
    \end{subfigure}
    \hfill
    \begin{subfigure}[b]{0.32\textwidth}
        \centering
        \begin{tikzpicture}
  \begin{feynman}

    \node[dot, label = $-x/2$] (x0);
    \vertex[right=0.75 cm of x0] (x) [label={[yshift=-0.5cm] $x_s$}] ;
    \vertex[right=3 cm of x] (xf) [label={$p_J$}] ; 
    \vertex[above = 1.25 cm of xf] (yf) [label={$k$}];
    
    \draw [fermion] (x0) --  (x);
    \draw [fermion] (x) -- (xf);
    \draw [gluon] (x) -- (yf);

    \node[below=1 cm of x0] (aux);
    \node[right=1.85 cm of aux, dot, label={[yshift=-0.75 cm] $x/2$}] (x0bar);
    \vertex[right= 1.25 cm of x0bar] (xbar) [label=$y_s$];
    \vertex[right=0.65 cm of xbar] (xfbar) [label={$p_J$}] ; 
    \vertex[below = 0.75 cm of xfbar] (yfbar) [label={[yshift=-0.5cm]$k$}];

    \draw[fermion] (xbar) -- (x0bar);
    \draw[fermion] (xfbar) -- (xbar);
    \draw[gluon] (xbar) -- (yfbar);

    \node[right= 0.35 cm of x] (aux2);
    \node[above=2 cm of aux2] (aux3);
    \node[below = 3 cm of aux2] (aux4) [label={[shift={(0.25, 0)}]$\tau_0$}];
    \node[right = 1.5 cm of aux3] (aux5);
    \node[right = 1.5 cm of aux4] (aux6) [label={[shift={(0.25, -0.07)}]$\tau_f$}];

    \draw[dashed] (aux3) -- (aux4);
    \draw[dashed] (aux5) -- (aux6);

  \end{feynman}
\end{tikzpicture}
        \caption{Bef-af radiation}
        \label{fig:NLODiagram_in_befaf}
    \end{subfigure}
    
\caption{\justifying Diagrams corresponding to the case when the jet is created inside the medium in the conjugated amplitude $2 \tau_0 < |x^+| < 2\tau_f$. The different diagrams correspond to the different situations depending on whether the gluon is radiated before (bef), inside (in) or after (af) the medium. We also show separately the two situations: when the gluon in the amplitude is radiated before the jet in the conjugate amplitude is created, and when it is radiated after.}
\label{fig:NLODiagrams_in}
\end{figure}
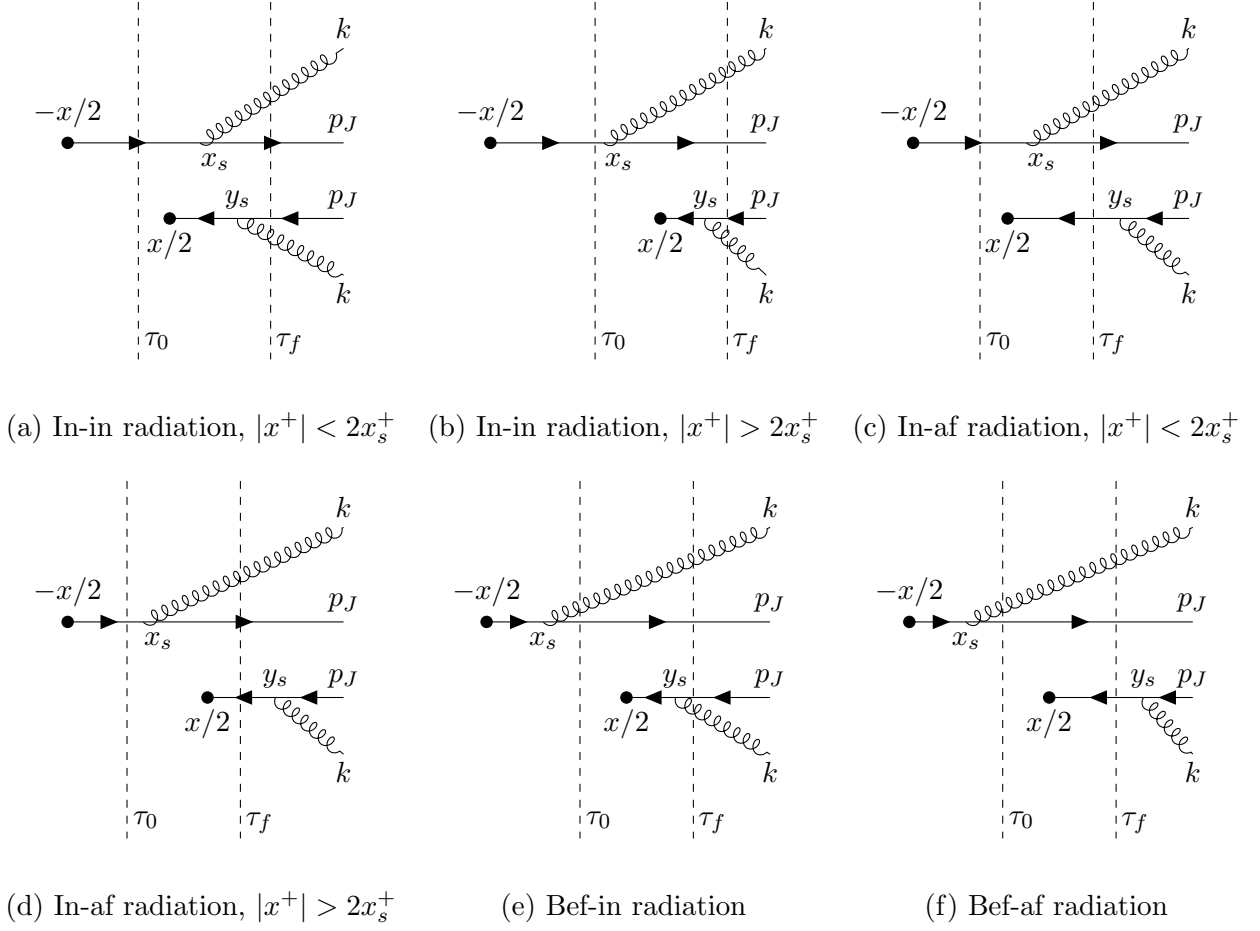

We now consider the case where the jet in the amplitude or in the conjugated amplitude is created inside the medium, $2\tau_0 < |x^+| <2\tau_f$. These diagrams are completely absent in the BDMPS-Z calculation \cite{Baier:1996kr, Baier:1998kq}, where $x^+ = 0$, and we expect their contribution to vanish if we integrate over the initial jet virtuality. The total spectrum for this region of the phase space can be decomposed as
\begin{align}
    dI_{2\tau_0<|x^+|<2\tau_f} & = dI_{2\tau_0<|x^+|<2\tau_f}^{in-in, |x^+|<2x_s^+} + dI_{2\tau_0<|x^+|<2\tau_f}^{in-in, |x^+|>2x_s^+} + dI_{2\tau_0<|x^+|<2\tau_f}^{in-af, |x^+| < 2x_s^+} \notag \\
    & + dI_{2\tau_0<|x^+|<2\tau_f}^{in-af, |x^+| > 2x_s^+} + dI_{2\tau_0<|x^+|<2\tau_f}^{bef-in} + dI_{2\tau_0<|x^+|<2\tau_f}^{bef-af}\, ,
\end{align}
with the corresponding diagrams shown in \cref{fig:NLODiagrams_in} for the quark case. All this diagrams contain different convolutions of in-medium and vacuum propagators, which are collected in \cref{app:convolutions}.

All these diagrams contain the convolution of the propagation of the jet in the amplitude before the jet in the conjugate amplitude has been created and the jet-dipole propagation after it has been created in both amplitudes. In the derivation of the jet function, these regions factorize so that we can write
\begin{align} \label{eq:PropagationFactorization}
    \langle W(x^+/2, \tau_0; \mathbf{x}) & W(\tau_f, x^+/2; \mathbf{x}) W^\dagger(\tau_f, x^+/2; \mathbf{x})\rangle \notag \\
    & = \langle W(x^+/2, \tau_0; \mathbf{x}) \rangle  \langle W(\tau_f, x^+/2; \mathbf{x}) W^\dagger(\tau_f, x^+/2; \mathbf{y}) \rangle \notag \\
    & = e^{-\frac{x^+/2-\tau_0}{2\lambda_R}} \langle W(\tau_f, x^+/2; \mathbf{x}) W^\dagger(\tau_f, x^+/2; \mathbf{y}) \rangle\, .
\end{align}
In the total absence of a QCD medium, 
\begin{align}
    \langle W(x^+/2, \tau_0; \mathbf{x}) & W(\tau_f, x^+/2; \mathbf{x}) W^\dagger(\tau_f, x^+/2; \mathbf{x})\rangle_{vac} = 1\, .
\end{align}
Note that subtracting this quantity from the jet function is not sufficient to eliminate all the vacuum-like radiation from the splitting kernel and the virtuality dependent spectrum will present infrared (IR) divergences. The presence of the medium allows for vacuum-like radiation in regions of the phase space where it was forbidden in pure vacuum: the soft gluon can be radiated without undergoing any interactions with medium constituents in both the amplitude and the conjugate amplitude, while the quark scatters with medium particles before $x^+/2$ in the amplitude. That is, the virtuality exchange between the medium and the jet will allow for this radiation which is still vacuum-like and does not share the structure of medium induced radiation. To completely cancel such vacuum radiation from the jet function, one must first factorize the propagation as in \cref{eq:PropagationFactorization} and then take the $\hat{q} \to 0$ limit in the remaining correlator
\begin{align} \label{eq:VacuumSUbtraction}
    \langle W(x^+/2, \tau_0; \mathbf{x}) & W(\tau_f, x^+/2; \mathbf{x}) W^\dagger(\tau_f, x^+/2; \mathbf{x})\rangle_{vac} \notag \\
    & = e^{-\frac{x^+/2-\tau_0}{2\lambda_R}} \langle W(\tau_f, x^+/2; \mathbf{x}) W^\dagger(\tau_f, x^+/2; \mathbf{y}) \rangle_{vac}\, ,
\end{align}
such that we will have quantities of the type $e^{-\Delta t/(2\lambda_R)} (\mathcal{K} - \mathcal{K}_0)$ which no longer contains the divergent vacuum term of the harmonic oscillator. In this work we will chose this second method to include only medium induced radiation, the spectrum will then not contain IR divergences.

We start considering the diagrams where the gluon in the amplitude is radiated after the jet in the conjugated amplitude is created. For the situation pictured in \cref{fig:NLODiagram_in_inin_a}, the spectrum reads
\begin{align} \label{eq:SpectrumInInin_a}
    k^+ \frac{dI_{2\tau_0<|x^+|<2\tau_f}^{in-in, |x^+| < 2x_s^+}}{dk^+ dm_I^2}  & = 
    \frac{\alpha_s C_R}{(k^+)^2} 2\Re \int_{2\tau_0}^{2\tau_f} \frac{dx^+}{2p_J^+ 2\pi} e^{i \frac{m_I^2}{2p_J^+}x^+} e^{-\frac{\frac{x^+}{2}-\tau_0}{2 \lambda_R}} 2 \Re \int_{x^+/2}^{\tau_f} dy_s^+  \notag \\
    & \hspace{-40 pt} \times \int_{x^+/2}^{y_s^+} dx_s^+ \nabla_\mathbf{u} \cdot \nabla_\mathbf{v} \left[\mathcal{K}(y_s^+, \mathbf{v}; x_s^+, \mathbf{u}; k^+) -  \mathcal{K}_0(y_s^+, \mathbf{v}; x_s^+, \mathbf{u}; k^+)\right]_{\mathbf{u=\mathbf{v=\mathbf{0}}}}\, ,
\end{align}
where the overall factor $2$ in front of the real part of the final result comes from considering both $x^+ >0$ and $x^+<0$, while the other factor of $2$ in front the real part of the $y_s^+$ integral comes from combining $y_s^+ > x_s^+$ with $y_s^+ < x_s^+$. For the homogeneous medium we can write
\begin{align} \label{eq:SpectrumInInin_a_Harmonic}
    k^+ \frac{dI_{2\tau_0<|x^+|<2\tau_f}^{in-in, |x^+|<2x_s^+}}{dk^+ dm_I^2} & = 
    \frac{2\alpha_s C_R}{\pi}  2 \Re \int_{2\tau_0}^{2\tau_f} \frac{dx^+}{2p_J^+ 2\pi} e^{i \frac{m_I^2}{2p_J^+}x^+} e^{-\frac{\Delta_0}{2 \lambda_R}} \log\left| \frac{\sin(\Omega \Delta_f )}{\Omega \Delta_f} \right|\, ,
\end{align}
where we defined, for brevity,
\begin{align} \label{eq:TimeIntervalVariables}
    \Delta_0 \equiv \frac{x^+}{2} - \tau_0 \qquad \text{and} \qquad \Delta_f \equiv \tau_f - \frac{x^+}{2}\, .
\end{align}
These definitions will be used in the derivations that follow.
The contribution of this diagram is therefore similar to the one of \cref{eq:SpectrumBeforeIninHarmonic} by replacing the time that the jet in the conjugated amplitude travels inside the medium $L^+ \to \tau_f - x^+/2$. The spectrum in \cref{eq:SpectrumInInin_a_Harmonic} is also suppressed by the probability that the color state of the quark in the amplitude is not changed before the jet in the conjugated amplitude is created, so that the net color state of the system remains a singlet.

In a similar way, for the diagram in \cref{fig:NLODiagram_in_inaf_a}, we find
\begin{align} \label{eq:SpectrumInInaf_a}
    & k^+ \frac{dI_{2\tau_0<|x^+|<2\tau_f}^{in-af, |x^+| < 2x_s^+}}{dk^+ dm_I^2} =
    \frac{\alpha_s C_R}{(k^+)^2}  2\Re  \int_{2\tau_0}^{2\tau_f} \frac{dx^+}{2p_J^+ 2\pi} e^{i \frac{m_I^2}{2p_J^+}x^+} e^{-\frac{\frac{x^+}{2}-\tau_0}{2 \lambda_R}} 2 \Re \int_{x^+/2}^{\tau_f} dx_s^+ \int_{\tau_f}^{+\infty} dy_s^+ \nabla_\mathbf{u} \cdot \nabla_\mathbf{v} \notag \\
    & \hspace{20 pt} \times \left[\int d^2\mathbf{x_f}\, \mathcal{K}_0(y_s^+, \mathbf{v}; \tau_f, \mathbf{x_f}; k^+)\mathcal{K}(\tau_f, \mathbf{x_f}; x_s^+, \mathbf{u}; k^+) -  \mathcal{K}_0(y_s^+, \mathbf{v}; x_s^+, \mathbf{u}; k^+)\right]_{\mathbf{u=\mathbf{v=\mathbf{0}}}}\, ,
\end{align}
which for the homogeneous medium reads
\begin{align} \label{eq:SpectrumInInaf_a_Harmonic}
    k^+ \frac{dI_{2\tau_0<|x^+|<2\tau_f}^{in-af, |x^+|<2x_s^+}}{dk^+ dm_I^2} =
    \frac{2\alpha_s C_R}{\pi} 2 & \Re \int_{2\tau_0}^{2\tau_f} \frac{dx^+}{2p_J^+ 2\pi} e^{i \frac{m_I^2}{2p_J^+}x^+} e^{-\frac{\Delta_0}{2 \lambda_R}} \log\left| \Omega \Delta_f \cot(\Omega \Delta_f) \right|\, .
\end{align}
Again, this contribution is similar to the one in \cref{eq:SpectrumBeforeInafHarmonic}, obtained by replacing the times that the jet in the conjugated amplitude travels inside the medium and adding the color state suppression factor.

We now consider the diagrams where the gluon is radiated in the amplitude before the jet in the conjugated amplitude is created. The first diagram to consider is the one pictured in \cref{fig:NLODiagram_in_inin_b}:
\begin{align} \label{eq:SpectrumInInin_b}
    k^+ \frac{dI_{2\tau_0<|x^+|<2\tau_f}^{in-in, |x^+| > 2x_s^+}}{dk^+ dm_I^2} &  = 
    \frac{\alpha_s C_R}{(k^+)^2} 2\Re \int_{2\tau_0}^{2\tau_f} \frac{dx^+}{2p_J^+ 2\pi} e^{i \frac{m_I^2}{2p_J^+}x^+} e^{-\frac{\frac{x^+}{2}-\tau_0}{2 \lambda_R}} 2 \Re \int_{\tau_0}^{x^+/2} dx_s^+  \notag \\
    & \times \int_{x^+/2}^{\tau_f} dy_s^+\, \nabla_\mathbf{u} \cdot \nabla_\mathbf{v} \left[\int d^2\mathbf{x_0}\, \mathcal{K}\left(y_s^+, \mathbf{v}; \frac{x^+}{2}, \mathbf{x_0}; k^+\right) \right. \notag \\
    & \left. \times \tilde{\mathcal{K}}\left(\frac{x^+}{2}, \mathbf{x_0}; x_s^+, \mathbf{u}; k^+\right) -  \mathcal{K}_0(y_s^+, \mathbf{v}; x_s^+, \mathbf{u}; k^+)\right]_{\mathbf{u=\mathbf{v=\mathbf{0}}}}\, ,
\end{align}
where $\tilde{\mathcal{K}}$ is the propagator we defined in \cref{eq:qgCorrelator}, corresponding to an harmonic oscillator of frequency $\tilde{\Omega} = \Omega/\sqrt{2}$. 
Note that now we only take the real part once instead of twice, as in \cref{eq:SpectrumInInin_a}, because the ordering of $x_s^+$ and $y_s^+$ is fixed by the sign of $x^+$.
Taking the convolution between the oscillators of frequencies $\Omega$ and $\tilde{\Omega}$, it gives
\begin{align} \label{eq:SpectrumInInin_b_Harmonic}
    k^+ \frac{dI_{2\tau_0<|x^+|<2\tau_f}^{in-in, |x^+|>2x_s^+}}{dk^+ dm_I^2} & = 
    \frac{2\alpha_s C_R}{\pi}  \Re \int_{2\tau_0}^{2\tau_f} \frac{dx^+}{2p_J^+ 2\pi} e^{i \frac{m_I^2}{2p_J^+}x^+} e^{-\frac{\Delta_0}{2 \lambda_R}} \notag \\
    & \times \log\left[ \frac{\Omega}{L^+} \Delta_0 \Delta_f \left( \cot(\Omega \Delta_f) + \frac{1}{\sqrt{2}} \cot(\frac{\Omega}{\sqrt{2}} \Delta_0) \right) \right]\, .
\end{align}
The interpretation of this spectrum is less straightforward as we have the combination of two different situations: the propagation of the $qg$ dipole in a triplet state when $\tau < x^+/2$ and the propagator of the $qgqg$ quadrupole in a singlet state when $\tau > x^+/2$, which can be formally reduced to a gluon dipole for soft radiation. Again, this diagram is suppressed by the probability that the net color state of the jet in the amplitude does not change before the jet in the conjugated amplitude is created.

We now have to consider the diagram in \cref{fig:NLODiagram_in_inaf_b}:
\begin{align} \label{eq:SpectrumInInaf_b}
    k^+ \frac{dI_{2\tau_0<|x^+|<2\tau_f}^{in-af, |x^+| > 2x_s^+}}{dk^+ dm_I^2} & =
    \frac{\alpha_s C_R}{(k^+)^2}  2\Re \int_{2\tau_0}^{2\tau_f} \frac{dx^+}{2p_J^+ 2\pi} e^{i \frac{m_I^2}{2p_J^+}x^+} e^{-\frac{\frac{x^+}{2}-\tau_0}{2 \lambda_R}} 2 \Re \int_{\tau_0}^{x^+/2} dx_s^+ \int_{\tau_f}^{+\infty} dy_s^+ \notag \\
    & \hspace{-20 pt} \times \nabla_\mathbf{u} \cdot \nabla_\mathbf{v} \left[\int d^2\mathbf{x_0} d^2\mathbf{x_f}\, \mathcal{K}_0\left(y_s^+, \mathbf{v}; \tau_f, \mathbf{x_f}; k^+\right) \mathcal{K}\left(\tau_f, \mathbf{x_f}; \frac{x^+}{2}, \mathbf{x_0}; k^+\right) \right. \notag \\
    & \hspace{-20 pt} \left. \times\, \tilde{\mathcal{K}}\left(\frac{x^+}{2}, \mathbf{x_0}; x_s^+, \mathbf{u}; k^+\right) -  \mathcal{K}_0(y_s^+, \mathbf{v}; x_s^+, \mathbf{u}; k^+)\right]_{\mathbf{u=\mathbf{v=\mathbf{0}}}}\, .
\end{align}
Here, we have the convolution of three propagators: one harmonic oscillator with frequency $\Omega$, another with $\tilde{\Omega} = \Omega / \sqrt{2}$, and the vacuum one. Performing the triple convolution gives
\begin{align} \label{eq:SpectrumInInaf_b_Harmonic}
    k^+ \frac{dI_{2\tau_0<|x^+|<2\tau_f}^{in-af, |x^+|>2x_s^+}}{dk^+ dm_I^2} & = 
    \frac{2\alpha_s C_R}{\pi} \Re \int_{2\tau_0}^{2\tau_f} \frac{dx^+}{2p_J^+ 2\pi} e^{i \frac{m_I^2}{2p_J^+}x^+} e^{-\frac{\Delta_0}{2 \lambda_R}} \notag \\
    & \times \log\left[ \frac{L^+}{\Delta_f} \frac{1}{\cot(\Omega \Delta_f)} \frac{\cot(\Omega \Delta_f) \cot(\frac{\Omega}{\sqrt{2}} \Delta_0) - \sqrt{2}}{\cot(\frac{\Omega}{\sqrt{2}} \Delta_0) + \sqrt{2} \cot(\Omega \Delta_f) } \right]\, ,
\end{align}
which again shows the same suppression factor as the previous diagrams.

The last two diagrams are also included in this category. For the one in \cref{fig:NLODiagram_in_befin}, the jet function reads
\begin{align} \label{eq:SpectrumInBefin}
    k^+ \frac{dI_{2\tau_0<|x^+|<2\tau_f}^{bef-in}}{dk^+ dm_I^2} & = \frac{\alpha_s C_R}{(k^+)^2} 2\Re \int_{2\tau_0}^{2\tau_f} \frac{dx^+}{2p_J^+ 2\pi} e^{i \frac{m_I^2}{2p_J^+}x^+} e^{-\frac{\frac{x^+}{2}-\tau_0}{2 \lambda_R}} 2 \Re \int^{\tau_0}_{-x^+/2} dx_s^+ \int^{\tau_f}_{x^+/2} dy_s^+ \notag \\
    & \times \nabla_\mathbf{u} \cdot \nabla_\mathbf{v} \left[\int d^2\mathbf{x_0} d^2\mathbf{x_i}\, \mathcal{K}\left(y_s^+, \mathbf{v}; \frac{x^+}{2}, \mathbf{x_0}; k^+\right) \tilde{\mathcal{K}}\left(\frac{x^+}{2}, \mathbf{x_0}; \tau_0, \mathbf{x_i}; k^+\right) \right. \notag \\
    & \left. \times\, \mathcal{K}_0\left(\tau_0, \mathbf{x_i}; x_s^+, \mathbf{u}; k^+\right) -  \mathcal{K}_0(y_s^+, \mathbf{v}; x_s^+, \mathbf{u}; k^+)\right]_{\mathbf{u=\mathbf{v=\mathbf{0}}}}\, ,
\end{align}
which after working out the propagators reads
\begin{align} \label{eq:SpectrumInBefin_Harmonic}
    & k^+ \frac{dI_{2\tau_0<|x^+|<2\tau_f}^{bef-in}}{dk^+ dm_I^2}  = \frac{2\alpha_s C_R}{\pi} \Re \int_{2\tau_0}^{2\tau_f} \frac{dx^+}{2p_J^+ 2\pi} e^{i \frac{m_I^2}{2p_J^+}x^+} e^{-\frac{\Delta_0}{2 \lambda_R}} \notag \\
    & \hspace{10 pt} \times \log\left[ \frac{L^+ x^+}{\Sigma_f\Delta_0}  \frac{\frac{1}{\sqrt{2}}\cot(\frac{\Omega}{\sqrt{2}}\Delta_0) + \cot(\Omega \Delta_f) + \frac{\Omega}{\sqrt{2}} \Sigma_0 \left( \cot(\frac{\Omega}{\sqrt{2}}\Delta_0) \cot(\Omega \Delta_f) - \frac{1}{\sqrt{2}} \right) }{\left(\cot(\Omega\Delta_f) + \frac{1}{\sqrt{2}} \cot(\frac{\Omega}{\sqrt{2}} \Delta_0)\right) \left(1 + \frac{\Omega}{\sqrt{2}}\Sigma_0 \cot(\frac{\Omega}{\sqrt{2}}\Delta_0) \right)  }  \right]\, ,
\end{align}
where we additionally defined
\begin{align} \label{eq:TimeSumVariables}
    \Sigma_0 \equiv \frac{x^+}{2} + \tau_0 \qquad \text{and} \qquad \Sigma_f \equiv \tau_f + \frac{x^+}{2}\, .
\end{align}

Finally, we consider the contribution of \cref{fig:NLODiagram_in_befaf}:
\begin{align} \label{eq:SpectrumInBefaf}
    k^+ \frac{dI_{2\tau_0<|x^+|<2\tau_f}^{bef-af}}{dk^+ dm_I^2} & = \frac{\alpha_s C_R}{(k^+)^2} 2\Re \int_{2\tau_0}^{2\tau_f} \frac{dx^+}{2p_J^+ 2\pi} e^{i \frac{m_I^2}{2p_J^+}x^+} e^{-\frac{\frac{x^+}{2}-\tau_0}{2 \lambda_R}} 2 \Re \int^{\tau_0}_{-x^+/2} dx_s^+ \int^{+\infty}_{\tau_f} dy_s^+ \notag \\
    & \hspace{-40 pt} \times \nabla_\mathbf{u} \cdot \nabla_\mathbf{v} \left[\int d^2\mathbf{x_0} d^2\mathbf{x_i} d^2\mathbf{x_f}\, \mathcal{K}_0\left(y_s^+, \mathbf{v}; \tau_f, \mathbf{x_f}; k^+\right) \mathcal{K}\left(\tau_f, \mathbf{x_f}; \frac{x^+}{2}, \mathbf{x_0}; k^+\right)  \right. \notag \\
    & \hspace{-40 pt} \left. \times\, \tilde{\mathcal{K}}\left(\frac{x^+}{2}, \mathbf{x_0}; \tau_0, \mathbf{x_i}; k^+\right) \mathcal{K}_0\left(\tau_0, \mathbf{x_i}; x_s^+, \mathbf{u}; k^+\right) -  \mathcal{K}_0(y_s^+, \mathbf{v}; x_s^+, \mathbf{u}; k^+)\right]_{\mathbf{u=\mathbf{v=\mathbf{0}}}}\, ,
\end{align}
which can be expressed for the homogeneous medium as
\begin{align} \label{eq:SpectrumInBefaf_Harmonic}
     &k^+ \frac{dI_{2\tau_0<|x^+|<2\tau_f}^{bef-af}}{dk^+ dm_I^2} = \frac{2\alpha_s C_R}{\pi} \Re \int_{2\tau_0}^{2\tau_f} \frac{dx^+}{2p_J^+ 2\pi} e^{i \frac{m_I^2}{2p_J^+}x^+} e^{-\frac{\Delta_0}{2 \lambda_R}}  \notag \\
     & \hspace{30 pt} \times \log\left[ \frac{\Sigma_f}{L^+}\, \frac{\cot(\frac{\Omega}{\sqrt{2}} \Delta_0) + \sqrt{2} \cot(\Omega \Delta_f)}{\cot(\frac{\Omega}{\sqrt{2}} \Delta_0) \cot(\Omega \Delta_f) - \sqrt{2}} \right. \notag \\
     & \hspace{30 pt} \times \left. \frac{\frac{1}{\sqrt{2}} \cot(\frac{\Omega}{\sqrt{2}} \Delta_0) \cot(\Omega \Delta_f) - \Sigma_0 \frac{\Omega}{\sqrt{2}}  \left( \cot(\frac{\Omega}{\sqrt{2}} \Delta_0) + \frac{1}{\sqrt{2}} \cot(\Omega \Delta_f) \right)-1}{\frac{1}{\sqrt{2}}\cot(\frac{\Omega}{\sqrt{2}} \Delta_0) + \cot(\Omega \Delta_f) + \Sigma_0 \frac{\Omega}{\sqrt{2}} \left( \cot(\frac{\Omega}{\sqrt{2}} \Delta_0) \cot(\Omega \Delta_f) - \frac{1}{\sqrt{2}}\right)}  \right]\, .
\end{align}

As all the diagrams show the same prefactors, including the color state suppression, we can combine them together into a single contribution. The expression for the individual diagrams are complicated but, as it happened in \cref{eq:SpectrumBefore}, large cancellations happen between them so that the total combined expression is simpler than the contribution of the individual diagrams. They must therefore be understood as a whole and not as individual terms that contribute separately to soft gluon radiation. The total spectrum for this region of the phase space is
\begin{align} \label{eq:SpectrumIn}
     &k^+ \frac{dI_{2\tau_0<|x^+|<2\tau_f}}{dk^+ dm_I^2}  = \frac{2\alpha_s C_R}{\pi} \Re \int_{2\tau_0}^{2\tau_f} \frac{dx^+}{2p_J^+ 2\pi} e^{i \frac{m_I^2}{2p_J^+}x^+} e^{-\frac{\Delta_0}{2 \lambda_R}} \left\lbrace 2 \log\left| \cos(\Omega \Delta_f)\right| \right.  \notag \\
     & \hspace{40 pt} \left.  + \log\left[ \frac{x^+ \Omega}{\cot(\Omega \Delta_f) \left(1 + \frac{\Omega}{\sqrt{2}}\Sigma_0 \cot(\frac{\Omega}{\sqrt{2}} \Delta_0) \right)} \left[\frac{1}{\sqrt{2}} \cot(\frac{\Omega}{\sqrt{2}} \Delta_0) \cot(\Omega \Delta_f) \right. \right. \right. \notag \\
     & \hspace{40 pt} \left. \left. \left.  - \frac{\Omega}{\sqrt{2}}\Sigma_0 \left(\cot(\frac{\Omega}{\sqrt{2}} \Delta_0) + \frac{1}{\sqrt{2}} \cot(\Omega \Delta_f) \right) -1
     \right]  \right]  \right\rbrace  \, . 
\end{align}
Note that the first logarithm contains only the modulus of the argument, as the combination of the cases $y_s^+ > x_s^+$ and $x_s^+ > y_s^+$ in \cref{fig:NLODiagram_in_inin_a} and \cref{fig:NLODiagram_in_inaf_a} states that we only need to take the real part of the logarithm. However, for the remaining diagrams the ordering of $y_s^+$ and $x_s^+$ is fixed by the sign of $x^+$, and we can only take the real part after combining the logarithm with the exponential factor depending on the virtuality. Time integration in \cref{eq:SpectrumIn} does not present a closed solution in terms of analytic functions. We will solve the integral numerically and present the results in \cref{sec:NLOResults}.

\subsubsection{Jet created after the medium in one of the amplitudes}

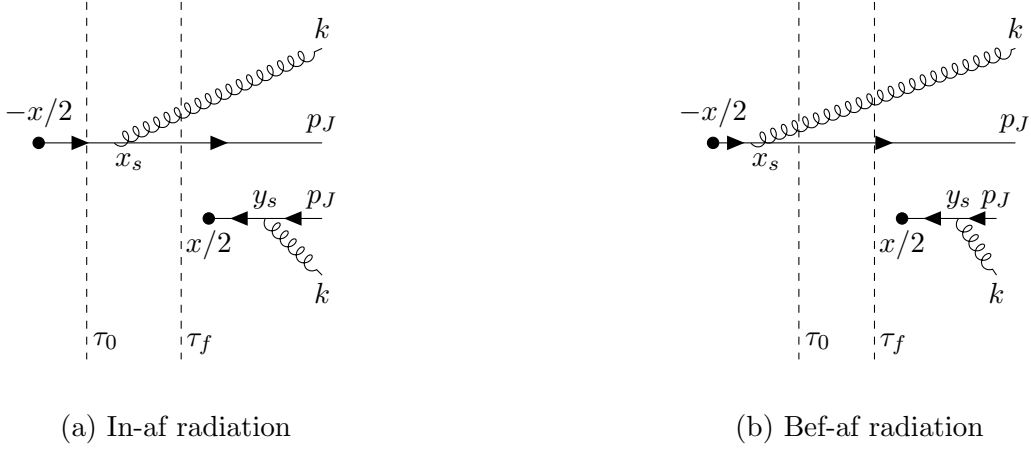
\begin{figure} [tp]
    \centering
    \begin{subfigure}[b]{0.45\textwidth}
        \centering
        \begin{tikzpicture}
  \begin{feynman}

    \node[dot, label = $-x/2$] (x0);
    \vertex[right=1 cm of x0] (x) [label={[yshift=-0.5cm, xshift=0.2cm] $x_s$}] ;
    \vertex[right=2.75 cm of x] (xf) [label={$p_J$}] ; 
    \vertex[above = 1.25 cm of xf] (yf) [label={$k$}];
    
    \draw [fermion] (x0) --  (x);
    \draw [fermion] (x) -- (xf);
    \draw [gluon] (x) -- (yf);

    \node[below=1 cm of x0] (aux);
    \node[right=2.25 cm of aux, dot, label={[yshift=-0.75 cm] $x/2$}] (x0bar);
    \vertex[right= 0.75 cm of x0bar] (xbar) [label=$y_s$];
    \vertex[right=0.75 cm of xbar] (xfbar) [label={$p_J$}] ; 
    \vertex[below = 0.75 cm of xfbar] (yfbar) [label={[yshift=-0.5cm] $k$}];

    \draw[fermion] (xbar) -- (x0bar);
    \draw[fermion] (xfbar) -- (xbar);
    \draw[gluon] (xbar) -- (yfbar);

    \node[right= -0.5 cm of x] (aux2);
    \node[above=2 cm of aux2] (aux3);
    \node[below = 3 cm of aux2] (aux4) [label={[shift={(0.25, 0)}]$\tau_0$}];
    \node[right = 1.25 cm of aux3] (aux5);
    \node[right = 1.25 cm of aux4] (aux6) [label={[shift={(0.25, -0.07)}]$\tau_f$}];

    \draw[dashed] (aux3) -- (aux4);
    \draw[dashed] (aux5) -- (aux6);

  \end{feynman}
\end{tikzpicture}
        \caption{In-af radiation}
        \label{fig:NLODiagram_after_inaf}
    \end{subfigure}
    \hfill
    \begin{subfigure}[b]{0.45\textwidth}
        \centering
        \begin{tikzpicture}
  \begin{feynman}

    \node[dot, label = $-x/2$] (x0);
    \vertex[right=0.5 cm of x0] (x) [label={[yshift=-0.5cm, xshift=0.2cm] $x_s$}] ;
    \vertex[right=3.5 cm of x] (xf) [label={$p_J$}] ; 
    \vertex[above = 1.25 cm of xf] (yf) [label={$k$}];
    
    \draw [fermion] (x0) --  (x);
    \draw [fermion] (x) -- (xf);
    \draw [gluon] (x) -- (yf);

    \node[below=1 cm of x0] (aux);
    \node[right=2.5 cm of aux, dot, label={[yshift=-0.75 cm] $x/2$}] (x0bar);
    \vertex[right= 0.75 cm of x0bar] (xbar) [label=$y_s$];
    \vertex[right=0.5 cm of xbar] (xfbar) [label={$p_J$}] ; 
    \vertex[below = 0.75 cm of xfbar] (yfbar) [label={[yshift=-0.5cm] $k$}];

    \draw[fermion] (xbar) -- (x0bar);
    \draw[fermion] (xfbar) -- (xbar);
    \draw[gluon] (xbar) -- (yfbar);

    \node[right= 0.5 cm of x] (aux2);
    \node[above=2 cm of aux2] (aux3);
    \node[below = 3 cm of aux2] (aux4) [label={[shift={(0.25, 0)}]$\tau_0$}];
    \node[right = 1 cm of aux3] (aux5);
    \node[right = 1 cm of aux4] (aux6) [label={[shift={(0.25, -0.07)}]$\tau_f$}];

    \draw[dashed] (aux3) -- (aux4);
    \draw[dashed] (aux5) -- (aux6);

  \end{feynman}
\end{tikzpicture}
        \caption{Bef-af radiation}
        \label{fig:NLODiagram_after_befaf}
    \end{subfigure}
    
\caption{\justifying Diagrams corresponding to the case when the jet is created after the medium in the conjugated amplitude $|x^+| > 2\tau_f$. The different diagrams correspond to the different situations depending on whether the gluon is radiated before (bef), inside (in), or after (af) the medium.}
\label{fig:NLODiagrams_after}
\end{figure}

Let us consider, finally, the case where the jet in either the amplitude or the conjugate amplitude is created after the medium, $2\tau_f < |x^+|$. Again, this diagrams are not present in the BDMPS-Z calculation and must vanish if we integrate over virtuality. By subtracting the vacuum contribution as explained in \cref{eq:VacuumSUbtraction}, there are only two diagrams contributing
\begin{align}
    dI_{2\tau_f < |x^+|} = dI_{2\tau_f < |x^+|}^{in-af} + dI_{2\tau_f < |x^+|}^{bef-af}\, ,
\end{align}
which are shown in \cref{fig:NLODiagrams_after}.
As it happened in the previous subsections, different convolutions of propagators appear, which are summarized in \cref{app:convolutions}.

For the diagram in \cref{fig:NLODiagram_after_inaf}, considering the cases $x^+ >0$ and $x^+<0$, the spectrum can be expressed as
\begin{align} \label{eq:SpectrumAfterInaf}
    & k^+ \frac{dI_{2\tau_f<|x^+|}^{in-af}}{dk^+ dm_I^2}  = 
    \frac{\alpha_s C_R}{(k^+)^2} 2\Re \int_{2\tau_f}^{+\infty} \frac{dx^+}{2p_J^+ 2\pi} e^{i \frac{m_I^2+i\varepsilon}{2p_J^+}x^+} \int_{\tau_0}^{\tau_f} dx_s^+ \int_{+x^+/2}^{+\infty} dy_s^+ e^{-\frac{L^+}{2\lambda_R}} \nabla_\mathbf{u} \cdot \nabla_\mathbf{v} \\
    & \hspace{20 pt} \times  \left[\int d^2 \mathbf{x_f} \mathcal{K}_0(y_s^+, \mathbf{v}; \tau_f, \mathbf{x_f}; k^+) \tilde{\mathcal{K}}(\tau_f, \mathbf{x_f}; x_s^+, \mathbf{u}; k^+) -  \mathcal{K}_0(y_s^+, \mathbf{v}; x_s^+, \mathbf{u}; k^+)\right]_{\mathbf{u=\mathbf{v=\mathbf{0}}}}\notag .
\end{align}
For the homogeneous medium, it follows that
\begin{align} \label{eq:SpectrumAfterInafHarmonic}
    k^+ \frac{dI_{2\tau_f<|x^+|}^{in-af}}{dk^+ dm_I^2} = \frac{2\alpha_s C_R}{\pi} e^{-\frac{L^+}{2 \lambda_R}} & \Re \int_{2\tau_f}^{+\infty} \frac{dx^+}{2p_J^+ 2\pi} e^{i \frac{m_I^2+i\varepsilon}{2p_J^+}x^+} \log\left[\frac{\frac{\Omega}{\sqrt{2}} \Delta_0}{\tan(\frac{\Omega} {\sqrt{2}} L^+) - \frac{\Omega}{\sqrt{2}} \Delta_f} \right]\, .
\end{align}
It contains the medium contribution to the gluon formation and the suppression factor accounting for the probability that the jet in the amplitude traverses the whole medium without finding its color state modified by the interaction.

The diagram in \cref{fig:NLODiagram_after_befaf} yields the spectrum
\begin{align} \label{eq:SpectrumAfterBefaf}
    k^+ \frac{dI_{2\tau_f<|x^+|}^{bef-af}}{dk^+ dm_I^2} & = \frac{\alpha_s C_R}{(k^+)^2}  2\Re \int_{2\tau_f}^{+\infty} \frac{dx^+}{2p_J^+ 2\pi} e^{i \frac{m_I^2+i\varepsilon}{2p_J^+}x^+} \int_{-x^+/2}^{\tau_0} dx_s^+ \int_{+x^+/2}^{+\infty} dy_s^+  \notag \\
    & \hspace{-10 pt}\times e^{-\frac{L^+}{2 \lambda_R}} \nabla_\mathbf{u} \cdot \nabla_\mathbf{v} \left[\int d^2 \mathbf{x_f}\, d^2 \mathbf{x_i}\, \mathcal{K}_0(y_s^+, \mathbf{v}; \tau_f, \mathbf{x_f}; k^+) \right. \notag \\
    & \hspace{-10 pt} \times \left. \tilde{\mathcal{K}}(\tau_f, \mathbf{x_f}; \tau_0, \mathbf{x_i}; k^+) \mathcal{K}_{0}(\tau_0, \mathbf{x_i}; x_s^+, \mathbf{u}; k^+) -  \mathcal{K}_0(y_s^+, \mathbf{v}; x_s^+, \mathbf{u}; k^+)\right]_{\mathbf{u=\mathbf{v=\mathbf{0}}}}\, ,
\end{align}
which after solving the path integrals yields
\begin{align} \label{eq:SpectrumAfterBefafHarmonic}
    k^+ & \frac{dI_{2\tau_f<|x^+|}^{bef-af}}{dk^+ dm_I^2} = 
    \frac{2\alpha_s C_R}{\pi} e^{\frac{-L^+}{2 \lambda_R}} \Re \int_{2\tau_f}^{+\infty} \frac{dx^+}{2p_J^+ 2\pi} e^{i \frac{m_I^2+i\varepsilon}{2p_J^+}x^+} \log\left[ \frac{x^+}{\Delta_0} \right. \notag \\
    & \left. \times  \left(1 - \Sigma_0 \frac{\Omega}{\sqrt{2}} \tan(\frac{\Omega}{\sqrt{2}} L^+) \right)  \frac{\tan(\frac{\Omega} {\sqrt{2}} L^+) - \frac{\Omega}{\sqrt{2}} \Delta_f}{\frac{\Omega}{\sqrt{2}} (x^+-L^+) + \left(1 + \frac{\Omega^2}{2} \Delta_f \Sigma_0 \right) \tan(\frac{\Omega}{\sqrt{2}} L^+) }\right]\, ,
\end{align}
which, as expected, contains the same suppression factor and a logarithm that accounts for the probability of emitting a gluon due to the interaction with the medium.

We can now combine the contributions of both diagrams together to obtain the total virtuality-dependent medium induced gluon radiation spectrum for the case where the jet is created after the medium in one of the amplitudes
\begin{align} \label{eq:SpectrumAfter}
    k^+ \frac{dI_{2\tau_f<|x^+|}}{dk^+ dm_I^2} & = \frac{2\alpha_s C_R}{\pi}  e^{\frac{-L^+}{2 \lambda_R}} \Re \int_{2\tau_f}^{+\infty} \frac{dx^+}{2p_J^+ 2\pi} e^{i \frac{m_I^2+i\varepsilon}{2p_J^+}x^+} \notag \\
    & \times \log\left[ \frac{ \frac{\Omega}{\sqrt{2}} x^+ \left( 1 - \Sigma_0 \frac{\Omega}{\sqrt{2}} \tan(\frac{\Omega}{\sqrt{2}} L^+) \right)}{\frac{\Omega}{\sqrt{2}} (x^+-L^+) + \left(1 + \frac{\Omega^2}{2} \Delta_f \Sigma_0 \right) \tan(\frac{\Omega}{\sqrt{2}} L^+)} \right]\, .
\end{align}
Again, there exist cancellations among the different diagrams contributing to the same region of the phase space.
This equation admits a closed representation in terms of analytic functions
\begin{align} \label{eq:SpectrumAfterIntegrated}
    & k^+ \frac{dI_{2\tau_f<|x^+|}}{dk^+ dm_I^2} = \frac{2\alpha_s C_R}{\pi}\, e^{\frac{-L^+}{2\lambda_R}} \left\lbrace \frac{\log2}{2} \left[
    \delta(m_I^2) -\frac{1}{\pi m_I^2} \sin(\frac{m_I^2}{p_J^+}\tau_f) \right] + \frac{1}{2\pi p_J^+} \Re i \frac{p_J^+}{m_I^2}   e^{i \frac{m_I^2}{p_J^+} \tau_f} \right. \notag \\
    & \hspace{20 pt} \left. \times \left[ \log\left( \frac{\tau_f \left(\tau_f + \tau_0 - \frac{1}{\Omega/\sqrt{2}\, \tan(L^+ \Omega/\sqrt{2})}\right)}{d_+ d_-} \right) + e^{-i \frac{m_I^2}{p_J^+}} E_1\left(-i \frac{m_I^2}{p_J^+} \tau_f \right) \right. \right. \notag \\
    & \hspace{20 pt} \left. \left. + e^{-i \frac{m_I^2}{p_J^+} \left(\tau_f + \tau_0 - \frac{1}{\Omega/\sqrt{2}\, \tan(L^+ \Omega/\sqrt{2})} \right) } E_1\left( -i\frac{m_I^2}{p_J^+} \left(\tau_f + \tau_0 - \frac{1}{\Omega/\sqrt{2}\, \tan(L^+ \Omega/\sqrt{2})} \right) \right) \right.\right. \notag \\
    & \hspace{20 pt} \left.\left.  - e^{i\frac{m_I^2}{p_J^+} d_+} E_1\left(i\frac{m_I^2}{p_J^+} d_+\right) - e^{i\frac{m_I^2}{p_J^+} d_-} E_1\left(i\frac{m_I^2}{p_J^+} d_-\right)
 \right] \right\rbrace\, ,
\end{align}
where $E_1$ is the exponential integral introduced in \cref{eq:ExponentialIntegral} and we have defined
\begin{align}
    d_\pm = \frac{2 - \frac{\Omega}{\sqrt{2}} (\tau_f + \tau_0) \tan(\frac{\Omega L^+}{\sqrt{2}}) \pm \sqrt{4(1+\tan^2(\frac{\Omega  L^+}{\sqrt{2}})) + \frac{\Omega^2}{2}(\tau_f + \tau_0)^2 \tan^2(\frac{\Omega L^+}{\sqrt{2}})}}{\sqrt{2}\Omega \tan(\frac{\Omega}{\sqrt{2}}L^+)}\,.
\end{align}
The analytic formula is complicated and has no simple interpretation. To access the information contained in this spectrum, it is useful to look at the numerical results which show how the jet function depends on the initial jet virtuality.
Numerical results for this contribution to the jet function will be presented in \cref{sec:NLOResults}.

\section{Numerical Results}
\label{sec:Results}

We present here the numerical results 
\footnote{Numerical code containing the results and numerical stability checks publicly available at \url{https://github.com/CLamas-r/Virtuality-dependent-jet-functions.git}} 
for the in-medium virtuality differential transverse momentum distribution at LO (as defined in \cref{sec:jetFuncLO}) and the soft gluon radiation spectrum at NLO (as defined in \cref{sec:jetFuncNLO}). 
We will focus on the dependence of the jet function on the initial jet virtuality $m_I^2$ or, equivalently, on the dimensionless ratio $\xi_J = \tau_0\, m_I^2/p_J^+$. This is similar to the ratio of the medium formation time $\tau_0$ and the jet formation time $t_J = p_J^+ / |m_I^2|$, but it is important to note that the jet virtuality, and therefore $\xi_J$, can be negative while the jet formation time is always positive. We show the results for the case where the jet is initiated by a quark, similar qualitative behavior is expected for the gluon case. Virtuality-dependent jet functions present oscillations which are more easily interpreted when the jet function is integrated over virtuality. We define the virtuality cumulative jet function as the integral over virtuality of the jet function in \cref{eq:Jk_3D} up to some cutoff $\mu^2$
\begin{align} \label{eq:VirtualityCumulantJetFunct}
    I(\mu^2) \equiv \int_{-\mu^2}^{\mu^2} dm_I^2 \frac{dI}{dm_I^2}\, .
\end{align}
This cumulative jet functions allow us to have a clear vision on which regions of the phase space contribute more to medium induced radiation and which of them are dominated by the vacuum parton shower. Results will be presented as a function of the cutoff $\mu^2$ and the dimensionless ratio $\xi_\mu = \tau_0 \mu^2/p_J^+$.

We choose a medium of a length $L^+ = 14 \fm$ which is created after a thermalization time $\tau_0 = 0.85 \fm$ (corresponding to $L=10 \fm$ and $\tau_0 = 0.6 \fm$ if we use coordinate time instead of light-cone time), traversed by a $p_J^+ = 20 \GeV$ jet. The jet quenching parameter $\hat{q}$ is usually extracted from data, and it can be estimated to be in the range $2 \lesssim \hat{q}/T^3 \lesssim 5$ \cite{Apolinario:2022vzg}. We will here pick $\hat{q}_F = 3 T^3 \approx 0.024 \GeV^3$, corresponding to $T \approx 0.2 \GeV$, a rather dilute QGP. For more dense plasmas, we expect the diagrams where the jet is created inside or after the medium in one of the amplitudes to be further suppressed, as given by the exponential factor containing $-1/\lambda_F$. For the adjoint representation, we perform Casimir scaling $\hat{q}_A = C_A/C_F\, \hat{q}_F$. Finally, we have to estimate the fundamental representation mean-free path. Using the relation between the jet quenching parameter and momentum broadening \cite{Baier:1996sk}
\begin{align}
    \hat{q}_F = \frac{\langle \mathbf{q}^2 \rangle_{coll}}{\lambda_F / \sqrt{2}}\, ,
\end{align}
and that the average transverse momentum transferred from the medium to the jet in a single collision, $\langle \mathbf{q}^2 \rangle_{coll}$, is of the order of the Debye mass with \cite{Caron-Huot:2008zna}
\begin{align}
    m_D^2 = 4\pi\alpha_s \left(1 + \frac{N_f}{6} \right) T^2 \simeq 0.23 \GeV^2\, ,
\end{align}
where $\alpha_s \approx 0.3$ and $N_f = 3$, the fundamental representation mean free path turns out to be
\begin{align}
    \lambda_F \approx 2.6 \fm \, ,
\end{align}
so that the propagating particle suffers on average five soft scatterings while traveling through the medium.

We show separately the results for the jet created before the medium in both amplitudes, inside the medium in one of them and after the medium in one of them. We also show the total combined contribution of these three regions. In the case of the cumulative jet functions we compare the results with the BDMPS-Z result for transverse momentum broadening \cite{Baier:1996sk} in the LO case and for the soft gluon emission spectrum with a delayed medium starting time \cite{Andres:2022bql} in the NLO case, to which we expect our results to converge when $\mu^2 \to \infty$. In principle, the factorized form of the heavy-ion collision cross section in \cref{eq:dsig_Fact} is only valid when $p_I^- = m_I^2/(2p_J^+) \ll p_I^+$ but we can still formally integrate our factorized jet function up to $m_I^2 \to \infty$ to recover the BDMPS-Z result. The region where medium induced radiation is relevant is expected to be precisely the region where our factorization formalism is applicable and the factorized form \cref{eq:dsig_Fact} is a valid description of the heavy-ion collision.

\subsection{LO jet function results} \label{sec:LOResults}

\begin{figure} [tp]
    \centering
    \begin{subfigure}[b]{0.49\textwidth}
        \centering
        \includegraphics[width=\linewidth]{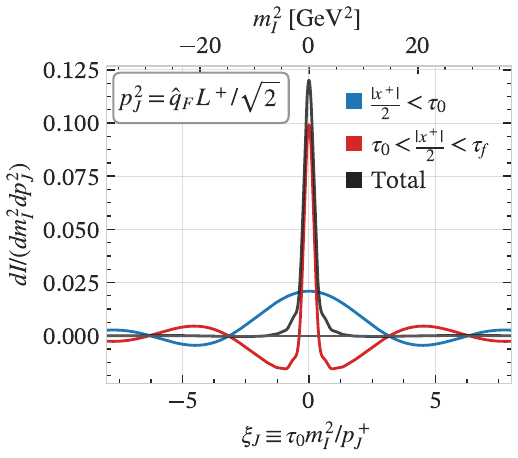}
        \caption{Virtuality differential}
        \label{fig:LOResultsDifferential}
    \end{subfigure}
    \hfill
    \begin{subfigure}[b]{0.49\textwidth}
        \centering
        \includegraphics[width=\linewidth]{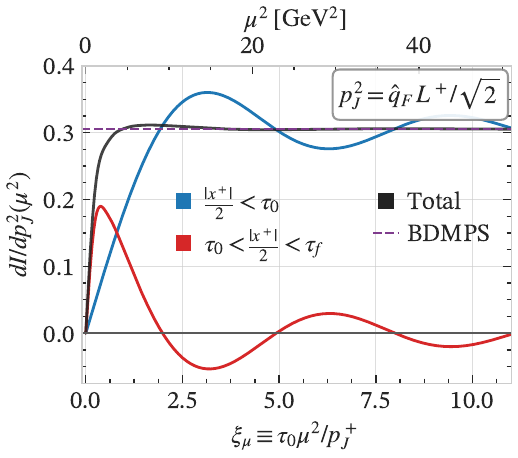}
        \caption{Virtuality cumulative}
        \label{fig:LOResultsIntegrated}
    \end{subfigure}
    
\caption{\justifying LO jet function results. The virtuality-differential jet function (left panel) and virtuality cumulative jet function (right panel) for transverse momentum broadening are presented. For each of them, the case where the jet is created before the medium in both amplitudes (blue line), and the case when the jet is created inside the medium in one of the amplitudes (red line), are isolated. The case where the jet is created after the medium in one of the amplitudes does not contribute to transverse momentum broadening. The total jet function is also shown (black line), together with the BDMPS-Z limit of the virtuality cumulant (purple dashed line).}
\label{fig:LOResults}
\end{figure}

We start presenting results for the transverse momentum broadening jet function in \cref{sec:jetFuncLO}. The total jet function only receives contributions from \cref{eq:LOBeforeResult} and \cref{eq:LOInResult}, while \cref{eq:LOAfterResult} does not contribute to transverse momentum broadening because there is no transverse momentum transfer from the medium to the jet, as given by the $\delta$-function on $p_J^2$. We fix $p_J^2 = \hat{q}_F L^+/\sqrt{2}$, parametrically corresponding to the expectation value of transverse momentum broadening and study how the different contributions to transverse momentum broadening depend on the initial jet virtuality.

Results for the LO jet function are presented in \cref{fig:LOResults}. 
Looking at \cref{fig:LOResultsDifferential}, we see how the virtuality differential jet function shows an oscillatory behavior for both the $|x^+| < 2\tau_0$ and the $2\tau_0 < |x^+| < 2\tau_f$ cases, peaking around $m_I^2$. Remarkably, the oscillations of both contributions generate a destructive interference, such that the total jet function converges to zero much faster than the individual contributions when $\tau_0/t_J > 1$, while barely oscillating. 

The interpretation of the results becomes clearer when we look to the cumulative jet function as shown in \cref{fig:LOResultsIntegrated}. We find that, if we restrict ourselves to jets with very low initial virtualities, the jet function is dominated by the case where the jet is created inside the medium in one of the amplitudes. However, as we open the phase space to include jets with higher virtualities, this contribution decreases and is rapidly overcomed by the case where the jet is created before the medium in both amplitudes. As we keep increasing $\mu^2$, the $2\tau_0 < |x^+| < 2\tau_f$ contribution vanishes as dictated by the uncertainty principle. If we do not impose any condition on the initial jet virtuality, then the jet must be created at the same point in the  amplitude and the conjugated amplitude, $\mu^2 x^+ \sim 1$, that is, before the medium is formed. Even thought this case does not contribute to the BDMPS-Z spectrum, it is of great relevance if we want to isolate jets only in a corner of the phase space, for example, by putting a cutoff on the maximum allowed virtuality, or classify them into classes according to their virtuality. The case $|x^+| < 2\tau_0$ converges in an oscillatory way to the BDMPS-Z result when $\mu^2 \to \infty$, giving the expected contribution. 

Let us now look at the total jet function. It exhibits rapid growth as we increase $\mu^2$, dominated by the contribution of $2\tau_0 < |x^+| < 2\tau_f$, but it rapidly saturates and converges to the BDMPS-Z result for $\xi_\mu \sim 1$. This is indicating that only jets with $t_J > \tau_0$ are contributing to transverse momentum broadening at LO, while the jets with $t_J < \tau_0$ do not produce an increase in the jet function. This result is consistent with expectations, as jets with a formation time smaller than the medium formation time must lose virtuality by radiation gluons before entering bulk matter and, therefore, cannot contribute to the LO spectrum where no gluon radiation is considered. Consequently, transverse momentum broadening effects are localized in the $p_I^- \ll p_I^+$ region, precisely where our factorization formula is applicable. They can therefore be studied separately from the hard vertex, where the jet is created, and then plugged into \cref{eq:dsig_Fact} to compute phenomenological observables.

It is important to note that the absence of transverse momentum broadening for high virtuality jets is a consequence of our calculation being performed at LO in $\alpha_s$. If the virtuality is high, then the quark radiates and the diagram no longer contributes at LO, but it will contribute at higher orders in the jet function. In practice, our calculation indicates that when $t_J \ll \tau_0$, the parent parton will generate a vacuum parton shower in the pre-hydrodynamic phase. This jet will then enter the medium and the different partons in the shower will interact with the nuclear matter, suffering transverse momentum broadening and medium induced energy loss.

\subsection{NLO jet function results} \label{sec:NLOResults}

We here present the results for the medium induced soft gluon radiation spectrum derived in \cref{sec:jetFuncNLO}. There are three separate contributions to this jet function, corresponding to the cases: when the jet is created before the medium in both amplitudes as in \cref{eq:SpectrumBefore}, when it is created inside the medium in one of the amplitudes as in \cref{eq:SpectrumIn}, and when it is created after the medium in one of the amplitudes as in \cref{eq:SpectrumAfter}. We study both the virtuality-differential spectrum and the virtuality cumulant up to some scale $\mu^2$. We compare the virtuality cumulant with the BDMPS-Z result \cite{Andres:2022bql}, which corresponds to the limit where we integrate over all the virtuality phase space, $\mu^2 \to \infty$. We will take the longitudinal momentum of the radiated gluon to be $k^+ = 2 \GeV$.

\begin{figure} [tp]
    \centering
    \begin{subfigure}[b]{0.49\textwidth}
        \centering
        \includegraphics[width=\linewidth]{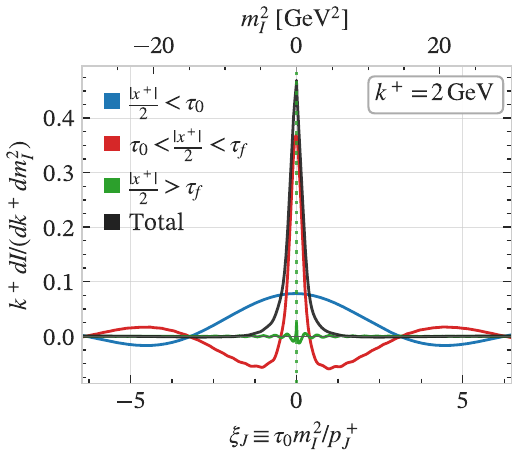}
        \caption{Virtuality differential}
        \label{fig:NLOResultsDifferential}
    \end{subfigure}
    \hfill
    \begin{subfigure}[b]{0.49\textwidth}
        \centering
        \includegraphics[width=\linewidth]{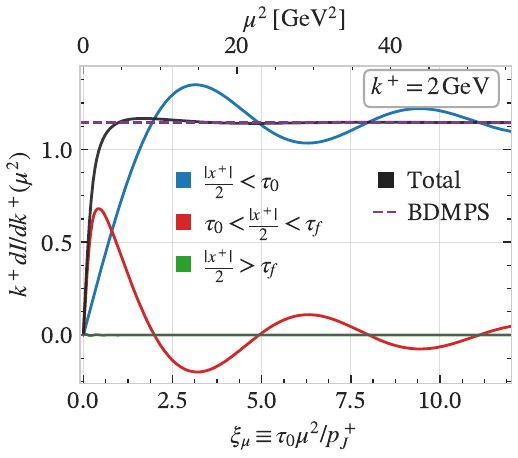}
        \caption{Virtuality cumulative}
        \label{fig:NLOResultsIntegrated}
    \end{subfigure}
    
\caption{\justifying NLO jet function results. The virtuality-differential jet function (left panel) and virtuality cumulative jet function (right-panel) for medium induced soft gluon emission are presented. For each of them, the case where the jet is created before the medium in both amplitudes (blue line), the case where the jet is created inside the medium in one of the amplitudes (red line), and the case where the jet is created after the medium in one of the amplitudes (green line) are isolated. The full $|x^+| > 2\tau_f$ result in \cref{eq:SpectrumAfter} is a distribution in $m_I^2$, it contains a delta function at $m_I^2=0$, while its regular continuum has a locally integrable logarithmic singularity there. The plot shows only the regular continuum while the delta contribution is indicated with a dashed line at $m_I^2 = 0$. The total jet function is shown (black line), together with the BDMPS-Z limit of the virtuality cumulant (purple dashed line).}
\label{fig:NLOResults}
\end{figure}

Results for the medium-induced soft gluon spectrum are shown in \cref{fig:NLOResults}, where in \cref{fig:NLOResultsDifferential} we plot the virtuality differential jet function and in \cref{fig:NLOResultsIntegrated} the virtuality cumulant, integrated as in \cref{eq:VirtualityCumulantJetFunct}. In the first place, we find that the case $|x^+| > 2\tau_f$ almost does not contribute to the total spectrum, as it is exponentially suppressed by the medium length. We can then conclude that the probability that a particle travels through the whole medium without getting its color state modified is negligible. For the cases $|x^+| < 2 \tau_0$ and $2\tau_0 < |x^+| < 2\tau_f$, the results are very similar to the LO case in \cref{sec:LOResults}. If we restrict ourselves to the low virtuality region of the phase space, $t_J > \tau_0$, the case where the jet is created inside the medium in one of the amplitudes will be dominating. However, as we open the phase space and allow for higher virtuality jets, the creation point of the jet is very localized and the contribution of this kind of diagram vanishes. The cumulative jet function for the case where the jet is created before the medium in both amplitudes converges to the BDMPS-Z result in an oscillatory way when $\mu^2 \to \infty$. A large part of these oscillations is canceled by the oscillatory behavior of the case $2\tau_0 < |x^+| < 2\tau_f$, so that the total jet function converges to the BDMPS-Z result much faster that the $|x^+| < 2 \tau_0$ case alone.

Remarkably, the total cumulative soft-gluon emission spectrum converges to the BDMPS-Z result already when $t_J \sim \tau_0$ and no longer grows or decreases significantly as we increase the available virtuality phase space. This indicates that the region where the jet formation time is shorter than the medium formation time does not contribute to the medium induced radiation spectrum at $\mathcal{O}(\alpha_s)$ and is dominated by a vacuum-like parton shower. This is an expected result, as in this case the high-virtuality quark cannot live long enough to enter the medium and must radiate vacuum gluons to lose part of its virtuality. On the contrary, the region where $t_J > \tau_0$ is populated by medium induced radiation, as in this case the jet can enter the medium without radiating and will be modified by the bulk. As we discussed in \cref{sec:LOResults}, this low virtuality region is precisely the region of the phase space which allows us to factorize the jet function from the hard process cross section in the impact-parameter-dependent cross section. Therefore, for the parameters considered here, the factorized formula in \cref{eq:dsig_Fact} can be used to compute jet quenching experimental observables, as the medium-induced gluon radiation spectrum receives its main contributions from the region $p_I^- \ll p_I^+$, where factorization works.

Again, it is important to remark that these results do not imply that high virtuality jets are not modified by the medium, but these modifications will appear at higher orders in $\alpha_s$. When $t_J \ll \tau_0$, the jet will radiate a gluon before entering the medium and then at $\mathcal{O}(\alpha_s)$ no more splittings are allowed. Consequently, medium induced radiation cannot appear at this order. At higher orders, the quark and the partons radiated during the pre-equilibrium phase are expected to subsequently enter the bulk and interact with the medium, potentially undergoing momentum broadening and medium-induced energy loss. The present NLO result therefore supports a physical picture in which high-virtuality jets first develop through vacuum-like radiation before their lower-virtuality constituents become sensitive to the medium. It should not be interpreted as implying the absence of medium-induced radiation from high-virtuality jets in an all-orders treatment.

\begin{figure} [tp]
    \centering
    \begin{subfigure}[b]{0.49\textwidth}
        \centering
        \includegraphics[width=\linewidth]{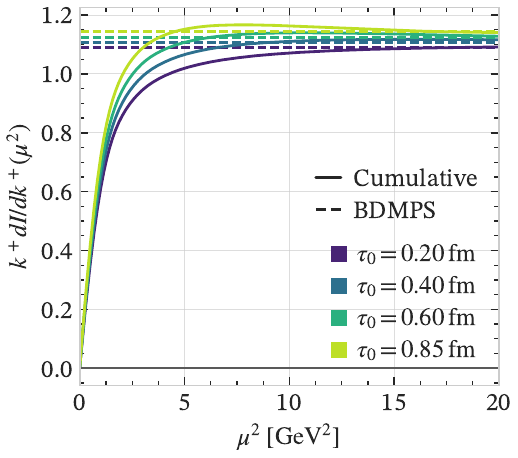}
        \caption{Dimensionful cutoff}
        \label{fig:tau0ScanDimful}
    \end{subfigure}
    \hfill
    \begin{subfigure}[b]{0.49\textwidth}
        \centering
        \includegraphics[width=\linewidth]{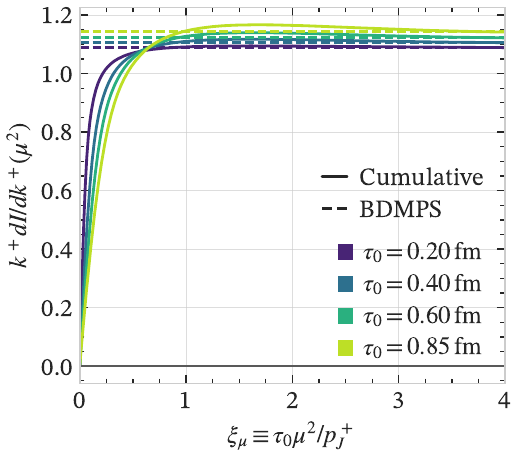}
        \caption{Dimensionless cutoff}
        \label{fig:tau0ScaneDimless}
    \end{subfigure}
    
\caption{\justifying Dependence of the NLO cumulative jet function with the medium formation time $\tau_0$. Results are shown for four different values of $\tau_0$, as a function of the dimensionful cutoff $\mu^2$ (left panel) and the dimensionless variable $\xi_\mu \equiv \tau_0 \mu^2 /p_J^+$. For each value of $\tau_0$ the corresponding BDMPS-Z result is also shown for comparison.}
\label{fig:tau0Scan}
\end{figure}

We can also study how the medium induced radiation spectrum depends on the QGP formation time $\tau_0$. Results for the total cumulative jet function are shown in \cref{fig:tau0Scan} for four different values of $\tau_0$. In \cref{fig:tau0ScanDimful}, we show the results as a function of the virtuality cutoff $\mu^2$, and in \cref{fig:tau0ScaneDimless}, as a function of the dimensionless variable $\xi_\mu$. We see how, for shorter $\tau_0$, the convergence to the BDMPS-Z result happens at larger values of $\mu^2$. This is expected because the condition for a parton to enter the medium before radiating, therefore contributing to the NLO spectrum, is that $t_J \sim p_J^+/|m_I^2| > \tau_0$. For smaller $\tau_0$, this condition is satisfied for larger values of $m_I^2$: if the medium is created rapidly, then the particle lives in vacuum for a very short time and its virtuality can be high. On the other hand, if the medium takes a long time to thermalize, the particle has to travel large distances through vacuum so it will radiate if the virtuality is not large. This explanation is also supported by \cref{fig:tau0ScaneDimless}, where we see that all the lines have reached approximately the BDMPS-Z value when $\tau_0 \mu^2 / p_J^+ \sim 1$, indicating that jets with $t_J < \tau_0$ are not contributing to the NLO spectrum and this region is dominated by the vacuum parton shower, independent of the value of $\tau_0$.

We now study what happens when we consider that the medium is created immediately after the collision, so that $\tau_0 = 0 \fm$. In this limit, the pre-equilibrium vacuum propagation disappears so we no longer have diagrams where the jet is created before the medium in both amplitudes. The case where the jet is created inside the medium in one of the amplitudes reduces to the BDMPS-Z result when we integrate over virtuality. The contribution of the term where the jet is created after the medium in one of the amplitudes is very small, as it was shown in \cref{fig:NLOResults}, because it is suppressed by $e^{-L^+/(2\lambda_F)}$ to preserve color neutrality, so the total contribution matches almost exactly the one given by the term where the jet is created inside the medium in one of the amplitudes. 

\begin{figure} [tp]
    \centering
    \begin{subfigure}[b]{0.49\textwidth}
        \centering
        \includegraphics[width=\linewidth]{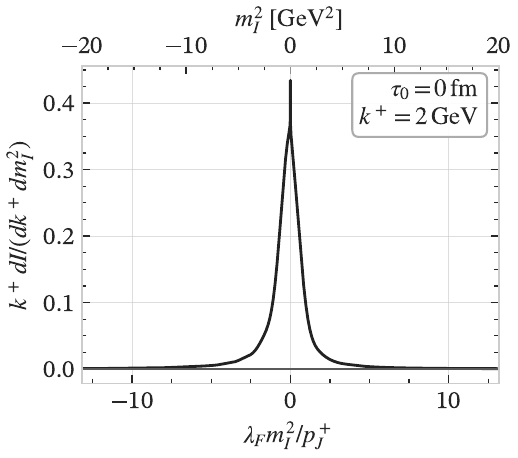}
        \caption{Virtuality differential}
        \label{fig:InmediateMediumDifferential}
    \end{subfigure}
    \hfill
    \begin{subfigure}[b]{0.49\textwidth}
        \centering
        \includegraphics[width=\linewidth]{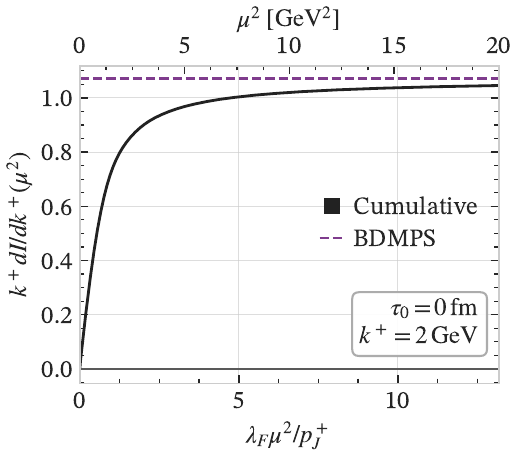}
        \caption{Virtuality cumulative}
        \label{fig:InmediateMediumCumulative}
    \end{subfigure}
    
\caption{\justifying NLO jet function for the case $\tau_0 = 0 \fm$. The virtuality differential (left panel) and virtuality cumulative (right panel) jet functions are shown. Results are presented as a function of virtuality $m_I^2$ (or the virtuality cutoff $\mu^2$) and the dimensionless ratio $\lambda_F m_I^2/p_J^+$ (or $\lambda_F \mu^2/p_J^+$). The corresponding BDMPS-Z result is also shown in the right panel for comparison.}
\label{fig:InmediateMedium}
\end{figure}

Results for the $\tau_0 = 0 \fm$ are shown in \cref{fig:InmediateMedium}, where in \cref{fig:InmediateMediumDifferential} we show the virtuality differential spectrum and in \cref{fig:InmediateMediumCumulative} the cumulative. At first sight, it can be appreciated how the BDMPS-Z spectrum contains contributions up to much larger values of the virtuality than before. This is expected because we no longer have a pre-thermalization vacuum propagation, so even high virtuality jets will enter the medium before radiating and they can therefore suffer medium induced radiation at $\mathcal{O}(\alpha_s)$. The relation between $t_J = p_J^+/|m_I^2|$ and $\lambda_F$ will determine if the dominant process is vacuum radiation or medium induced radiation. In an all-orders description, the ratio $t_J/\lambda_F$ provides an estimate of the number of soft scatterings that can occur over the characteristic vacuum formation time. If $t_J\gg\lambda_F$, the parton can undergo several interactions with the medium during this time, and medium effects are expected to become increasingly important. Conversely, if $t_J\ll\lambda_F$, vacuum-like evolution occurs on a timescale shorter than the mean time between successive soft scatterings. Nevertheless, the ratio $t_J/\lambda_F$ alone does not fully determine the relative importance of vacuum-like and medium-induced radiation, which also depends on their corresponding formation times and splitting rates.

\begin{figure} [tp]
    \centering
    \begin{subfigure}[b]{0.49\textwidth}
        \centering
        \includegraphics[width=\linewidth]{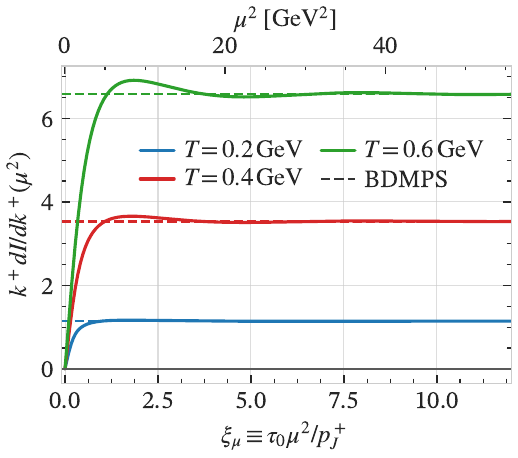}
        \caption{Cumulative jet function}
        \label{fig:TemperatureDependenceTotal}
    \end{subfigure}
    \hfill
    \begin{subfigure}[b]{0.49\textwidth}
        \centering
        \includegraphics[width=\linewidth]{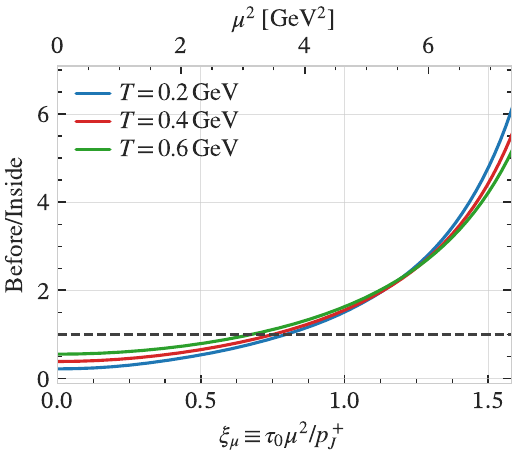}
        \caption{Before/inside ratio}
        \label{fig:TemperatureDependenceRatio}
    \end{subfigure}
    
\caption{\justifying Temperature dependence of the jet function. In the left panel, the total cumulative jet function is shown as a function of the virtuality cutoff $\mu^2$ for three different values of the temperature, together with the corresponding BDMPS-Z results (dashed lines) for comparison. In the right panel, the ratio between the case where the jet is created before the medium in both amplitudes, $|x^+| < 2\tau_0$, and the case where it is created inside the medium in one of the amplitudes, $2\tau_0 < |x^+| < 2\tau_f$, is shown as a function of the cutoff $\mu^2$ for the same three different temperatures. The dashed line in the right panel indicates the value Before/Inside $=1$, separating the regions where the inside case dominates from those where the before case does.}
\label{fig:TemperatureDependence}
\end{figure}

At the end, we study the dependence of the virtuality cumulative jet function with the medium temperature. In \cref{fig:TemperatureDependenceTotal}, we show the total cumulative jet function for different values of the temperature, with the dependence of $\hat{q}_R$ and $\lambda_R$ on the temperature explained at the beginning of \cref{sec:Results}. As expected, the medium-induced radiation spectrum increases with temperature, as does the jet quenching parameter $\hat{q}_R$. We find the convergence to the BDMPS-Z result to be almost independent of temperature, first reaching the asymptotic value for $\xi_\mu \sim 1$ and then slightly oscillating around this value. We then conclude that jets with $t_J < \tau_0$ do not contribute significantly to the NLO spectrum, independently of the temperature, and the virtuality dependence of the jet function does not depend strongly on the medium characteristic scale $T$ but only on the vacuum propagation time. This is one of the limitations of the BDMPS-Z formalism, where $\delta(x^+)$ in the field correlator of \cref{eq:field averages} causes the $p^-$ exchange between the medium and the jet not to be regulated by the in-medium scales. For a more realistic model where the medium potential not only depends on $\mathbf{p}$ but also on $p^-$, we expect this virtuality dependence of the jet function to change with the medium temperature or, equivalently, with the jet quenching parameter.

In \cref{fig:TemperatureDependenceRatio}, we study how the relative importance of the different contributions to the total jet function changes with temperature. In \cref{fig:NLOResultsIntegrated} we showed the contribution of the case where the jet is created after the medium in one of the amplitudes to be largely suppressed, so we present here results for the ratio between the $|x^+| < 2\tau_0$ and the $2\tau_0 < |x^+| < 2\tau_f$ contributions. Remarkably, despite the fact that the $\exp(-\Delta t/(2\lambda_F))$ suppression factor increases with temperature, we do not find a strong dependence of the ratio with $T$. This result suggests that most of the contribution of the case where the jet is created inside the medium in one of the amplitudes comes from the region $|x^+| \sim 2\tau_0$, while larger values of $|x^+|$ are suppressed even for small temperatures. In fact, for small $\mu^2$, such that $\Delta x^+$ is large and the jets can be created well inside the medium, the ``Inside" contribution is more dominant for lower temperatures, as the region $|x^+| > 2\tau_0$ will be less suppressed by the color neutrality factor. However, as we increase $\mu^2$, the uncertainty principle automatically selects jets with $|x^+| \lesssim 2\tau_0$, so that the region that is very suppressed for large $T$ and less suppressed for small $T$ does no longer contribute and the `Inside" contribution becomes more relevant for higher temperatures.

The results in this section motivate the following qualitative picture of jet evolution in a heavy-ion collision. Initially, a high-energy, high-virtuality parton propagates through vacuum or through the pre-equilibrium stages and develops predominantly through vacuum-like radiation. This evolution may be accompanied by medium-induced radiation if pre-equilibrium nuclear matter is present. As the jet propagates, its constituents lose virtuality through both vacuum-like emissions and interactions with the background medium. At sufficiently low virtuality, medium-induced processes are expected to become increasingly important, although vacuum-like splittings remain present.

\section{Conclusions and outlook} \label{sec:Conclusions}

In this work we have taken the very first step towards the construction of a factorization formalism to coherently describe the virtuality evolution of jets propagating through a QCD medium. Using the Glauber model to describe the colliding nuclei, we derived a factorized expression for the impact-parameter-dependent cross section in heavy-ion collisions, where the hard cross section that creates the jet is separated from the jet function that contains the all the information about the evolution of the jet inside the medium. A distinctive feature of this construction is that the jet function remains differential in the virtuality of the parton emerging from the hard interaction, $m_I^2$. This dependence provides access to information that is integrated out in the standard BDMPS-Z formalism and makes it possible to investigate which regions of the virtuality phase space are more relevant for medium induced radiation.

We used the BDMPS-Z formalism to evaluate the jet function at fixed order in $\alpha_s$, as a function of the initial jet virtuality $m_I^2$. At leading order we computed the virtuality-differential transverse momentum broadening distribution for a high-energy parton. At next-to-leading order we compute the corresponding medium-induced soft-gluon spectrum. Remarkably, due to the uncertainty principle, $\Delta x^+ \Delta p_I^- \sim 1$, if virtuality is not integrated over, the jet does not need to be created at the same point in the amplitude and the conjugate amplitude. Therefore, even if the hard process happens on average at a time $X^+ = 0$, before the medium is created, the jet in one of the amplitudes can still be created inside the medium or even after the medium. We studied separately the contributions of these regions and then combined them together to obtain the total jet function.

To study the relation between our factorized formula and the BDMPS-Z formalism, we introduced a virtuality cumulant jet function by integrating the virtuality differential jet function up to some cutoff $\mu^2$ (see \cref{eq:VirtualityCumulantJetFunct}). For a medium formed at a time $\tau_0$, the total cumulant approaches the BDMPS-Z result when
\begin{align}
    \xi_\mu = \tau_0 \mu^2 /p_J^+ \gtrsim 1\, ,
\end{align}
both at LO and NLO, indicating that the integrated jet function receives its main contribution from the $t_J > \tau_0$ region of the phase space, with $t_J = p_J^+/|m_I^2|$ the jet formation time. At LO the interpretation is that, when $t_J < \tau_0$, the quark can not live long enough to enter medium and must radiate gluons during the vacuum pre-equilibrium propagation. Therefore, this diagram does not contribute to the spectrum at $\mathcal{O}(\alpha_s^0)$. The interpretation is similar at NLO: the high virtuality jet radiates before entering the medium, and then subsequent medium induced radiation will appear at higher orders in $\alpha_s$. 

These fixed-order results motivate the expectation that, in an all-orders calculation, a high-virtuality jet with $t_J \ll \tau_0$ will initially undergo a vacuum like evolution before the medium is formed, which can be resummed in a parton shower. The partons in the shower will then enter the medium and will experience transverse momentum broadening and energy loss. If instead of having vacuum propagation we consider that the jet propagates through and interacts with pre-equilibrium matter in the early stages, the vacuum parton-shower needs to be combined with the medium-induced energy loss in the pre-equilibrium phase. The ratio $t_J/\lambda_F$ gives a qualitative estimate of how many soft interactions can occur over the characteristic vacuum formation time. Nevertheless, this ratio alone does not determine the relative importance of vacuum-like and medium-induced radiation, which in an all-orders description will also depend on their respective formation times and splitting rates.

We also studied the dependence of the virtuality cumulative jet function on the thermalization time of the medium $\tau_0$, finding that a shorter thermalization time allows for higher virtuality jets to enter the medium without radiating as expected. The shape of the spectrum as a function of the dimensionless variable $\xi_\mu \equiv \tau_0 \mu^2 / p_J^+$ is not significantly sensitive to the value of $\tau_0$, indicating that the relevant variable is the ratio between $t_J$ and $\tau_0$. We studied in more detail the case $\tau_0 = 0 \fm$, where the medium is created immediately after the collision. In this case the contribution associated to the creation before the medium disappears, and the term in which the jet is created inside the medium in one of the amplitudes gives the BDMPS-Z result after $m_I^2$ integration. The spectrum receives contributions from a wider virtuality range due to the absence of vacuum-propagation interval, so the jet in one of the amplitudes will always enter the medium before radiating.

Finally, the temperature dependence of the virtuality cumulant was studied, which changes both the jet quenching parameter and the mean free path. The normalization of the medium-induced spectrum increases with temperature, as given by the increase of the jet quenching parameter, whereas the virtuality scale at which the asymptotic BDMPS-Z result is reached remains approximately $\xi_\mu \sim 1$ independently of the temperature. This behavior exposes an important limitation of the BDMPS-Z model when it is used to study virtuality-dependent quantities.

In the BDMPS-Z formalism the medium is represented by a stochastic background field whose correlations are local in light-cone time, as in \cref{eq:field averages}. The factor $\delta(x^+-y^+)$ has a Fourier transform that is independent of the conjugate momentum $q^-$. Consequently, the medium correlator has no characteristic scale that sets the typical $q^-$ exchange between the jet and the medium. Although the BDMPS-Z formalism successfully describes transverse momentum broadening and medium-induced radiation within its domain of validity, this approximation may prevent it from providing a dynamical description of the virtuality evolution of partons inside QCD matter.

The initial-virtuality dependence obtained in this work originates instead from the phase relating $m_I^2$ to the relative creation point of the jet in the amplitude and conjugate amplitude. It allows us to determine which initial virtualities contribute to a fixed-order medium-induced process, but it does not yet describe the continuous change of the parton virtuality along its path through the medium. Following the virtuality dynamically requires a more general medium correlator with finite support in $\Delta x^+ = x^+ - y^+$, and therefore a nontrivial dependence on $q^-$. Combined with an all-orders treatment of the multiple emissions, this extension could ultimately allow the derivation of an evolution kernel in both light-cone time and virtuality.

The present calculation therefore presents a formalism that can be systematically improved to consistently construct medium-induced parton showers, providing a systematic framework to justify and distinguish the different phenomenological treatments of vacuum-like and medium-induced radiation in QCD media, such as those in \cite{Schenke:2009gb, Ovanesyan:2011xy, Mehtar-Tani:2011hma, Zapp:2012ak, Casalderrey-Solana:2014bpa, He:2015pra, Cao:2017qpx, Caucal:2018dla, Putschke:2019yrg, Duan:2026nvr}. The combination of the virtuality dependence derived here with a medium model that resolves light-cone energy exchange could provide a systematic connection between the hard production process, the early high-virtuality vacuum-like parton shower and the medium-modified evolution. Additionally, the jet functions computed here can be plugged back into \cref{eq:dsig_Fact} to compute physical observables in phenomenological calculations.

\section*{Acknowledgments}

This work is supported by the European Research Council under project ERC-2018-ADG-835105 YoctoLHC; by Maria de Maeztu excellence unit grant CEX2023-001318-M and project PID2023-152762NB-I00 funded by MICIU/AEI/10.13039/501100011033; and by ERDF/EU. It has received funding from Xunta de Galicia (CIGUS Network of Research Centres). C.L. and B.W. are also supported by Xunta de Galicia under the ED431F 2023/10 project. In addition,  C.L. also acknowledges support from the Ministerio de Ciencia e Innovación through the predoctoral fellowship PRE2022-102748 funded by MCIN/AEI/10.13039/501100011033; and B.W. acknowledges the support of the Ram\'{o}n y Cajal program with the Grant No. RYC2021-032271-I.

\appendix

\section{Convolutions of propagators} \label{app:convolutions}

In the NLO calculation convolutions of propagators in different regions of the phase space appear frequently. We here present the results for this integrations that are used in the calculation, within the harmonic oscillator approximation for an homogeneous medium. All this convolutions can be expressed in a common form as
\begin{align}\label{eq:GeneralConvolution}
    \mathcal{K}_T(t_2, \mathbf{x}_2; t_1, \mathbf{x}_1; k^+) = \frac{k^+}{2\pi i \beta} \exp{i \frac{k^+}{2\beta} (\gamma\, \mathbf{x}_2^2 + \alpha\, \mathbf{x}_1^2 - 2 \mathbf{x}_1 \cdot \mathbf{x}_2)} \, .
\end{align}
The value of the coefficients $\alpha$, $\beta$ and $\gamma$ will depend on the specific case we consider.

Let us start considering the convolutions that appear in the case where the jet is created before the medium in both amplitudes. For the inside-inside case only one propagator appears so there are no convolutions. For the inside-after and before-inside cases one has the convolution of a vacuum propagator and a dressed propagator
\begin{align} \label{eq:ConvolutionBeforeInaf}
    \mathcal{K}_{in-af}(t_2, \mathbf{x_2}; t_1; \mathbf{x_1}; k^+) & \equiv \int d^2 \mathbf{x_a} \mathcal{K}_0(t_2, \mathbf{x_2}; t_a, \mathbf{x_a}; k^+) \mathcal{K}(t_a, \mathbf{x_a}; t_1, \mathbf{x_1}; k^+)\, ,
\end{align}
which in both cases can be obtain from \cref{eq:GeneralConvolution} with
\begin{align} \label{eq:ParametersBeferoeInaf}
    \alpha = c - \Delta_2, \qquad \beta = \frac{s}{\Omega} + \Delta_2 c, \qquad \gamma = c\, ,
\end{align}
where we defined
\begin{align} \label{eq:IntervalsBeferoeInaf}
    \Delta_1 \equiv t_a - t_1, & \qquad \Delta_2 \equiv t_2 - t_a \qquad \text{and} \qquad
    c \equiv \cos(\Omega \Delta_1),  \qquad s \equiv \sin(\Omega \Delta_1)\, .
\end{align}
In the case of the before-after contribution, we have the convolution of the dressed propagator with two vacuum propagators
\begin{align} \label{eq:ConvolutionBeforeBefaf}
    \mathcal{K}_{bef-af} & (t_2, \mathbf{x_2}; t_1; \mathbf{x_1}; k^+) \notag \\
    & \equiv \int d^2 \mathbf{x_a} d^2\mathbf{x_b} \mathcal{K}_0(t_2, \mathbf{x_2}; t_a, \mathbf{x_a}; k^+) \mathcal{K}(t_a, \mathbf{x_a}; t_b, \mathbf{x_b}; k^+) \mathcal{K}_0(t_b, \mathbf{x_b}; t_1, \mathbf{x_1}; k^+) \, ,
\end{align}
where the gaussian parameters now are
\begin{align} \label{eq:ParametersBeferoeBefaf}
    \alpha = c - \Delta_2 \Omega s, \qquad \beta = \frac{s}{\Omega} + (\Delta_2 + \Delta_1) c - \Delta_1 \Delta_2 \Omega s, \qquad \gamma = c - \Delta_1\Omega s\, ,
\end{align}
together with the definitions
\begin{align} \label{eq:IntervalsBeferoeBefaf}
    \Delta_1 \equiv t_b - t_1, & \quad \Delta_2 \equiv t_2 - t_a, \quad L^+ = t_a - t_b \quad \text{and} \quad
    c \equiv \cos(\Omega L^+),  \quad s \equiv \sin(\Omega L^+)\, .
\end{align}

We now move to the case where the jet is created inside the medium in one of the amplitudes. We now have a new type of propagator appearing the convolutions, corresponding to an harmonic oscillator of frequency $\widetilde{\Omega} = \Omega/\sqrt{2}$. For the diagrams with $|x^+| < x_s^+$ no new convolutions appear, for the inside-inside case only one propagator contributes and for the inside-after case the convolution is given by \cref{eq:ConvolutionBeforeInaf}. The situation changes when the radiation in the amplitude happens before the jet in the conjugated amplitude s created, $|x^+| > x_s^+$. In this case, for the inside-inside diagram,
\begin{align} \label{eq:ConvolutionsInsideInin}
    \widetilde{\mathcal{K}}_{in-in}(t_2, \mathbf{x_2}; t_1; \mathbf{x_1}; k^+) & \equiv \int d^2 \mathbf{x_a} \mathcal{K}(t_2, \mathbf{x_2}; t_a, \mathbf{x_a}; k^+) \widetilde{\mathcal{K}}(t_a, \mathbf{x_a}; t_1, \mathbf{x_1}; k^+)\, ,
\end{align}
we have
\begin{align} \label{eq:ParametersInsideInin}
    \alpha = c_1 c_2 - \frac{\widetilde{\Omega}}{\Omega} s_1 s_2, \qquad \beta = \frac{c_2 s_1}{\widetilde{\Omega}} + \frac{c_1 s_2}{\Omega}, \qquad \gamma = c_1 c_2 - \frac{\Omega}{\widetilde{\Omega}} s_1 s_2\, ,
\end{align}
with the same $\Delta_1$ and $\Delta_2$ as in \cref{eq:IntervalsBeferoeInaf} and 
\begin{align} \label{eq:IntervalsInsideInIn}
    c_1 \equiv \cos(\widetilde{\Omega} \Delta_1),  \qquad s_1 \equiv \sin(\widetilde{\Omega} \Delta_1), \qquad c_2 \equiv \cos(\Omega \Delta_2), \qquad s_2 \equiv \sin(\Omega \Delta_2)\, .
\end{align}
For the in-after case we have to include also a vacuum propagator
\begin{align} \label{eq:ConvolutionInsideInaf}
    \widetilde{\mathcal{K}}_{in-af}& (t_2, \mathbf{x_2}; t_1; \mathbf{x_1}; k^+) \notag \\
    & \equiv \int d^2 \mathbf{x_a} d^2\mathbf{x_b}\, \mathcal{K}_0(t_2, \mathbf{x_2}; t_a, \mathbf{x_a}; k^+) \mathcal{K}(t_a, \mathbf{x_a}; t_b, \mathbf{x_b}; k^+) \widetilde{\mathcal{K}}(t_b, \mathbf{x_b}; t_1, \mathbf{x_1}; k^+) \, ,
\end{align}
where now
\begin{align} \label{eq:ParametersInsideInaf}
    \alpha & = c_1 c_2 - \frac{\widetilde{\Omega}}{\Omega} s_1 s_2 - b(\Omega c_1 s_2 + \widetilde{\Omega} c_2 s_1), \qquad \beta = \frac{c_2 s_1}{\widetilde{\Omega}} + \frac{c_1 s_2}{\Omega} + b \left( c_1 c_2 - \frac{\Omega}{\widetilde{\Omega}} s_1 s_2 \right), \notag \\
    \gamma & = c_1 c_2  - \frac{\Omega}{\widetilde{\Omega}} s_1 s_2\, ,
\end{align}
with
\begin{align} \label{eq:IntervalsInsideInAf}
    \Delta_1 = t_b - t_1, \qquad \Delta_2 = t_a - t_b, \qquad b = t_2 - t_a \, ,
\end{align}
and the same definitions for the sines and the cosines as in \cref{eq:ParametersInsideInin}. Using symmetry arguments it is easy to show that the before-inside case is equivalent to the inside after case replacing $\Omega \leftrightarrow \widetilde{\Omega}$ and $\Delta_1 \leftrightarrow \Delta_2$. Finally, we must consider the before--after term, which contains the
convolution of four propagators
\begin{align} \label{eq:ConvolutionInsideBefaf}
\widetilde{\mathcal{K}}_{\mathrm{bef-af}}(t_2,\mathbf{x}_2;t_1 & ,\mathbf{x}_1; k^+)  \equiv
\int d^2\mathbf{x}_a\, d^2\mathbf{x}_b\, d^2\mathbf{x}_c\, \mathcal{K}_0 \left(t_2,\mathbf{x}_2; t_a,\mathbf{x}_a; k^+ \right) \notag\\
& \times \mathcal{K} \left(t_a,\mathbf{x}_a; t_b,\mathbf{x}_b; k^+ \right) \widetilde{\mathcal{K}} \left(t_b,\mathbf{x}_b; t_c,\mathbf{x}_c; k^+ \right) \mathcal{K}_0 \left(t_c,\mathbf{x}_c; t_1,\mathbf{x}_1; k^+\right)\, .
\end{align}
The parameters of the resulting gaussian now read
\begin{align} \label{eq:ParametersInsideBefaf}
    \alpha & = c_1 c_2 - \frac{\widetilde{\Omega}}{\Omega} s_1 s_2 - b(\Omega c_1 s_2 + \widetilde{\Omega} c_2 s_1)\, , \notag \\
    \beta & = \frac{c_2 s_1}{\widetilde{\Omega}} + \frac{c_1 s_2}{\Omega} + b \left( c_1 c_2 - \frac{\Omega}{\widetilde{\Omega}} s_1 s_2 \right) + a \left( c_1 c_2 - \frac{\widetilde{\Omega}}{\Omega} s_1 s_2 \right), \notag \\
    \gamma & = c_1 c_2  - \frac{\Omega}{\widetilde{\Omega}} s_1 s_2 - a(\Omega c_1 s_2 + \widetilde{\Omega} c_2 s_1)\, ,
\end{align}
and
\begin{align} \label{eq:IntervalsInsideInAf}
    a = t_c - t_1, \qquad \Delta_1 = t_b - t_c, \qquad \Delta_2 = t_a - t_b, \qquad b = t_2 - t_a \, ,
\end{align}
and the sines and cosines are those defined in \cref{eq:ParametersInsideInin}.

Finally, we have to consider the case where the jet in the conjugated amplitude is created outside the medium. In this case, the in-after convolution is identical to that in \cref{eq:ConvolutionBeforeInaf} and the before-after convolution identical to that in \cref{eq:ConvolutionBeforeBefaf} after replacing $\Omega \to \widetilde{\Omega}$.

\bibliographystyle{JHEP}
\bibliography{jets.bib}
\end{document}